\documentclass[longauth]{aa}

\usepackage{graphicx}
\usepackage{natbib}
\usepackage{scalerel}

\usepackage[table]{xcolor}

\usepackage{txfonts}
\usepackage[pdfencoding=auto,psdextra]{hyperref}
\hypersetup{
    colorlinks=true,
    linkcolor=blue,
    filecolor=magenta,      
    urlcolor=blue,
    citecolor=blue
}
\makeatletter
\renewcommand*\aa@pageof{, page \thepage{} of \pageref*{LastPage}}
\makeatother

\usepackage[utf8]{inputenc}
\usepackage{ulem}
\usepackage[switch, modulo]{lineno}

\usepackage{euclid}
\usepackage{booktabs}
\usepackage{multirow}
\newcommand{\dz}{\mathrm{d}z}

\def\l{\langle}
\def\r{\rangle}
\def\np{n_{\rm p}}
\def\wsp{w_{\rm sp}}
\def\bp{b_{\rm p}}
\def\bs{b_{\rm s}}

\def\drp{{\rm d}r_{\rm p}}
\def\rp{r_{\rm p}}
\def\rpmin{r_{\rm p}^{\rm min}}
\def\rpmax{r_{\rm p}^{\rm max}}

\usepackage{multirow}

\newcommand{\smallO}[1]{\ensuremath{\mathop{}\mathopen{}{\scriptstyle\mathcal{O}}\mathopen{}\left(#1\right)}}
\newcommand{\bigO}[1]{\ensuremath{\mathop{}\mathopen{}{ \Large\mathcal{O}}\mathopen{}\left(#1\right)}}

\begin{document}
%
%
   \title{\Euclid\/: Photometric redshift calibration performance with the clustering-redshifts technique in the Flagship 2 simulation \thanks{This paper is published on behalf of the Euclid Consortium}}

\newcommand{\orcid}[1]{} 
\author{W.~d'Assignies\orcid{0000-0002-9719-1717}\thanks{\email{wdoumerg@ifae.es}}\inst{\ref{aff1}}
\and M.~Manera\orcid{0000-0003-4962-8934}\inst{\ref{aff2},\ref{aff1}}
\and C.~Padilla\orcid{0000-0001-7951-0166}\inst{\ref{aff1}}
\and O.~Ilbert\orcid{0000-0002-7303-4397}\inst{\ref{aff3}}
\and H.~Hildebrandt\orcid{0000-0002-9814-3338}\inst{\ref{aff4}}
\and L.~Reynolds\orcid{0000-0002-4732-3718}\inst{\ref{aff1},\ref{aff5}}
\and J.~Chaves-Montero\orcid{0000-0002-9553-4261}\inst{\ref{aff1}}
\and A.~H.~Wright\orcid{0000-0001-7363-7932}\inst{\ref{aff4}}
\and P.~Tallada-Cresp\'{i}\orcid{0000-0002-1336-8328}\inst{\ref{aff6},\ref{aff7}}
\and M.~Eriksen\orcid{0000-0003-0601-0990}\inst{\ref{aff1},\ref{aff7}}
\and J.~Carretero\orcid{0000-0002-3130-0204}\inst{\ref{aff6},\ref{aff7}}
\and W.~Roster\orcid{0000-0002-9149-6528}\inst{\ref{aff8}}
\and Y.~Kang\orcid{0009-0000-8588-7250}\inst{\ref{aff9}}
\and K.~Naidoo\orcid{0000-0002-9182-1802}\inst{\ref{aff10}}
\and R.~Miquel\orcid{0000-0002-6610-4836}\inst{\ref{aff1},\ref{aff11}}
\and B.~Altieri\orcid{0000-0003-3936-0284}\inst{\ref{aff12}}
\and A.~Amara\inst{\ref{aff13}}
\and S.~Andreon\orcid{0000-0002-2041-8784}\inst{\ref{aff14}}
\and N.~Auricchio\orcid{0000-0003-4444-8651}\inst{\ref{aff15}}
\and C.~Baccigalupi\orcid{0000-0002-8211-1630}\inst{\ref{aff16},\ref{aff17},\ref{aff18},\ref{aff19}}
\and D.~Bagot\inst{\ref{aff20}}
\and M.~Baldi\orcid{0000-0003-4145-1943}\inst{\ref{aff21},\ref{aff15},\ref{aff22}}
\and A.~Balestra\orcid{0000-0002-6967-261X}\inst{\ref{aff23}}
\and S.~Bardelli\orcid{0000-0002-8900-0298}\inst{\ref{aff15}}
\and P.~Battaglia\orcid{0000-0002-7337-5909}\inst{\ref{aff15}}
\and A.~Biviano\orcid{0000-0002-0857-0732}\inst{\ref{aff17},\ref{aff16}}
\and E.~Branchini\orcid{0000-0002-0808-6908}\inst{\ref{aff24},\ref{aff25},\ref{aff14}}
\and M.~Brescia\orcid{0000-0001-9506-5680}\inst{\ref{aff26},\ref{aff27}}
\and S.~Camera\orcid{0000-0003-3399-3574}\inst{\ref{aff28},\ref{aff29},\ref{aff30}}
\and V.~Capobianco\orcid{0000-0002-3309-7692}\inst{\ref{aff30}}
\and C.~Carbone\orcid{0000-0003-0125-3563}\inst{\ref{aff31}}
\and V.~F.~Cardone\inst{\ref{aff32},\ref{aff33}}
\and S.~Casas\orcid{0000-0002-4751-5138}\inst{\ref{aff34}}
\and F.~J.~Castander\orcid{0000-0001-7316-4573}\inst{\ref{aff35},\ref{aff36}}
\and M.~Castellano\orcid{0000-0001-9875-8263}\inst{\ref{aff32}}
\and G.~Castignani\orcid{0000-0001-6831-0687}\inst{\ref{aff15}}
\and S.~Cavuoti\orcid{0000-0002-3787-4196}\inst{\ref{aff27},\ref{aff37}}
\and K.~C.~Chambers\orcid{0000-0001-6965-7789}\inst{\ref{aff38}}
\and A.~Cimatti\inst{\ref{aff39}}
\and C.~Colodro-Conde\inst{\ref{aff40}}
\and G.~Congedo\orcid{0000-0003-2508-0046}\inst{\ref{aff41}}
\and C.~J.~Conselice\orcid{0000-0003-1949-7638}\inst{\ref{aff42}}
\and L.~Conversi\orcid{0000-0002-6710-8476}\inst{\ref{aff43},\ref{aff12}}
\and Y.~Copin\orcid{0000-0002-5317-7518}\inst{\ref{aff44}}
\and F.~Courbin\orcid{0000-0003-0758-6510}\inst{\ref{aff45},\ref{aff11}}
\and H.~M.~Courtois\orcid{0000-0003-0509-1776}\inst{\ref{aff46}}
\and M.~Crocce\orcid{0000-0002-9745-6228}\inst{\ref{aff35},\ref{aff36}}
\and A.~Da~Silva\orcid{0000-0002-6385-1609}\inst{\ref{aff47},\ref{aff48}}
\and H.~Degaudenzi\orcid{0000-0002-5887-6799}\inst{\ref{aff9}}
\and S.~de~la~Torre\inst{\ref{aff3}}
\and G.~De~Lucia\orcid{0000-0002-6220-9104}\inst{\ref{aff17}}
\and M.~Douspis\orcid{0000-0003-4203-3954}\inst{\ref{aff49}}
\and X.~Dupac\inst{\ref{aff12}}
\and A.~Ealet\orcid{0000-0003-3070-014X}\inst{\ref{aff44}}
\and S.~Escoffier\orcid{0000-0002-2847-7498}\inst{\ref{aff50}}
\and M.~Farina\orcid{0000-0002-3089-7846}\inst{\ref{aff51}}
\and F.~Faustini\orcid{0000-0001-6274-5145}\inst{\ref{aff32},\ref{aff52}}
\and S.~Ferriol\inst{\ref{aff44}}
\and F.~Finelli\orcid{0000-0002-6694-3269}\inst{\ref{aff15},\ref{aff53}}
\and P.~Fosalba\orcid{0000-0002-1510-5214}\inst{\ref{aff36},\ref{aff35}}
\and S.~Fotopoulou\orcid{0000-0002-9686-254X}\inst{\ref{aff54}}
\and M.~Frailis\orcid{0000-0002-7400-2135}\inst{\ref{aff17}}
\and E.~Franceschi\orcid{0000-0002-0585-6591}\inst{\ref{aff15}}
\and M.~Fumana\orcid{0000-0001-6787-5950}\inst{\ref{aff31}}
\and S.~Galeotta\orcid{0000-0002-3748-5115}\inst{\ref{aff17}}
\and K.~George\orcid{0000-0002-1734-8455}\inst{\ref{aff55}}
\and B.~Gillis\orcid{0000-0002-4478-1270}\inst{\ref{aff41}}
\and C.~Giocoli\orcid{0000-0002-9590-7961}\inst{\ref{aff15},\ref{aff22}}
\and P.~G\'omez-Alvarez\orcid{0000-0002-8594-5358}\inst{\ref{aff56},\ref{aff12}}
\and J.~Gracia-Carpio\inst{\ref{aff8}}
\and A.~Grazian\orcid{0000-0002-5688-0663}\inst{\ref{aff23}}
\and F.~Grupp\inst{\ref{aff8},\ref{aff55}}
\and W.~Holmes\inst{\ref{aff57}}
\and I.~M.~Hook\orcid{0000-0002-2960-978X}\inst{\ref{aff58}}
\and A.~Hornstrup\orcid{0000-0002-3363-0936}\inst{\ref{aff59},\ref{aff60}}
\and K.~Jahnke\orcid{0000-0003-3804-2137}\inst{\ref{aff61}}
\and M.~Jhabvala\inst{\ref{aff62}}
\and B.~Joachimi\orcid{0000-0001-7494-1303}\inst{\ref{aff63}}
\and E.~Keih\"anen\orcid{0000-0003-1804-7715}\inst{\ref{aff64}}
\and S.~Kermiche\orcid{0000-0002-0302-5735}\inst{\ref{aff50}}
\and A.~Kiessling\orcid{0000-0002-2590-1273}\inst{\ref{aff57}}
\and B.~Kubik\orcid{0009-0006-5823-4880}\inst{\ref{aff44}}
\and M.~K\"ummel\orcid{0000-0003-2791-2117}\inst{\ref{aff55}}
\and M.~Kunz\orcid{0000-0002-3052-7394}\inst{\ref{aff65}}
\and H.~Kurki-Suonio\orcid{0000-0002-4618-3063}\inst{\ref{aff66},\ref{aff67}}
\and O.~Lahav\orcid{0000-0002-1134-9035}\inst{\ref{aff63}}
\and A.~M.~C.~Le~Brun\orcid{0000-0002-0936-4594}\inst{\ref{aff68}}
\and S.~Ligori\orcid{0000-0003-4172-4606}\inst{\ref{aff30}}
\and P.~B.~Lilje\orcid{0000-0003-4324-7794}\inst{\ref{aff69}}
\and V.~Lindholm\orcid{0000-0003-2317-5471}\inst{\ref{aff66},\ref{aff67}}
\and I.~Lloro\orcid{0000-0001-5966-1434}\inst{\ref{aff70}}
\and G.~Mainetti\orcid{0000-0003-2384-2377}\inst{\ref{aff71}}
\and D.~Maino\inst{\ref{aff72},\ref{aff31},\ref{aff73}}
\and E.~Maiorano\orcid{0000-0003-2593-4355}\inst{\ref{aff15}}
\and O.~Mansutti\orcid{0000-0001-5758-4658}\inst{\ref{aff17}}
\and S.~Marcin\inst{\ref{aff74}}
\and O.~Marggraf\orcid{0000-0001-7242-3852}\inst{\ref{aff75}}
\and K.~Markovic\orcid{0000-0001-6764-073X}\inst{\ref{aff57}}
\and M.~Martinelli\orcid{0000-0002-6943-7732}\inst{\ref{aff32},\ref{aff33}}
\and N.~Martinet\orcid{0000-0003-2786-7790}\inst{\ref{aff3}}
\and F.~Marulli\orcid{0000-0002-8850-0303}\inst{\ref{aff76},\ref{aff15},\ref{aff22}}
\and R.~Massey\orcid{0000-0002-6085-3780}\inst{\ref{aff77}}
\and D.~C.~Masters\orcid{0000-0001-5382-6138}\inst{\ref{aff78}}
\and E.~Medinaceli\orcid{0000-0002-4040-7783}\inst{\ref{aff15}}
\and S.~Mei\orcid{0000-0002-2849-559X}\inst{\ref{aff79},\ref{aff80}}
\and M.~Melchior\inst{\ref{aff81}}
\and Y.~Mellier\inst{\ref{aff82},\ref{aff83}}
\and M.~Meneghetti\orcid{0000-0003-1225-7084}\inst{\ref{aff15},\ref{aff22}}
\and E.~Merlin\orcid{0000-0001-6870-8900}\inst{\ref{aff32}}
\and G.~Meylan\inst{\ref{aff84}}
\and A.~Mora\orcid{0000-0002-1922-8529}\inst{\ref{aff85}}
\and M.~Moresco\orcid{0000-0002-7616-7136}\inst{\ref{aff76},\ref{aff15}}
\and L.~Moscardini\orcid{0000-0002-3473-6716}\inst{\ref{aff76},\ref{aff15},\ref{aff22}}
\and C.~Neissner\orcid{0000-0001-8524-4968}\inst{\ref{aff1},\ref{aff7}}
\and S.-M.~Niemi\orcid{0009-0005-0247-0086}\inst{\ref{aff86}}
\and S.~Paltani\orcid{0000-0002-8108-9179}\inst{\ref{aff9}}
\and F.~Pasian\orcid{0000-0002-4869-3227}\inst{\ref{aff17}}
\and K.~Pedersen\inst{\ref{aff87}}
\and V.~Pettorino\inst{\ref{aff86}}
\and S.~Pires\orcid{0000-0002-0249-2104}\inst{\ref{aff88}}
\and G.~Polenta\orcid{0000-0003-4067-9196}\inst{\ref{aff52}}
\and M.~Poncet\inst{\ref{aff20}}
\and L.~A.~Popa\inst{\ref{aff89}}
\and L.~Pozzetti\orcid{0000-0001-7085-0412}\inst{\ref{aff15}}
\and F.~Raison\orcid{0000-0002-7819-6918}\inst{\ref{aff8}}
\and R.~Rebolo\orcid{0000-0003-3767-7085}\inst{\ref{aff40},\ref{aff90},\ref{aff91}}
\and A.~Renzi\orcid{0000-0001-9856-1970}\inst{\ref{aff92},\ref{aff93}}
\and J.~Rhodes\orcid{0000-0002-4485-8549}\inst{\ref{aff57}}
\and G.~Riccio\inst{\ref{aff27}}
\and E.~Romelli\orcid{0000-0003-3069-9222}\inst{\ref{aff17}}
\and M.~Roncarelli\orcid{0000-0001-9587-7822}\inst{\ref{aff15}}
\and E.~Rossetti\orcid{0000-0003-0238-4047}\inst{\ref{aff21}}
\and R.~Saglia\orcid{0000-0003-0378-7032}\inst{\ref{aff55},\ref{aff8}}
\and Z.~Sakr\orcid{0000-0002-4823-3757}\inst{\ref{aff94},\ref{aff95},\ref{aff96}}
\and D.~Sapone\orcid{0000-0001-7089-4503}\inst{\ref{aff97}}
\and B.~Sartoris\orcid{0000-0003-1337-5269}\inst{\ref{aff55},\ref{aff17}}
\and J.~A.~Schewtschenko\orcid{0000-0002-4913-6393}\inst{\ref{aff41}}
\and P.~Schneider\orcid{0000-0001-8561-2679}\inst{\ref{aff75}}
\and T.~Schrabback\orcid{0000-0002-6987-7834}\inst{\ref{aff98}}
\and A.~Secroun\orcid{0000-0003-0505-3710}\inst{\ref{aff50}}
\and E.~Sefusatti\orcid{0000-0003-0473-1567}\inst{\ref{aff17},\ref{aff16},\ref{aff18}}
\and G.~Seidel\orcid{0000-0003-2907-353X}\inst{\ref{aff61}}
\and M.~Seiffert\orcid{0000-0002-7536-9393}\inst{\ref{aff57}}
\and S.~Serrano\orcid{0000-0002-0211-2861}\inst{\ref{aff36},\ref{aff99},\ref{aff35}}
\and P.~Simon\inst{\ref{aff75}}
\and C.~Sirignano\orcid{0000-0002-0995-7146}\inst{\ref{aff92},\ref{aff93}}
\and G.~Sirri\orcid{0000-0003-2626-2853}\inst{\ref{aff22}}
\and A.~Spurio~Mancini\orcid{0000-0001-5698-0990}\inst{\ref{aff100}}
\and L.~Stanco\orcid{0000-0002-9706-5104}\inst{\ref{aff93}}
\and J.~Steinwagner\orcid{0000-0001-7443-1047}\inst{\ref{aff8}}
\and D.~Tavagnacco\orcid{0000-0001-7475-9894}\inst{\ref{aff17}}
\and A.~N.~Taylor\inst{\ref{aff41}}
\and H.~I.~Teplitz\orcid{0000-0002-7064-5424}\inst{\ref{aff78}}
\and I.~Tereno\orcid{0000-0002-4537-6218}\inst{\ref{aff47},\ref{aff101}}
\and N.~Tessore\orcid{0000-0002-9696-7931}\inst{\ref{aff63}}
\and S.~Toft\orcid{0000-0003-3631-7176}\inst{\ref{aff102},\ref{aff103}}
\and R.~Toledo-Moreo\orcid{0000-0002-2997-4859}\inst{\ref{aff104}}
\and F.~Torradeflot\orcid{0000-0003-1160-1517}\inst{\ref{aff7},\ref{aff6}}
\and A.~Tsyganov\inst{\ref{aff105}}
\and I.~Tutusaus\orcid{0000-0002-3199-0399}\inst{\ref{aff95}}
\and L.~Valenziano\orcid{0000-0002-1170-0104}\inst{\ref{aff15},\ref{aff53}}
\and J.~Valiviita\orcid{0000-0001-6225-3693}\inst{\ref{aff66},\ref{aff67}}
\and T.~Vassallo\orcid{0000-0001-6512-6358}\inst{\ref{aff55},\ref{aff17}}
\and G.~Verdoes~Kleijn\orcid{0000-0001-5803-2580}\inst{\ref{aff106}}
\and Y.~Wang\orcid{0000-0002-4749-2984}\inst{\ref{aff78}}
\and J.~Weller\orcid{0000-0002-8282-2010}\inst{\ref{aff55},\ref{aff8}}
\and G.~Zamorani\orcid{0000-0002-2318-301X}\inst{\ref{aff15}}
\and E.~Zucca\orcid{0000-0002-5845-8132}\inst{\ref{aff15}}
\and M.~Bolzonella\orcid{0000-0003-3278-4607}\inst{\ref{aff15}}
\and C.~Burigana\orcid{0000-0002-3005-5796}\inst{\ref{aff107},\ref{aff53}}
\and L.~Gabarra\orcid{0000-0002-8486-8856}\inst{\ref{aff108}}
\and J.~Mart\'{i}n-Fleitas\orcid{0000-0002-8594-569X}\inst{\ref{aff109}}
\and I.~Risso\orcid{0000-0003-2525-7761}\inst{\ref{aff110}}
\and V.~Scottez\inst{\ref{aff82},\ref{aff111}}
\and M.~Viel\orcid{0000-0002-2642-5707}\inst{\ref{aff16},\ref{aff17},\ref{aff19},\ref{aff18},\ref{aff112}}}
										   
\institute{Institut de F\'{i}sica d'Altes Energies (IFAE), The Barcelona Institute of Science and Technology, Campus UAB, 08193 Bellaterra (Barcelona), Spain\label{aff1}
\and
Serra H\'unter Fellow, Departament de F\'{\i}sica, Universitat Aut\`onoma de Barcelona, E-08193 Bellaterra, Spain\label{aff2}
\and
Aix-Marseille Universit\'e, CNRS, CNES, LAM, Marseille, France\label{aff3}
\and
Ruhr University Bochum, Faculty of Physics and Astronomy, Astronomical Institute (AIRUB), German Centre for Cosmological Lensing (GCCL), 44780 Bochum, Germany\label{aff4}
\and
Departament de F\'{\i}sica, Universitat Aut\`onoma de Barcelona, 08193 Bellaterra (Barcelona), Spain\label{aff5}
\and
Centro de Investigaciones Energ\'eticas, Medioambientales y Tecnol\'ogicas (CIEMAT), Avenida Complutense 40, 28040 Madrid, Spain\label{aff6}
\and
Port d'Informaci\'{o} Cient\'{i}fica, Campus UAB, C. Albareda s/n, 08193 Bellaterra (Barcelona), Spain\label{aff7}
\and
Max Planck Institute for Extraterrestrial Physics, Giessenbachstr. 1, 85748 Garching, Germany\label{aff8}
\and
Department of Astronomy, University of Geneva, ch. d'Ecogia 16, 1290 Versoix, Switzerland\label{aff9}
\and
Institute of Cosmology and Gravitation, University of Portsmouth, Portsmouth PO1 3FX, UK\label{aff10}
\and
Instituci\'o Catalana de Recerca i Estudis Avan\c{c}ats (ICREA), Passeig de Llu\'{\i}s Companys 23, 08010 Barcelona, Spain\label{aff11}
\and
ESAC/ESA, Camino Bajo del Castillo, s/n., Urb. Villafranca del Castillo, 28692 Villanueva de la Ca\~nada, Madrid, Spain\label{aff12}
\and
School of Mathematics and Physics, University of Surrey, Guildford, Surrey, GU2 7XH, UK\label{aff13}
\and
INAF-Osservatorio Astronomico di Brera, Via Brera 28, 20122 Milano, Italy\label{aff14}
\and
INAF-Osservatorio di Astrofisica e Scienza dello Spazio di Bologna, Via Piero Gobetti 93/3, 40129 Bologna, Italy\label{aff15}
\and
IFPU, Institute for Fundamental Physics of the Universe, via Beirut 2, 34151 Trieste, Italy\label{aff16}
\and
INAF-Osservatorio Astronomico di Trieste, Via G. B. Tiepolo 11, 34143 Trieste, Italy\label{aff17}
\and
INFN, Sezione di Trieste, Via Valerio 2, 34127 Trieste TS, Italy\label{aff18}
\and
SISSA, International School for Advanced Studies, Via Bonomea 265, 34136 Trieste TS, Italy\label{aff19}
\and
Centre National d'Etudes Spatiales -- Centre spatial de Toulouse, 18 avenue Edouard Belin, 31401 Toulouse Cedex 9, France\label{aff20}
\and
Dipartimento di Fisica e Astronomia, Universit\`a di Bologna, Via Gobetti 93/2, 40129 Bologna, Italy\label{aff21}
\and
INFN-Sezione di Bologna, Viale Berti Pichat 6/2, 40127 Bologna, Italy\label{aff22}
\and
INAF-Osservatorio Astronomico di Padova, Via dell'Osservatorio 5, 35122 Padova, Italy\label{aff23}
\and
Dipartimento di Fisica, Universit\`a di Genova, Via Dodecaneso 33, 16146, Genova, Italy\label{aff24}
\and
INFN-Sezione di Genova, Via Dodecaneso 33, 16146, Genova, Italy\label{aff25}
\and
Department of Physics "E. Pancini", University Federico II, Via Cinthia 6, 80126, Napoli, Italy\label{aff26}
\and
INAF-Osservatorio Astronomico di Capodimonte, Via Moiariello 16, 80131 Napoli, Italy\label{aff27}
\and
Dipartimento di Fisica, Universit\`a degli Studi di Torino, Via P. Giuria 1, 10125 Torino, Italy\label{aff28}
\and
INFN-Sezione di Torino, Via P. Giuria 1, 10125 Torino, Italy\label{aff29}
\and
INAF-Osservatorio Astrofisico di Torino, Via Osservatorio 20, 10025 Pino Torinese (TO), Italy\label{aff30}
\and
INAF-IASF Milano, Via Alfonso Corti 12, 20133 Milano, Italy\label{aff31}
\and
INAF-Osservatorio Astronomico di Roma, Via Frascati 33, 00078 Monteporzio Catone, Italy\label{aff32}
\and
INFN-Sezione di Roma, Piazzale Aldo Moro, 2 - c/o Dipartimento di Fisica, Edificio G. Marconi, 00185 Roma, Italy\label{aff33}
\and
Institute for Theoretical Particle Physics and Cosmology (TTK), RWTH Aachen University, 52056 Aachen, Germany\label{aff34}
\and
Institute of Space Sciences (ICE, CSIC), Campus UAB, Carrer de Can Magrans, s/n, 08193 Barcelona, Spain\label{aff35}
\and
Institut d'Estudis Espacials de Catalunya (IEEC),  Edifici RDIT, Campus UPC, 08860 Castelldefels, Barcelona, Spain\label{aff36}
\and
INFN section of Naples, Via Cinthia 6, 80126, Napoli, Italy\label{aff37}
\and
Institute for Astronomy, University of Hawaii, 2680 Woodlawn Drive, Honolulu, HI 96822, USA\label{aff38}
\and
Dipartimento di Fisica e Astronomia "Augusto Righi" - Alma Mater Studiorum Universit\`a di Bologna, Viale Berti Pichat 6/2, 40127 Bologna, Italy\label{aff39}
\and
Instituto de Astrof\'{\i}sica de Canarias, V\'{\i}a L\'actea, 38205 La Laguna, Tenerife, Spain\label{aff40}
\and
Institute for Astronomy, University of Edinburgh, Royal Observatory, Blackford Hill, Edinburgh EH9 3HJ, UK\label{aff41}
\and
Jodrell Bank Centre for Astrophysics, Department of Physics and Astronomy, University of Manchester, Oxford Road, Manchester M13 9PL, UK\label{aff42}
\and
European Space Agency/ESRIN, Largo Galileo Galilei 1, 00044 Frascati, Roma, Italy\label{aff43}
\and
Universit\'e Claude Bernard Lyon 1, CNRS/IN2P3, IP2I Lyon, UMR 5822, Villeurbanne, F-69100, France\label{aff44}
\and
Institut de Ci\`{e}ncies del Cosmos (ICCUB), Universitat de Barcelona (IEEC-UB), Mart\'{i} i Franqu\`{e}s 1, 08028 Barcelona, Spain\label{aff45}
\and
UCB Lyon 1, CNRS/IN2P3, IUF, IP2I Lyon, 4 rue Enrico Fermi, 69622 Villeurbanne, France\label{aff46}
\and
Departamento de F\'isica, Faculdade de Ci\^encias, Universidade de Lisboa, Edif\'icio C8, Campo Grande, PT1749-016 Lisboa, Portugal\label{aff47}
\and
Instituto de Astrof\'isica e Ci\^encias do Espa\c{c}o, Faculdade de Ci\^encias, Universidade de Lisboa, Campo Grande, 1749-016 Lisboa, Portugal\label{aff48}
\and
Universit\'e Paris-Saclay, CNRS, Institut d'astrophysique spatiale, 91405, Orsay, France\label{aff49}
\and
Aix-Marseille Universit\'e, CNRS/IN2P3, CPPM, Marseille, France\label{aff50}
\and
INAF-Istituto di Astrofisica e Planetologia Spaziali, via del Fosso del Cavaliere, 100, 00100 Roma, Italy\label{aff51}
\and
Space Science Data Center, Italian Space Agency, via del Politecnico snc, 00133 Roma, Italy\label{aff52}
\and
INFN-Bologna, Via Irnerio 46, 40126 Bologna, Italy\label{aff53}
\and
School of Physics, HH Wills Physics Laboratory, University of Bristol, Tyndall Avenue, Bristol, BS8 1TL, UK\label{aff54}
\and
Universit\"ats-Sternwarte M\"unchen, Fakult\"at f\"ur Physik, Ludwig-Maximilians-Universit\"at M\"unchen, Scheinerstrasse 1, 81679 M\"unchen, Germany\label{aff55}
\and
FRACTAL S.L.N.E., calle Tulip\'an 2, Portal 13 1A, 28231, Las Rozas de Madrid, Spain\label{aff56}
\and
Jet Propulsion Laboratory, California Institute of Technology, 4800 Oak Grove Drive, Pasadena, CA, 91109, USA\label{aff57}
\and
Department of Physics, Lancaster University, Lancaster, LA1 4YB, UK\label{aff58}
\and
Technical University of Denmark, Elektrovej 327, 2800 Kgs. Lyngby, Denmark\label{aff59}
\and
Cosmic Dawn Center (DAWN), Denmark\label{aff60}
\and
Max-Planck-Institut f\"ur Astronomie, K\"onigstuhl 17, 69117 Heidelberg, Germany\label{aff61}
\and
NASA Goddard Space Flight Center, Greenbelt, MD 20771, USA\label{aff62}
\and
Department of Physics and Astronomy, University College London, Gower Street, London WC1E 6BT, UK\label{aff63}
\and
Department of Physics and Helsinki Institute of Physics, Gustaf H\"allstr\"omin katu 2, 00014 University of Helsinki, Finland\label{aff64}
\and
Universit\'e de Gen\`eve, D\'epartement de Physique Th\'eorique and Centre for Astroparticle Physics, 24 quai Ernest-Ansermet, CH-1211 Gen\`eve 4, Switzerland\label{aff65}
\and
Department of Physics, P.O. Box 64, 00014 University of Helsinki, Finland\label{aff66}
\and
Helsinki Institute of Physics, Gustaf H{\"a}llstr{\"o}min katu 2, University of Helsinki, Helsinki, Finland\label{aff67}
\and
Laboratoire d'etude de l'Univers et des phenomenes eXtremes, Observatoire de Paris, Universit\'e PSL, Sorbonne Universit\'e, CNRS, 92190 Meudon, France\label{aff68}
\and
Institute of Theoretical Astrophysics, University of Oslo, P.O. Box 1029 Blindern, 0315 Oslo, Norway\label{aff69}
\and
SKA Observatory, Jodrell Bank, Lower Withington, Macclesfield, Cheshire SK11 9FT, UK\label{aff70}
\and
Centre de Calcul de l'IN2P3/CNRS, 21 avenue Pierre de Coubertin 69627 Villeurbanne Cedex, France\label{aff71}
\and
Dipartimento di Fisica "Aldo Pontremoli", Universit\`a degli Studi di Milano, Via Celoria 16, 20133 Milano, Italy\label{aff72}
\and
INFN-Sezione di Milano, Via Celoria 16, 20133 Milano, Italy\label{aff73}
\and
University of Applied Sciences and Arts of Northwestern Switzerland, School of Computer Science, 5210 Windisch, Switzerland\label{aff74}
\and
Universit\"at Bonn, Argelander-Institut f\"ur Astronomie, Auf dem H\"ugel 71, 53121 Bonn, Germany\label{aff75}
\and
Dipartimento di Fisica e Astronomia "Augusto Righi" - Alma Mater Studiorum Universit\`a di Bologna, via Piero Gobetti 93/2, 40129 Bologna, Italy\label{aff76}
\and
Department of Physics, Institute for Computational Cosmology, Durham University, South Road, Durham, DH1 3LE, UK\label{aff77}
\and
Infrared Processing and Analysis Center, California Institute of Technology, Pasadena, CA 91125, USA\label{aff78}
\and
Universit\'e Paris Cit\'e, CNRS, Astroparticule et Cosmologie, 75013 Paris, France\label{aff79}
\and
CNRS-UCB International Research Laboratory, Centre Pierre Bin\'etruy, IRL2007, CPB-IN2P3, Berkeley, USA\label{aff80}
\and
University of Applied Sciences and Arts of Northwestern Switzerland, School of Engineering, 5210 Windisch, Switzerland\label{aff81}
\and
Institut d'Astrophysique de Paris, 98bis Boulevard Arago, 75014, Paris, France\label{aff82}
\and
Institut d'Astrophysique de Paris, UMR 7095, CNRS, and Sorbonne Universit\'e, 98 bis boulevard Arago, 75014 Paris, France\label{aff83}
\and
Institute of Physics, Laboratory of Astrophysics, Ecole Polytechnique F\'ed\'erale de Lausanne (EPFL), Observatoire de Sauverny, 1290 Versoix, Switzerland\label{aff84}
\and
Telespazio UK S.L. for European Space Agency (ESA), Camino bajo del Castillo, s/n, Urbanizacion Villafranca del Castillo, Villanueva de la Ca\~nada, 28692 Madrid, Spain\label{aff85}
\and
European Space Agency/ESTEC, Keplerlaan 1, 2201 AZ Noordwijk, The Netherlands\label{aff86}
\and
DARK, Niels Bohr Institute, University of Copenhagen, Jagtvej 155, 2200 Copenhagen, Denmark\label{aff87}
\and
Universit\'e Paris-Saclay, Universit\'e Paris Cit\'e, CEA, CNRS, AIM, 91191, Gif-sur-Yvette, France\label{aff88}
\and
Institute of Space Science, Str. Atomistilor, nr. 409 M\u{a}gurele, Ilfov, 077125, Romania\label{aff89}
\and
Consejo Superior de Investigaciones Cientificas, Calle Serrano 117, 28006 Madrid, Spain\label{aff90}
\and
Universidad de La Laguna, Departamento de Astrof\'{\i}sica, 38206 La Laguna, Tenerife, Spain\label{aff91}
\and
Dipartimento di Fisica e Astronomia "G. Galilei", Universit\`a di Padova, Via Marzolo 8, 35131 Padova, Italy\label{aff92}
\and
INFN-Padova, Via Marzolo 8, 35131 Padova, Italy\label{aff93}
\and
Institut f\"ur Theoretische Physik, University of Heidelberg, Philosophenweg 16, 69120 Heidelberg, Germany\label{aff94}
\and
Institut de Recherche en Astrophysique et Plan\'etologie (IRAP), Universit\'e de Toulouse, CNRS, UPS, CNES, 14 Av. Edouard Belin, 31400 Toulouse, France\label{aff95}
\and
Universit\'e St Joseph; Faculty of Sciences, Beirut, Lebanon\label{aff96}
\and
Departamento de F\'isica, FCFM, Universidad de Chile, Blanco Encalada 2008, Santiago, Chile\label{aff97}
\and
Universit\"at Innsbruck, Institut f\"ur Astro- und Teilchenphysik, Technikerstr. 25/8, 6020 Innsbruck, Austria\label{aff98}
\and
Satlantis, University Science Park, Sede Bld 48940, Leioa-Bilbao, Spain\label{aff99}
\and
Department of Physics, Royal Holloway, University of London, TW20 0EX, UK\label{aff100}
\and
Instituto de Astrof\'isica e Ci\^encias do Espa\c{c}o, Faculdade de Ci\^encias, Universidade de Lisboa, Tapada da Ajuda, 1349-018 Lisboa, Portugal\label{aff101}
\and
Cosmic Dawn Center (DAWN)\label{aff102}
\and
Niels Bohr Institute, University of Copenhagen, Jagtvej 128, 2200 Copenhagen, Denmark\label{aff103}
\and
Universidad Polit\'ecnica de Cartagena, Departamento de Electr\'onica y Tecnolog\'ia de Computadoras,  Plaza del Hospital 1, 30202 Cartagena, Spain\label{aff104}
\and
Centre for Information Technology, University of Groningen, P.O. Box 11044, 9700 CA Groningen, The Netherlands\label{aff105}
\and
Kapteyn Astronomical Institute, University of Groningen, PO Box 800, 9700 AV Groningen, The Netherlands\label{aff106}
\and
INAF, Istituto di Radioastronomia, Via Piero Gobetti 101, 40129 Bologna, Italy\label{aff107}
\and
Department of Physics, Oxford University, Keble Road, Oxford OX1 3RH, UK\label{aff108}
\and
Aurora Technology for European Space Agency (ESA), Camino bajo del Castillo, s/n, Urbanizacion Villafranca del Castillo, Villanueva de la Ca\~nada, 28692 Madrid, Spain\label{aff109}
\and
INAF-Osservatorio Astronomico di Brera, Via Brera 28, 20122 Milano, Italy, and INFN-Sezione di Genova, Via Dodecaneso 33, 16146, Genova, Italy\label{aff110}
\and
ICL, Junia, Universit\'e Catholique de Lille, LITL, 59000 Lille, France\label{aff111}
\and
ICSC - Centro Nazionale di Ricerca in High Performance Computing, Big Data e Quantum Computing, Via Magnanelli 2, Bologna, Italy\label{aff112}}     
%
%
   \abstract{
Aims: The precision of cosmological constraints from imaging surveys hinges on an accurately estimated redshift distribution $n(z)$ of the tomographic bins, especially their mean redshifts. We assess the effectiveness of the clustering-redshifts technique in constraining \Euclid tomographic redshift bins to meet the target uncertainty of  $ \sigma(\langle z \rangle)< 0.002 (1 + z)$. We inferred these mean redshifts from the small-scale angular clustering of \Euclid galaxies, which were distributed into bins with spectroscopic samples localised in narrow redshift slices.

Methods: We generated spectroscopic mocks from the Flagship2 simulation for the Baryon Oscillation Spectroscopic Survey (BOSS), the Dark Energy Spectroscopic Instrument (DESI), and the \textit{Euclid} Near-Infrared Spectrometer and Photometer (NISP) spectroscopic survey. We evaluated and optimised the clustering-redshifts pipeline, and we introduced a new method for measuring the photometric galaxy bias (clustering), which is the primary limitation of this technique. 

Results: We have successfully constrained the means and standard deviations of the redshift distributions for all of the tomographic bins (with a maximum photometric redshift of 1.6). We achieved precision beyond the required thresholds. We have identified the main sources of bias, particularly the impact of the one-halo galaxy distribution, which imposed the minimal separation scale to be larger than 1.5 Mpc for evaluating cross-correlations. These results demonstrate that clustering-redshifts can meet the precision requirements for \Euclid, and we highlighted several avenues for future improvements.
}
%
%
%
%
   \titlerunning{\Euclid\/: Clustering redshifts}
   \authorrunning{W.  d'Assignies et al.}
   
   \maketitle
%
%
%
\nolinenumbers
\section{\label{sc:Intro}Introduction}

The \Euclid space telescope is currently mapping the positions and shapes of billions of galaxies. This provides  data that are critical to understand the large-scale structure of the Universe, in particular, the mysterious nature of dark matter and dark energy \citep{Euclid_overview}. Through its imaging survey, \Euclid measures the flux of galaxies in the visible and near-infrared wavelengths using broadband filters. Additionally, it determines spectroscopic redshifts for a subset of galaxies via slitless spectroscopy, targeting emission-line galaxies (ELGs). These data will enable the precise determination of cosmological parameters and models, primarily through the study of galaxy clustering, galaxy-galaxy lensing, and cosmic shear \citep{Blanchard-EP7}, with galaxy samples divided into approximate line-of-sight tomographic bins.

It is essential to accurately estimate the redshift distributions of these galaxy samples to interpret  cosmological measurements correctly  \citep[see e.g.][]{Hurterer2006,photo-z-perf_cosmo, KIDS_redshift_dis, Stolzner_GaussianFitting}. The vast amount of forthcoming data means that it is not feasible to obtain spectroscopic redshifts for every individual galaxy because spectroscopy of large samples is time-consuming and costly.
Photometric surveys provide redshift estimates for each galaxy based on multi-band photometry of that galaxy, a technique called photometric redshift, or photo-$z$. A large variety of photo-$z$ methods exists \citep[e.g.][]{photo-z_HSC,Salvato_2019,Ilbert-EP11,Desprez-EP10}.
The photometric information can also be used to produce redshift estimates using scheme based on a self-organising map (hereafter, SOM), which allows a more general control over all the known potential sources of uncertainties that affect the estimates \citep[][]{Wright_2020,Wright_2020b,3sdir, DESY3_MAGLIM_z,Campos2024, kids2025_zcal,Roster25}.
The degeneracies between colours and redshift, however, and unrepresentative spectroscopic samples for training and calibration ultimately limit the performance of photometric methods \citep{Wright_2020, Hartley2020}. 

Clustering-based redshift estimation methods offer another alternative to infer redshift distributions \citep[see e.g.][]{newman2008, Menard2013,clust_z_mcQuinn_white,theWizz, Scottez, Gatti_DESY1,van_den_Busch_2020, KIDS_redshift_dis, Gatti_Giulia_DESY3, Cawton2022, HSC_clustering_z,kids2025_zcal}. Unlike photo-$z$, clustering-redshifts techniques statistically infer the redshift properties for entire bins instead of individual galaxies. Furthermore, this approach is independent of photometric uncertainties, observations over smaller deep fields, or representative spectroscopic samples.

The clustering-redshifts method relies on angular cross-correlations between spectroscopic samples, with secure redshifts, and photometric samples. Samples that overlap in redshift trace the same underlying dark matter field and are consequently correlated in their position. The amplitude of this angular correlation provides insight into their redshift overlap, which results in the measurement of the redshift distribution of the photometric samples. 
This amplitude is degenerate with galaxy biases (of both samples), however, which makes the measurement of these biases, and their redshift evolution, a critical step  \citep{clust_z_mcQuinn_white, Gatti_DESY1,Naidoo23}.
While the galaxy bias for the spectroscopic sample can be directly constrained through auto-correlation functions, it is more challenging to estimate the galaxy bias for the photometric sample \citep{van_den_Busch_2020, Cawton2022}, and is sometimes evaluated using simulations \citep{Gatti_Giulia_DESY3}.

The precision of the clustering-redshifts method depends on the redshift and sky overlap of the photometric and spectroscopic data, but also on range of the clustering angular scale. To achieve the primary science goal of \Euclid,  the uncertainty on the mean redshift, $\sigma(\langle z\rangle)$, for each tomographic bin must remain below $0.002(1+z)$ at 68$\%$ confidence \citep{Laureijs11}. \cite{Naidoo23} found that clustering-redshifts would be able to reach the statistical uncertainties required by \Euclid for a sufficiently large sky overlap, typically several hundreds of deg$^2$, when analysing scales from 100 kpc to 1 Mpc, although systematic biases limit the accuracy. In particular, there were some unknown residual biases for high-redshift bins with their methods or simulated data.

This paper aims to re-evaluate the potential of photometric-spectroscopic cross-correlations as a core component of the redshift calibration for \Euclid. Our method may be seen as a continuation, and a substantial improvement from \cite{Naidoo23}. 
We used $5000$ deg$^2$ of simulated data from the \Euclid Flagship2 simulation \citep{FS_2024}. Our goal is to evaluate the uncertainties associated with the clustering-redshifts calibration through a realistic framework, and to determine the limitation of the method. We created spectroscopic mock samples for the Dark Energy Spectroscopic Instruments (DESI), or alternatively, for the 4-metre Multi-Object Spectroscopic Telescope (4MOST), the Baryon Oscillation Spectroscopic Survey (BOSS), and \Euclid. We assessed the potential systematic effects on the measurement, introduced a new method to measure photometric galaxy bias, optimised the clustering-$z$ pipeline, and derived methods for inferring the redshift constraints from the amplitude of two-point correlations. We assumed the same flat $\Lambda$CDM cosmology as was used for the simulation, with  $\Omega_{\rm m}=0.319$, $\Omega_{\rm b}=0.049$, $\Omega_{\Lambda}=0.681$,\, $A_{\rm s}=2.1 \times 10^{-9}$,\, $n_{\rm s}=0.96$, $\sigma_8=0.813$, and $h=0.67$. Our code is publicly available.\footnote{\url{https://github.com/wdassignies/Clustering_z.git}}

This paper is organised as follows. In Sect. \ref{sc:Data} we describe the Flagship2 simulation and the mocks we generated to mimic \Euclid and spectroscopic data. In Sect. \ref{sc:Method} we give a detailed overview of angular clustering, its application to clustering redshifts, and potential sources of errors. In Sect. \ref{sc:Result} we present the different tests we ran to optimise the pipeline and address potential sources of bias. Finally in Sect. \ref{sc:forecast} we apply the results pipeline to realistic \Euclid tomographic bins and compare our results with previous work and the requirements.  

 \section{\label{sc:Data} Simulated data}

To assess the performance of the \Euclid clustering-redshifts method, we  constructed simulated datasets that were as similar as possible to the observational datasets. In this section, we describe how these samples can be generated from the Flagship simulation. 
\begin{figure*}
    \centering
    \includegraphics[width=1\linewidth]{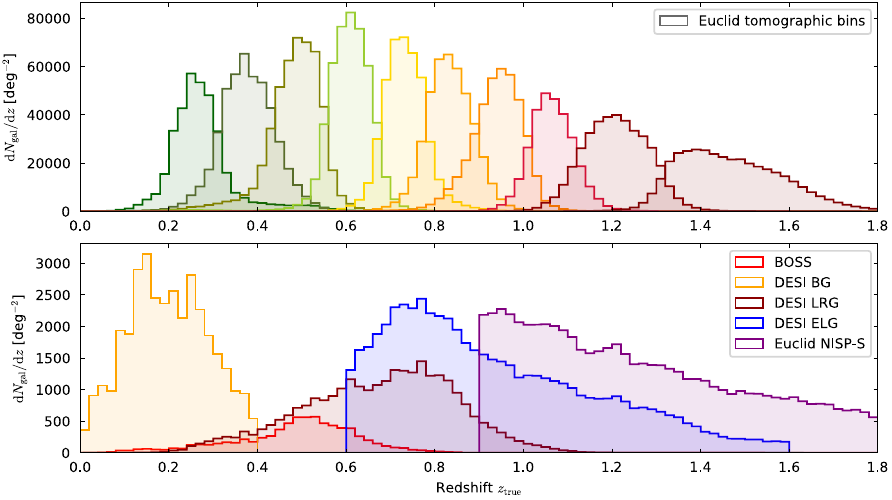}
    \caption{\textit{Top}: Simulated true-redshift distribution of each of the first ten tomographic bins ($z_{\rm photo}<1.6$). \textit{Bottom}: Simulated redshift distribution of the spectroscopic samples from the BOSS, DESI, and \Euclid surveys. The lack of spectroscopic samples for $z_{\rm spec}>1.8$ imposes the $z_{\rm photo}<1.6$ condition for clustering-redshifts calibration.}
    \label{fig:TB_spec_samples}
\end{figure*}
\subsection{The Flagship simulation and survey mocks}
For this study, we used the second edition of the Flagship catalogue (version 2.1.10), which was presented in detail in \cite{FS_2024}. The \Euclid Flagship $N$-body dark matter simulation comprises a simulation box measuring $3600$ $\hMpc$  on each side, containing 4 trillion particles. This simulation is the largest $N$-body simulation 
performed to date and encompasses a cosmological volume comparable to what the telescope will survey. It enables precise resolution of 16 billion dark matter haloes, which host even the faintest galaxies that \Euclid intends to observe. They are identified with the \textsc{rockstar} halo finder \citep{Rockstar_2013}. From this dark matter simulation, a synthetic galaxy catalogue was generated with a prescription including the halo occupation distribution (HOD) and abundance matching (AM) techniques, as well as empirical relations between galaxy properties \citep{FS_2024}.

Each galaxy entry in the catalogue is associated with observed fluxes for multiple survey bands, observed spectroscopic redshift  ($z_{\rm spec}$, including peculiar velocity), photometric redshift estimates ($z_{\rm p}$), true redshift ($z_{\rm true}$), true  and magnified coordinates, lensing convergence $\kappa$ and shear $\gamma=\gamma_1+{\rm i}\gamma_2$. 
Source photometry is provided in terms of fluxes $F$ (in erg s$^{-1}$ cm$^{-2}$ Hz$^{-1}$) rather than in magnitude $m$. 
We generate magnified fluxes as in \cite{Lepori24}, using the relation 
\begin{equation} 
{F}_{\rm magn}=\frac{F}{(1-\kappa)^2-\gamma_1^2-\gamma_2^2}\,.
\end{equation}
We do not include the Doppler effect, which is usually a very small correction, and note that the effect of magnification is achromatic: that is, it is the same for for the different bands and it cancels for colours.

We need mock samples  \Euclid-like photometric galaxies, \Euclid-like Near-Infrared Spectrometer and Photometer Survey (NISP-S) galaxies, and other spectroscopic tracers \citep{BOSS_color,DESI_validation}: BOSS-like low-redshift (LOWZ) and constant mass (CMASS) galaxies; DESI-like bright galaxies (BGs), luminous red galaxies (LRGs), and ELGs. 
The selection of each sample is described in the following sections. Differences in the redshift distributions or colours of observational and simulated galaxy samples may stem from the incompleteness of the target sample or limitations in the simulation’s ability to replicate accurate colour distributions. As in \cite{Naidoo23}, systematic issues such as variable photometry, variable survey depth, and complex masks are not addressed. 
We generated one version of these samples with unmagnified fluxes, which will be our default samples, and one version with the same cuts on magnified photometry. We produced numerous fast generations of galaxy catalogues, varying sample definitions and areas,  using the web application CosmoHub \citep{Carretero_cosmo_hub,cosmohub}. Our photometric and spectroscopic samples are plotted in Fig. \ref{fig:TB_spec_samples}. 
\label{sec:mocks}

\subsection{Euclid photometric sample}
\label{sc:data_euclid_photo}

The photometric sample was constructed by imposing an upper limit on the $\IE$ magnitude \citep{EuclidSkyVIS}, set at 24.5, to emulate the expected performance of \Euclid \citep{Laureijs11}. Several algorithms exist for evaluating the photometric redshift. We used the Deep-$z$ estimate which was run for the full catalogue. 
The Deep-$z$ photo-$z$ method, introduced for the Physics of the Accelerating Universe (PAU) survey in \cite{deep-z},  uses a linear neural network with a mixture density network for predicting the redshift distributions.  
The network has not been pretrained on simulations, avoiding potential problems with over-fitting by using too similar simulations. The network was trained with $20\,000$ simulated galaxies, magnitude-limited to $i_{\rm LSST} < 23$ without colour selection. The training used the  Legacy Survey of Space and Time \citep[LSST;][]{lsst_book} bands plus \Euclid near-infrared (NIR) bands, a batch size of 200 and $1000$ epochs with varying learning rates.  
The photometric noise added to noiseless fluxes corresponds to the depth forecast in the southern hemisphere for the \Euclid DR3 time.  The limiting magnitudes at 10$\sigma$ are 24.4, 25.6, 25.7, 25.0, 24.3, 24.3 for the $ugrizy$ bands of LSST, and 25.0, 23.5, 23.5, 23.5 for the $\IE,\, \YE,\, \JE,\, \HE $ bands of \Euclid \citep{EuclidSkyVIS,EuclidSkyNISP}.

LSST photometry will not be available over the whole \Euclid footprint, and therefore the photometry will be different in the northern region where photometry will be provided by  UNIONS  \citep{Gwyn_UNIONS}. 
However, for clustering redshifts, we do not depend on photo-$z$ quality except for the tomographic bin definitions and photometric bias correction. Spectroscopy from DESI and BOSS-eBOSS will be available in the north, and 4MOST will provide DESI-like samples in the south. Thus, photometric angular anisotropies will be tested in practice.

We selected galaxies with photo-$z$ between 0.2 and 1.6. We did not consider galaxies with a photo-$z$ higher than 1.6 because we do not have large spectroscopic samples at redshifts greater than 1.8.
Future studies could test the possibility of using quasar galaxy samples from BOSS, eBOSS,  DESI, and 4MOST \citep{eboss_qso,desi_qso} for high-$z$ calibration. BOSS and eBOSS quasars were used for the clustering-redshifts DES calibration, but as a complement to ELGs and LRGs data sets for $z<1.1$ \citep{Gatti_Giulia_DESY3,DESY3_MAGLIM_z}. The low density of this sample, and clustering properties at high-$z$, require a dedicated study. Nonetheless, eBOSS QSOs will be used for the high-$z$ calibration of the final DES data for $0.8<z<2.2$ \citep{WZ_DES_dassignies}; thus one expects that DESI QSOs could be used up to $z=3.5$ for \Euclid. Generating DESI QSO mocks and testing clustering-redshifts with them is left for future work. 

We produced ten tomographic bins with approximately equal numbers of galaxies, through photo-$z$ cuts.
The fiducial \Euclid survey at data release 3 plans to use 13 bins instead of ten, and we can assume that three of these bins will contain galaxies with $z_{\rm p}>1.6$. The true redshift distributions (that we want to measure) of our simulated tomographic bins are reported in Fig. \ref{fig:TB_spec_samples}.

\subsection{Euclid NISP sample}
We defined mock NISP-S samples in Flagship2 by selecting galaxies with ${\rm H}\,\alpha$ fluxes greater than $2\times 10^{-16}$erg ${\rm cm}^{-2}\,{\rm s}^{-1}$ \citep{Laureijs11}, in the expected redshift range,  with
\begin{align}
    &\texttt{logf\_halpha\_model3\_ext}>-15.7\,,\\
    &0.9<z_{\rm spec}<1.8\,.
\end{align}
We obtained a galaxy density of 1990 deg$^{-2}$. Unlike the other spectroscopic samples, this dataset is expected to suffer from a significant level of contamination by interlopers \citep[][]{Q1_inter,EP-Risso}. These interlopers are galaxies whose true redshifts differ substantially from their estimated values, leading to incorrect associations in redshift space. Although our analysis did not employ realistic mocks that include such contaminants, we adopted the approach of \citet{Contarini22}: we uniformly down-sampled the galaxy catalogue, retaining only $60\%$ of the originally selected galaxies. This down-sampled catalogue is then assumed to have $100\%$ purity, effectively removing the impact of interlopers in our modelling. Thus, our fiducial density is $1200$ deg$^{-2}$ as the sample from \cite{Naidoo23}, which was generated with a brighter cut on the ${\rm H}\,\alpha$ flux. This is a pessimistic sample from the precision point of view \citep[i.e. fewer spectra means larger uncertainty on the redshift distribution reconstruction; cf.][]{clust_z_mcQuinn_white}, but optimistic in terms of accuracy, because our sample is interloper free \citep{interloper}. Interlopers may significantly degrade the performance of clustering-redshifts if not taken into consideration, as they are not necessarily distributed randomly in redshift. Indeed, clustering-redshifts might be helpful to quantify the interlopers' properties such as their fraction and redshift distribution, by cross-correlating them with spectroscopic samples which do not suffer from the same misidentification. We will explore these two aspects in future work.

\subsection{BOSS LOWZ and CMASS samples}
We replicated the BOSS colour-magnitude selections specified in \cite{BOSS_color} in Flagship, as  in \cite{Naidoo23}. We first define 
\begin{align}
    &c_\parallel=0.7(g-r)+1.2\,(r-i-0.18)\,,\\
    &c_\perp=(r-i)-(g-r)/4-0.18\,,\\
    &d_\perp=(r-i)-(g-r)/8\,.
\end{align}
There are two spectroscopic BOSS LRG samples: LOWZ and CMASS. They correspond to adjacent redshift intervals $0.15 \leq z_{\rm spec} \leq 0.43$ and $0.43<z_{\rm spec} \leq 0.7$, respectively, with true densities of about 30 deg$^{-2}$ and 120 deg$^{-2}$, respectively. 

The LOWZ sample is defined by
\begin{align}
    &\vert c_\perp \vert< 0.2\,,\\
    &16<r<19.5\,,\\
    &r<13.6+c_\parallel/0.3\,,
\end{align}
and the CMASS sample by
\begin{align}
    &d_\perp>0.55\,,\\
    &17.5<i<19.9\,,\\
    &i<19.86+1.6\,(d_\perp-0.8)\,.
\end{align}
We got LOWZ- and CMASS- like objects, with densities of 48.5 deg$^{-2}$ and 164.3 deg$^{-2}$. We then applied sparse sampling to get the desired densities.
We refer to the joint LOWZ+CMASS sample as the BOSS sample in our study.

\subsection{DESI samples}

We generated DESI BG, LRG, and ELG mocks that mimic the sample characteristics reported in \cite{DESI_validation}. We associated these galaxy samples to the DESI survey, but similar samples are expected to be observed by 4MOST \citep{4MOST_verdier}. For BGs, and ELGs, we used the same cuts as \cite{Naidoo23}, which were based on \cite{DESI_2016}. For LRGs, we used a different selection to generate a sample covering the entire $z<1.1$ redshift range. 
As we previously mentioned in Sect. \ref{sec:mocks}, there is a mismatch in the Flagship2 colour distribution and data expectation, and using the same colour cuts can lead to some significant differences. We find this issue to be particularly problematic for LRGs. Therefore we did not apply the same selection as \cite{DESI_validation}, but rather we tried to closely reproduce the redshift distribution and galaxy properties with similar but modified cuts.

We selected the BG-like galaxies with
\begin{align}
    &r<19.5\,,\\
    &z_{\rm spec}<0.4\,.
\end{align}
We adopted for LRGs  the tailored cuts
\begin{align}
    &18<z<21\,,\\
    &-6.5<r-z-0.4\,z<-5.8\,,\\
    &\texttt{color\_kind}=0  \text{ (i.e. red sequence)}\,.
\end{align}
Finally, we selected ELGs with 
\begin{align}
    &r<23.4\,,\\
    &r-z>0.3\,,\\
    &g-r<0.7\,,\\
    &0.6<z_{\rm spec}<1.6\,,\\
    &\texttt{color\_kind}\neq 0,\\
    &\texttt{logf\_o2\_model1\_ext}>-16\,.
\end{align}
These cuts provided BG-, LRG-, and ELG-like objects with densities of  850, 909, and $4158$ deg$^{-2}$. We achieved the fiducial densities 700, 550, and $1140$ deg$^{-2}$ with sparse sampling.

\section{\label{sc:Method}Theory and method}

In this section, we provide a detailed description of the clustering-redshifts method. 
Our approach begins with an overview of galaxy clustering in Sect. \ref{sc:meth0_gal-clust}.
In Sect. \ref{sc:meth_redshift_dist} we explain how the clustering equations are simplified for some particular redshift distributions. 
We then apply this formalism to the clustering-redshifts context in Sect. \ref{sc:method_clust_z}.  In Sect. \ref{sc:method_data}, we explain how clustering-redshifts are evaluated with data in practice.  Finally, in Sect. \ref{sc:metho_Fitting} we explain our strategy to deduce from the data-vectors constraints on the redshift distribution moments. In particular, in Sect. \ref{sc:requirement}, we state the \Euclid requirements for the $n_{\rm p}$ measurements, which must be fulfilled in order to provide robust constraints on cosmological parameters.

\subsection{\label{sc:meth0_gal-clust} Angular galaxy clustering}

In this section we provide a general introduction to photometric angular clustering, as a basis for the development of our clustering-redshifts method. It draws inspiration from many previous weak-lensing and clustering-$z$ works such as \cite{Davis2018}, \cite{Y3_DES_Krause}, \cite{Gatti_DESY1,Gatti_Giulia_DESY3}, \cite{Cawton2022}, and \cite{lsst_bias_magn_Sanchez}.

\subsubsection{Galaxy density contrast}
The observed angular galaxy count of a sample $a$ (photometric or spectroscopic) is
\begin{equation}
    N_{a}^{\rm obs}(\Vec{\theta})=\overline{N}_{a}\left(1+\Delta_{a}^{\rm obs}(\Vec{\theta})\right)\,,
\end{equation}
where $\overline{N}_{a}$ is the mean (observed) galaxy count, and $\Delta_{a}^{\rm obs}$ is the observed projected galaxy density contrast.  It can be expressed as  \citep{Y3_DES_Krause}
\begin{equation}
    \Delta_{a}^{\rm obs}(\Vec{\theta})\approx \Delta_{a}^{\rm D}(\Vec{\theta})+ \Delta_{a}^{\rm RSD}(\Vec{\theta})+ \Delta_{a}^{\mu}(\Vec{\theta})\,,
\end{equation}
where $\mu $ refers to magnification, RSD to redshift space distortion (in the linear regime\footnote{In the linear regime, RSD contribute to the projected galaxy density contrast through the apparent large-scale flow, of galaxies across the redshift boundaries of the samples. The redshift $z$ we use to parametrise the $D$ and $\mu$  terms includes the peculiar velocity, and is referred to as observed redshift.}), and $\Delta_{a}^{\rm D}$ to the line of sight projection of the three-dimensional (3D) galaxy density
 contrast $\delta_{a}(z,\,\Vec{\theta})$ :
\begin{equation}
    \Delta_{a}^{\rm D}(\Vec{\theta})=\int \dz \;n_{a}(z)\,\delta_{a}(z,\Vec{\theta})\,, \label{eq:Delta_a_i}
\end{equation}
where $n_{a}(z)$ is the redshift distribution of the observed sample, normalised to unity, and function of the observed redshift.
All these quantities are associated with observations, and they depend on survey properties (depth, masks etc.), but for clarity, we drop the ‘‘obs’’ index for the rest of our work, as we do not consider observational systematics. We also do not model relativistic corrections.

\subsubsection{Angular two-point correlation function}\label{sec:anglular_2p}

The angular two-point correlation function of samples $a$ and $b$ (with possibly  $a=b$) from the observed projected density contrast can be written as
\begin{equation}
\begin{aligned}
    w_{ab}^{\rm full}(\theta)&=\langle \Delta^{\rm D}_a\Delta^{\rm D}_b \rangle(\theta)+\langle \Delta^{\rm RSD}_a\Delta^{\rm D}_b \rangle(\theta)+\langle \Delta^{\rm D}_a\Delta^{\rm RSD}_b \rangle(\theta)\\
    &\hspace{2cm}+\langle \Delta^{\rm D}_a\Delta^{\mu}_b \rangle(\theta)+\langle \Delta^{\mu}_a\Delta^{D}_b \rangle(\theta)\,.\label{eq:wab_full}
\end{aligned}
\end{equation}
where $\langle \ldots \rangle$ indicates an average over angular scales, and $\theta$ is now a scalar.\footnote{Our Universe and its observation is assumed to be isotropic. Although the real isotropy of our Universe seems a reasonable hypothesis, observation may not always support it. For example, closer to the Galactic plane, there is more star contamination, leading to angular anisotropy. The \Euclid observation strategy might also induce some angular anisotropy. This could potentially introduce bias on the cosmological parameters, especially for experiments such as \Euclid or LSST \citep{anisotropic_redshift_distribution}. Indeed, clustering-redshifts may be a valuable tool to investigate this issue by splitting the samples into independent sky patches.} 
The dominant corrections (RSD$\times$D; $\mu \,\times$D) are already small with respect to the dominant term (D$\times$D), and we decided to omit even smaller corrections \citep{Y3_DES_Krause,Gatti_Giulia_DESY3}.

We will mainly focus on the ${\rm D}\times{\rm D} $ correlation. Since it corresponds to the leading order, we omit its $\rm D$ index in the rest of the article. The clustering two-point correlation can be expressed as  
\begin{align}
    w_{ab}(\theta)&=\langle \Delta^{\rm D}_a\Delta^{\rm D}_b \rangle(\theta)\\
    &=\iint \dz_1 \,\dz_2\, n_a(z_1)\,n_b(z_2)\, \langle \delta_a\,\delta_b \rangle(z_1,z_2,\theta)\\
    &=\iint \dz_1\, \dz_2\, n_a(z_1)\,n_b(z_2)\, \xi_{ab}(z_1,z_2,\theta)\,,
\end{align}
where we have defined the 3D cross-correlation function $ \xi_{ab}(z_a,z_b,\theta)$, and used $\l \delta_x \r =0$. 
While we have introduced the correlations as a function of sky separation $\theta$, we can also express them as functions of the perpendicular comoving separation $\rp$, defined as
\begin{equation}
    \rp=\theta \; \chi(z)\,,
\end{equation}
in the flat sky approximation, and small angle limit, where $\chi$ is the comoving radial distance (equal to the comoving angular-diameter distance with no spatial curvature). Working with $\rp$ instead of $\theta$ is relevant when making particular scale choices for evaluating the correlation, but the Python package we use, \texttt{TreeCorr} \citep{Treecorr} has angular separation as the default convention.

\subsubsection{Moving to matter statistics with galaxy bias}\label{sec:galaxy_bias}
For large-scale correlations, the linear galaxy bias provides a relation between the galaxy overdensity and the matter density field $\delta_{\rm m}$
\begin{equation}
    \delta_{\rm g}(\Vec{\theta},\,z)=b_{\rm g}(z)\,\delta_{\rm m}(\Vec{\theta},z)\,, \label{eq:linear_bias}
\end{equation}
where $b_{\rm g}$ is the galaxy sample bias, and $\delta_{\rm m}$ is the matter density field. Under this approximation, the angular correlation function becomes,
\begin{align}
    w_{ab}(\theta)=\iint \dz_1\, \dz_2\, b_a(z_1)\,b_b(z_2)\,{n}_a(z_1)\,{n}_b(z_2)\,\xi_{\rm m}(z_1,z_2,\theta)\,,\label{eq:linear_bias_corr}
\end{align}
where $\xi_{\rm m}$ is the 3D matter two-point correlation.

Following the literature, we used the linear bias expression of Eq. \eqref{eq:linear_bias} by default. It is worth noting, however, that the linear bias expression is expected to break at scales around 6 $\hMpc$. In addition, the next-order expression would extend this scale range to around 4 $\hMpc$ \citep{scales_bias}. Thus, none of the equations (linear or non-linear e.g.  App. \ref{app:bias_next_order}) are properly adapted to the standard clustering-redshifts scale ranges, (0.1--1) $\hMpc$, (0.5--1.5) $\hMpc$ or (1--5) $\hMpc$. However, linear bias has been tested to give accurate enough modelling for clustering-redshifts analysis \citep{Gatti_DESY1,Gatti_Giulia_DESY3,van_den_Busch_2020,Cawton2022,Naidoo23}. As a consequence, it is crucial to quantify and address potential deviations introduced in the reconstructed $n(z)$. We will pay particular attention to the galaxy bias issue, nonlinear contributions, and the scale range choice.

\subsubsection{Magnification} \label{sec:magn}
In this section, we describe the two magnification terms in Eq. \eqref{eq:wab_full},
\begin{align}
    M_{ab}(\rp)&=\langle \Delta^{\rm D}_{a}\Delta^{\mu}_{b} \rangle(\rp)+\langle \Delta^{\mu}_{a}\Delta^{D}_{b} \rangle(\rp)\,.\label{eq:2magn_terms}
\end{align}
Magnification is a lensing effect due to the gravitational bending of distant source light by the matter between those sources and the observer \citep{Y3_DES_Krause, magnification_DESY3}. 
It introduces an angular position correlation for galaxies distant in redshift because of the lensing of a high-$z$ galaxy sample by the matter traced by a lower-$z$ sample.\footnote{The magnification of the two samples by the lower-$z$ matter is negligible as it is a smaller correction  ($\mu \times \mu $ cf. Sect. \ref{sec:anglular_2p}).} Fortunately, this effect, which breaks the naive hypothesis that in clustering-redshifts only galaxies close in redshift are correlated, can be modelled.  

Magnification impacts the apparent position of galaxies and their density. In regions of positive convergence,\footnote{Positive convergence is induced by galaxy over-density. Correlation with voids would probe the negative convergence regime.} the apparent separation between any two points on a source plane is enlarged, leading the telescope to capture a larger apparent solid angle, consequently reducing the apparent galaxy density locally. Another effect of positive convergence is an increase in distant source flux received by the telescope, potentially resulting in a change in the number of galaxies passing through the selection function.

Magnification was taken into account for clustering-$z$ in DES, through direct modelling of its effect, and simulation-based estimation of magnification coefficients \citep{Gatti_Giulia_DESY3, Cawton2022}. 
Complete inclusion of magnification should include position magnification, flux magnification, and its impact on photo-$z$, which can be significant (Legnani et al., in prep.), but we restricted this work to position and flux magnification. We detail how magnification can be modelled in the App. \ref{app:magnification}. 
Since the tomographic bins are expected to be narrower for \Euclid than for DES, the effect of magnification is expected to be smaller, as illustrated by the toy examples in App.~\ref{app:magnification}. Therefore, we do not include magnification in our default modelling. We will reassess the need to incorporate magnification in future work, based on the properties of the data bins.

\subsubsection{RSD}
\label{sec:rsd}
In the linear regime, RSD contributes to the projected galaxy density contrast through the apparent large-scale flow of galaxies across the redshift boundaries of the bins \citep{Y3_DES_Krause}. Equivalently, the redshift distribution of galaxies as a function of the observed redshift is slightly different from the distribution as a function of the Hubble redshift. The latter should be used, however, as the modelling assumes isotropy through the correlation function $\xi$. Since the photometric tomographic bins are quite large, it would mainly affect the spectroscopic sample which is divided into narrow redshift slices; the narrower the slices, the larger the effect is.  Therefore, only the term $\langle \Delta^{\rm RSD}_{\rm s}  \Delta^{\rm D}_{\rm p}\rangle(\theta)$ may contribute to the two-point correlation of Eq. \eqref{eq:wab_full}. We tested in Sect. \ref{sub:Res_indep} the bias induced by the use of the spec-$z$ instead of true-$z$ for the spectroscopic slicing, in the Flagship simulation, since peculiar velocities are included, cf. Sect. \ref{sec:mocks}.

\subsection{ Impact of the galaxy redshift distribution and its modelling}\label{sc:meth_redshift_dist}
A tomographic bin will be referred to as a sample p, without referring to the tomographic ID, as the latter is always explicit. 
Different spectroscopic samples are considered in this article from different simulated surveys and tracers, such as DESI ELGs. For a given survey tracer, we will additionally distribute the sample into redshift slices, with spectroscopic redshift cuts. For a given redshift slice, the galaxies will be referred to as ${\rm s}_j$, or $\{{\rm s},\, z_j\}$ with $z_j$ the mean of the redshift slice. 
We explore two cases for the modelling of the distribution and two-point correlations with a sliced spectroscopic sample. 

\subsubsection{Ideal case of a Dirac galaxy distribution}
\label{sec:dirac}
The situation often presented in clustering-redshifts literature \citep[e.g.][]{Gatti_Giulia_DESY3, DESY3_MAGLIM_z} is the one of a sample localised in such a small redshift slice that it can be approximated by a Dirac distribution
\begin{equation}
    n_a(z)\approx \delta_{\rm D}(z-z_i)\,.
\end{equation}
The cross-correlation with a second sample is then simplified to
\begin{align}
    w_{ab}^{\rm Dirac}(\theta)&=b_a(z_i)\int \dz \;b_b(z)\,{n}_b(z)\,\xi_{\rm m}(z_i,\,z,\,\theta),\label{eq:linear_bias_corr_Dirac1}\\
    &\approx b_a(z_i)\,b_b(z_i)\, n_b(z_i)\, \xi_{\rm m}(z_i,\theta)\,,\label{eq:linear_bias_corr_Dirac2}
\end{align}
with 
\begin{equation}
    \xi_{\rm m}(z_i,\theta)=\int \dz \;\xi_{\rm m}(z_i,\,z,\,\theta)\,.\label{eq:wdm_simple}
\end{equation} 
To go from Eq. \eqref{eq:linear_bias_corr_Dirac1} to Eq. \eqref{eq:linear_bias_corr_Dirac2}, we assume the redshift evolution of the bias and distribution of the second sample is small in the support of $\xi_{\rm m}$ (i.e. in a small redshift interval centred in $z_i$).

\subsubsection{Realistic case: Spectroscopic sample into small redshift slices}
\label{sec:real_spec_bins}
Using the Limber approximation, Eq. \eqref{eq:linear_bias_corr} reduces to,
\begin{align}
    w_{ab}^{\rm Limber-full-z}(\theta)=\int \dz\; b_a(z)\,b_b(z)\,{n}_a(z)\,{n}_b(z)\,\xi_{\rm m}(z,\,\theta)\,,\label{eq:wm_limber}
\end{align}
where $\xi_{\rm m}$ is the same as in Eq. \eqref{eq:wdm_simple}. In the standard scenario of clustering redshifts, the spectroscopic galaxy sample is localised within small redshift slices $z_i\pm \Delta z_i/2$, so that the distribution can be approximated as uniform
\begin{equation}
    n_a(z)=\begin{cases} 
1/\Delta z_i & \text{if } \vert z-z_i\vert <\Delta z_i/2\,,  \\
0 & \text{ else,}
\end{cases}\label{eq:nz_real}
\end{equation}
where  $\Delta z_i$ is typically between 0.01 and 0.05. This modelling is of course only approximated true in practice, and would lead to some deviations which we will evaluate. 
In particular, under the assumption of constant galaxy bias across the redshift bin, the spectroscopic auto-correlation ($b=a$) becomes
\begin{equation}
    w_{aa}^{\rm Limber-1-bin}(\theta)\approx\frac{b_a^2(z_i)}{\Delta z^2}\int_{-\Delta z_i/2}^{\Delta z_i/2}\dz\; \xi_{\rm m}(z_i+z,\,\theta)\,.\label{eq:bias_waa}
\end{equation}
Thus, given a cosmology, $w_{aa}$ provides a straightforward measurement of the spectroscopic galaxy bias. We note that this equation diverges as $\Delta z \rightarrow 0$, and that, more generally, both our modelling and the Limber approximation are expected to become less accurate for narrower redshift bins and on larger angular scales \citep{how_accurate_is_limber}.

For $b\neq a$, when the redshift evolution of the biases $b_a$ and  $b_b$ and the redshift distribution $n_b$ is neglected, we obtain 
\begin{equation}
    w_{ab}^{\rm Limber-1-bin}(\theta)=\frac{b_a(z_i)}{\Delta z_i}\,b_b(z_i)\,{n}_b(z_i)\,\int_{-\Delta z_i/2}^{\Delta z_i/2}  \dz\;\xi_{\rm m}(z_i+z,\,\theta)\,. \label{eq:wm_bin}
\end{equation}
The superscript Limber-1-bin refers to the integration of the matter distribution over a redshift bin equal to the spectroscopic slice; we uses the Limber approximation too.  If one further neglects the matter correlation variation over $\Delta z_i$, one recovers the Dirac approximation Eq. \eqref{eq:linear_bias_corr_Dirac2}; 
this is the approach of \cite{van_den_Busch_2020}, for example.  

A limitation of Eq. \eqref{eq:wm_bin} is the assumption that the redshift distribution $n_b$ is constant across the bin. In practice, $n_b$ could be larger near the bin edges and smaller near the centre (or the opposite), leading to an underestimation (resp. overestimation) of correlations with galaxies localised near $z_i\pm \Delta z_i$.
To address these issues without resorting to Eq. \eqref{eq:wm_limber}, we propose a correction by explicitly incorporating contributions from galaxies outside the spectroscopic redshift slice (three-bins):
\begin{align}
    w_{ab}^{\rm 3-bins}(\theta)=\frac{b_a(z_i)}{\Delta z_i}\sum_{\delta z\in\{0,\,-\Delta z_i,\, \Delta z_i\}}& b_b(z_i+\delta z)\,{n}_b(z_i+\delta z)\label{eq:wm_bins}\\
    &\times\int_{-\Delta z_i/2}^{\Delta z_i/2} \dz\;\xi_{\rm m}^{\,\delta z}(z_i+z,\,\theta)\,,\nonumber
\end{align}
 We introduce the phased correlations
\begin{equation}
    \xi_{\rm m}^{\,\delta z }(z,\,\theta)=\int_{-\Delta z_i/2}^{\Delta z_i/2} \dz' \;\xi_{\rm m}(z,\, z+z'+ \delta z,\,\theta)\,,\label{eq:wdm_phase}
\end{equation}
which account for the correlation of two bins of widths $\Delta z_i$, separated by $\delta z$. Here, the three-bins modelling does not uses the Limber approximation.
One can write Eq. \eqref{eq:wm_bins} as 
\begin{align}
    w_{ab}&(\theta)\approx b_a(z_i)\, \int_{-\Delta z_i/2}^{\Delta z_i/2} \dz\;\xi_{\rm m}^{\,0}(z_i+z,\,\theta)\label{eq:nb_eta_3.2.3}\\
    &\times \sum_{j\in \{-1,0,1\}}  b_b(z_i+j\,\Delta z)\, n_b(z_i+j\,\Delta z) \;\eta_j(z_i,\theta)
    \,,\nonumber
\end{align}
with $\eta_0=1$ and
\begin{align}
    \eta_\pm(z_i,\, \theta)=\frac{\int_{-\Delta z_i/2}^{\Delta z_i/2} \dz\;\xi_{\rm m}^{\pm \Delta z_i}(z_i+z,\,\theta)}{\int_{-\Delta z_i/2}^{\Delta z_i/2} \dz\;\xi_{\rm m}^{\,0}(z_i+z,\,\theta)}\,,
\end{align}
where $\xi_{\rm m}^{\,0,\pm\Delta z_i}$ is defined in Eq. \eqref{eq:wdm_phase}.
This formulation allows for a linear inference of $\Vec{n}$ from the two-point correlation, which can be implemented in practice, unlike the Limber-full-$z$ case where $n_{\rm p}(z)$ is integrated over, complicating the process.

\subsubsection{Comparison of the correlation modelling}\label{sec:comparison_corr}

\begin{figure}
    \centering
    \includegraphics[width=1\linewidth]{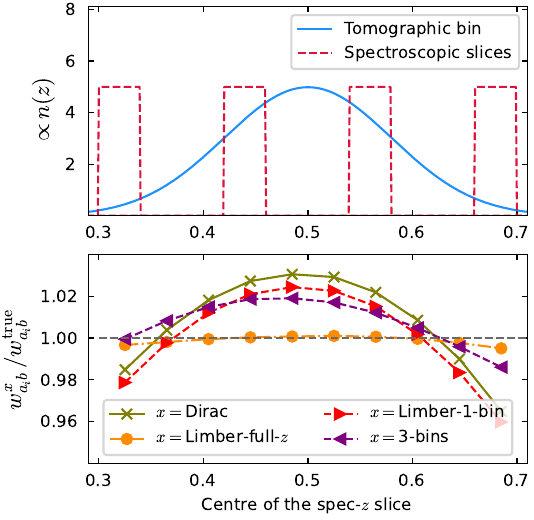}
    \caption{\textit{Top}: Redshift distribution of sample $b$ in blue, and four out of the ten samples $a_i$ in red. \textit{Bottom}: Deviation of approximated $w_{a_ib}$ with respect to the exact one for the ten cross-correlations.}
    \label{fig:Toy_model_wab}
\end{figure}

In the previous sections, we introduced four different approximations of the two-point correlation function described by Eq. \eqref{eq:linear_bias_corr}: $w_{ab}^{\rm Dirac}$, $w_{ab}^{\rm Limber}$, $w_{ab}^{\rm 1-bin}$, and $w_{ab}^{\rm 3-bins}$.
In this section, we investigate the differences between them with a toy model.  
We consider 10 samples $a_i$, each with a step-$n(z)$ distribution as given by Eq. \eqref{eq:nz_real}, centred at $z_i=0.325 +  0.05\, i$, with $i=0,1\dots,9$, and $\Delta z_i = 0.05$. 
For the redshift distribution of sample $b$ we take a Gaussian $(\mu,\, \sigma)=(0.5,0.08)$. This reproduces an idealistic case scenario of clustering-$z$ calibration, with sample $b$ representing the photometric data, and samples $a_i$ representing the spectroscopic data. Several of these $n(z)$s are plotted in the upper panel of Fig. \ref{fig:Toy_model_wab}. We assume a constant galaxy bias $b_a(z)=b_b(z)= 1$ for all samples, and evaluate correlation functions for $r_{\rm p}=2$ Mpc (a scale standard for clustering-$z$ analysis). We numerically evaluated the functions with the \texttt{CCL} library\footnote{\url{https://github.com/LSSTDESC/CCL}} \citep{CCL}. We used the default configuration, which is based on predictions from \texttt{CLASS} \citep{Diego_Blas_2011_class} and \texttt{halofit} \citep{halofit} for the nonlinear power spectrum. We evaluated the functions at scales where we know the predictions are not completely precise because of the galaxy-halo connection, but as will be explicit in Sect. \ref{sc:method_clust_z}, this is expected to have little impact on real data.  

In the lower panel of Fig. \ref{fig:Toy_model_wab}, we show the deviation of the four approximations to the ‘‘exact’’ correlation Eq. \eqref{eq:linear_bias_corr}, for the samples $a_i,\, b$. The level of the deviations depends on the $n_b(z)$ where the correlation is evaluated: we observe a larger deviation near the peak and for the tails. We evaluated the root mean square (RMS)  of the deviation to 1  and got 0.022 (Dirac), 0.021 (Limber-1-bin), 0.013 (3-bins), and 0.002 (Limber-full-$z$). We can see that the Dirac and one-bin approximations are very close. Taking into consideration the correlations from bin neighbours with three-bins, we get an improvement of $40 \%$. Finally, the best approximation is the Limber-full-$z$ case [i.e. integrating $n^2(z)$ instead of assuming some effective values], with a deviation of 0.2$\%$ for individual points, and an RMS smaller by an order of magnitude to the Dirac and one-bin approximations.

In clustering-$z$ studies, however, the redshift distribution $n_{\rm p}(z)$ is an unknown quantity that we aim to constrain from the measurement of $w_{ab}$. As a result, the Limber-full-$z$  case is challenging to apply in practice, as the unknown $n_{\rm p}(z)$ is integrated over. In contrast, for the Dirac and one-bin cases, $n_{\rm p}(z_i)$ enters the calculation linearly, making it easier to extract information. The same applies to the three-bins approximation, which provides a better approximation, as we have demonstrated.

\subsection{Application to clustering redshifts}\label{sc:method_clust_z}

In the context of clustering redshifts, a photometric sample has an unknown redshift distribution $n_{\rm p}(z)$. To reconstruct it, we measure the cross-correlations with spectroscopic samples, distributed into fine redshift slices (usually $0.02\leq \Delta z\leq 0.05$). 
The distribution of these spectroscopic samples is assumed to be constant in each of these smaller slices (see Eq. \ref{eq:nz_real}). The amplitude of the photometric distribution at redshift $z_i$ can then be inferred linearly from the cross-correlation function between a spectroscopic bin centred in $z_i$ and the photometric sample [noted $\wsp(z_i,\,\theta)$ instead of $w_{{\rm s}_i{\rm p}}(\theta)$],  assuming the Dirac modelling Eq. \eqref{eq:linear_bias_corr_Dirac2}: 
\begin{equation}
    \np(z_i)=\frac{\wsp(z_i,\,\theta)}{b_{\rm s}(z_i)\,b_{\rm p}(z_i)\,\xi_{\rm m}(z_i,\,\theta)}\,. \label{eq:eq_wsp_n_b}
\end{equation}
One can write a similar expression for the one-bin and three-bins modelling, as described in Sect. \ref{sec:comparison_corr}.

The spectroscopic bias can be measured with the spectroscopic auto-correlation, following Eq. \eqref{eq:bias_waa},
\begin{equation}
    \np(z_i)=\frac{\wsp(z_i,\,\theta)}{\sqrt{\Delta z\, w_{\rm ss}(z_i,\,\theta)\,\xi_{\rm m}(z_i,\,\theta)}\,b_{\rm p}(z_i)}\,.\label{eq:npz=wsp/bpbs}
\end{equation}
The $n_{\rm p}(z)$ is still degenerate with the photometric galaxy bias. In the next section, we describe different possibilities for correcting this degeneracy.

\subsubsection{Correcting the photometric galaxy bias}
\label{sec:correcting_bias}
It is more complicated to measure the photometric bias than the spectroscopic bias since the sample cannot be split into small redshift slices and Eq. \eqref{eq:bias_waa} cannot be applied. Here we introduce different methods.

\begin{itemize}
    \item M0: No correction at all. We assume there is no $b_{\rm r}$ and $b_{\rm p}$ redshift evolutions  and renormalise the measured redshift distribution. 
    \item M1: Spectroscopic bias only. We assume  there is no  $\bp$ redshift evolution and renormalise the measured redshift distribution.
    \item M2: The photometric bias is measured for the whole tomographic bin. Hence we do not take into account the redshift evolution within the tomographic bin, but we are increasing the uncertainty on the $\np(z_i)$, and thus we are more conservative.
    \item M3: The photometric bias is measured for relatively large photometric bins ($\Delta z_{\rm photo}=0.1$), and the redshift evolution is deduced from a polynomial fit. The photo-$z$ spreading of the bin needs to be corrected before the fit (more details in Sect. \ref{sec:moredetails_M3M4}).
    \item M4: The photometric bias is measured for small photometric bins ($\Delta z_{\rm photo}=0.02$), and the redshift evolution is deduced from a polynomial fit. The photo-$z$ spreading of the bin needs to be corrected before the fit (cf. Sect. \ref{sec:moredetails_M3M4}).
    \item M5: Full correction. The two biases are measured using the true redshift (simulation only). 

\end{itemize}
Methods 0-1-2-5 were already tested in \cite{Naidoo23}. Methods 3 and 4 are new in that context, but similar to methods from \cite{Cawton2022} and \cite{van_den_Busch_2020}.
Ideally, we would like to measure $w_{\rm ss}$ and $w_{\rm pp}$ over the same redshift slices: $\left[z_i \pm \Delta z/2\right[$. In doing so, we would get 
\begin{equation}
    \np(z_i)=\frac{\wsp(z_i,\,\theta)}{\Delta z\,\sqrt{ w_{\rm ss}(z_i,\,\theta)\,w_{\rm pp}(z_i,\,\theta)}}\,.\label{eq:npz=wsp/wsswpp}
\end{equation}
We note that we derive this equation with the linear bias assumption. Indeed we show that it remains valid at the next-order of the bias expansion in App. \ref{app:bias_next_order}.
This ideal situation corresponds to M5, the photometric sample being distributed into redshift slices with the true redshift variable. This method can only be evaluated via simulation. The aim of methods 3 and 4 is to estimate $w_{\rm pp}$ but with a different binning, and taking into account the photo-$z$ spreading. With M2, we don't correct for the redshift evolution, but we aim to take into account the variance induced by the photometric galaxy bias \citep{Naidoo23}.

\subsubsection{More details about the M3 and M4 corrections}
\label{sec:moredetails_M3M4}
With methods 3 and 4, the photometric sample is divided into bins (denoted $j$) based on photo-$z$ cuts. These differ from the spectroscopic distribution described in Eq. \eqref{eq:nz_real}, therefore requiring correction for the photo-$z$ spreading. In other words, a photometric bin cannot be approximated by a redshift slice.

In practice, we measure the $n(z)$ of these bins with the clustering-redshifts M1 method (cf. Sect. \ref{sec:correcting_bias}). This method assumes negligible photometric redshift evolution within each small bin. Then, we want to convert our imperfect-bin measurement to what we would measure for an ideal case with step $n(z)$. Using the Limber case (cf. Sect. \ref{sec:comparison_corr}), we introduce the correction to the two-point auto-correlation function
\begin{equation}
    w_{\rm pp}^{\rm corr}(z_j,\,\theta)=w_{\rm pp}^{\rm meas}(z_j,\,\theta)\; \frac{{(\Delta z)^{-2}}\int_{-\Delta z/2}^{\Delta z/2} \dz \;\xi_{\rm m}(z_j+z,\,\theta)}{\int_0^{\infty} \dz\; n_{{\rm p}_j}^2(z)\, \xi_{\rm m}(z,\,\theta)}\,,\label{eq:full_corr}
\end{equation}
where the $\Delta z$ corresponds to the spectroscopic redshift bin of the $w_{\rm sp}$ correlation. We refer to this as the ‘‘full’’ correction, as it accounts for the redshift evolution of the matter correlation function across the bin. We use \texttt{CCL} to evaluate $\xi_{\rm m}$ at small scales. 

A simpler approximation has been implemented for the DES redMaGiC and Maglim samples \citep{Cawton2022,DESY3_MAGLIM_z}. The correction was modelled with
\begin{equation}
    w_{\rm pp}^{\rm corr}(z_j,\,\theta)=w_{\rm pp}^{\rm meas}(z_j,\,\theta)\;\frac{(\Delta z)^{-1}}{\int_0^\infty \dz\, n_i^2(z)}\,,\label{eq:partial_corr}
\end{equation}
where $n_i(z)$ was estimated for the redMaGiC subsample having a spec-$z$  and a photo-$z$ \citep{Gatti_Giulia_DESY3}, or with clustering-redshifts \citep{Cawton2022}. This is referred to as the ‘‘partial’’ correction, as it makes additional simplifying assumptions compared to the ‘‘full’’ correction in Eq. \eqref{eq:full_corr}.

The ‘‘width-only’’ correction applies when using different bin widths for the auto- and cross-correlation functions, while still neglecting the effects of photo-$z$ spreading across the bin. This correction simplifies to
\begin{equation}
    w_{\rm pp}^{\rm corr}(z_j,\,\theta)=w_{\rm pp}^{\rm meas}(z_j,\,\theta)\;\frac{(\Delta z)_{\rm meas}}{(\Delta z)_{\wsp}}\,. \label{eq:width_corr}
\end{equation}
In summary, the ‘‘full’’ correction accounts for the redshift evolution of the matter correlation function, the ‘‘partial’’ correction introduces additional approximations by assuming a simplified $n(z)$, and the ‘‘width-only’’ correction applies when focusing solely on differences in bin widths while neglecting photo-$z$ spreading.
We illustrate these methods and apply them to clustering-redshifts in Sect. \ref{sub:Res_bias_corr}. 

These two methods for correcting photometric galaxy bias share similarities with the method in \cite{van_den_Busch_2020}.  We do not assume a bias model of the form $B_{\rm p}(z)\propto (1+z)^\alpha$, however. Instead, we directly measure and correct $w_{\rm pp}(z)$ by constraining the photo-$z$ spreading using clustering-$z$ for smaller bins. 
Theoretically, the two approaches could be combined, which would eliminate any systematic effects continuous on the redshift, without assuming a particular model.

\subsubsection{Data vectors reduced to scalars with some weighting}
\label{sec:weighting_w}
Using Eq. \eqref{eq:npz=wsp/bpbs}, one can extract a measurement of $\np$ for different $r_{\rm p}$ (or equivalently $\theta$). To optimise the sensitivity, it is standard to reduce all the data vectors to  scalars with a scale weighting $W_w(r_{\rm p})$,
\begin{equation}
    \overline{w}_{ab}(z_i)=\int_{\rpmin}^{\rpmax} \mathrm{d}\rp \;W_w(r_{\rm p})\,w_{ab}(r_{\rm p},\,z_i)\,.\label{eq:weight_w}
\end{equation}
We introduce the scale-weighted matter correlation
\begin{equation}
    \overline{\xi}_{\rm m}(z)= \int_{\rpmin}^{\rpmax} \mathrm{d}\rp \;W_w(r_{\rm p})\,\xi_{\rm m}(r_{\rm p},\,z)\,.
\end{equation}
It is easy to show that Eq. \eqref{eq:npz=wsp/bpbs} or Eq. \eqref{eq:npz=wsp/wsswpp} are still valid with integrated quantities. 
The scale weighting is usually chosen as a power law $W_w(\rp)\propto \rp^\gamma$, where the function is normalised to unity \citep[e.g. ][]{Gatti_DESY1,van_den_Busch_2020,Naidoo23}. Choosing $\gamma<0\ $ increases the impact of small scales. They are expected to be less noisy but also more sensitive to non-linearities. Inversely, the measurement is not biased at large scales (i.e. the linear bias is a good model), but uncertainties are usually larger. Another reason to use small scales instead of large ones is to keep the calibration independent from the two-point weak-lensing analysis: using smaller and different scales, one can safely neglect the covariance between calibration and the observables. Nonetheless, these small scales are not used for the cosmological analysis for physically motivated reasons (non-linearities), which in theory should apply to clustering-redshifts as well \citep{scales_bias}. Hence, the scale range is an important parameter and we will investigate this scale-weighting procedure in detail in Sect. \ref{sc:Result}.

Another possibility of weighting is to combine the different $n_{\rm p}(z_i\vert \rp)$, with a particular weighting $W_n(\rp)$:
\begin{equation}
    \overline{n}_{\rm p}(z_i)=\int_{\rpmin}^{\rpmax} \drp \;W_n(r_{\rm p})\,n_{\rm p}(z_i\vert r_{\rm p})\,.\label{eq:weight_n}
\end{equation}
In other words, we integrate directly $n(z\vert\, \rp)$, which is  $w_{\rm sp}$ for M0, $w_{\rm sp}/\sqrt{w_{\rm ss}}$ for M1 or $w_{\rm sp}/\sqrt{w_{\rm ss}\, w_{\rm pp}}$ for M2-3-4-5.

\subsection{Estimating the correlations with pair counts}\label{sc:method_data}
We evaluated position-position correlations by counting pairs separated by a certain distance divided by the expectation of uncorrelated samples (random) to properly normalise the correlation.
We  chose the Landy--Szalay (LS) estimator \citep{Landy_Szalay},
\begin{align}
    w_{ab}^{\rm data}(\rp )&=\frac{{\rm D}_a{\rm D}_b(\rp)-{\rm R}_a{\rm D}_b(\rp)-{\rm D}_a{\rm R}_b(\rp)+{\rm R}_a{\rm R}_b(\rp)}{{\rm R}_a{\rm R}_b(\rp)}\,,\label{eq:DD_RD_DR_RR/RR}
\end{align}
where ${\rm D}_a{\rm D}_b,\, {\rm D}_a{\rm R}_b,\,{\rm R}_a{\rm D}_b,$ and $ {\rm R}_a{\rm R}_b $ are the standard data-data, data-random, random-data, random-random pair counts of samples $a$ and $b$. 
We note that we omitted the galaxy number normalisation factors which correct for the randoms being ten times more numerous than data. Further details are given in App. \ref{App:pair_count}. As we were working with a simulation omitting observational systematics and masks, we generated random catalogues homogeneously over the footprint, taking 50 times more randoms than data.

\subsubsection{Estimator weighting scheme }
\label{sec:estimator_average}
Given the weighting procedure Eq. \eqref{eq:weight_w}, it is natural to weight the estimator as
\begin{equation}
    \Tilde{w}_{ab}^{(1)}(z_i)=\int_{\rpmin}^{\rpmax}\mathrm{d}\rp\; W(\rp)\frac{({\rm D}_a-{\rm R}_a)({\rm D}_b-{\rm R}_b)(\rp)}{{\rm R}_a{\rm R}_b(\rp)}\,,\label{eq:weightproc1}
\end{equation}
where the numerator is a concise notation for the numerator of Eq. \eqref{eq:DD_RD_DR_RR/RR}.
In the context of clustering redshifts, this weighted estimator was first introduced by \cite{Menard2013}, and later used by \cite{Cawton2022}.

 Another weighting scheme was introduced by \cite{Schmidt2013}, and was used by  \cite{Davis2018}, \cite{Gatti_DESY1}, \cite{van_den_Busch_2020}, and \cite{Naidoo23}. The principle is to scale-average data-data and random-random pair counts: 
\begin{equation}
    \Tilde{w}_{ab}^{(2)}(z_i)=\frac{\int_{\rpmin}^{\rpmax}\mathrm{d}\rp \;W(\rp)\,({\rm D}_a-{\rm R}_a)({\rm D}_b-{\rm R}_b)(\rp)}{\int_{\rpmin}^{\rpmax}\mathrm{d}\rp\; W(\rp)\,{\rm R}_a{\rm R}_b(\rp)}\,.\label{eq:weightproc2}
\end{equation}
Schmidt's motivation to do so was to assume that the ratio of the averaged quantities would lead to higher signal-to-noise (S/N) than the averaged ratio (which is expected for a constant ratio, which is not the case here). 
With this second estimator what we have is a re-weighting of the correlation functions with a new effective weighting, as
\begin{align}
   \Tilde{w}_{ab}^{(2)}(z_i)= &\iint \dz_1\,\dz_2\; n_a(z_1)\,n_b(z_2) \label{eq:second_estim_calcul}\\
\times&\frac{\int_{\rpmin}^{\rpmax}{\rm d}\rp\; W(\rp)\,\rp\,\Delta \rp \,\eta_{\rm mask}(\rp)\,\l \delta_a\,\delta_b\r_{\rp,\,z_1,\,z_2}}{\int_{\rpmin}^{\rpmax}{\rm d}\rp\; W(\rp)\,\rp\,\Delta \rp\,\eta_{\rm mask}(\rp)}\,,\nonumber
\end{align}
where $\eta_{\rm mask}(\rp)$ is a geometrical factor due to the masking of some sky area (cf. App. \ref{App:pair_count}).
In particular, the effective scale weighting is
\begin{align}
    W_{\rm eff}(\rp)&\propto W(\rp)\,\rp\,\Delta \rp\, \eta_{\rm mask}(\rp)  \label{eq:Weff}\\
&\propto W(\rp)\,\rp^2\, \eta_{\rm mask}\text{ \hspace{0.3cm } for a logarithmic binning,}\\
&\propto \rp^{\gamma +2}\, \eta_{\rm mask}\text{ \hspace{0.74cm } for a power law weighting }.
\end{align}
This effective weighting is still re-normalised to unity as required by the denominator of Eq. \eqref{eq:second_estim_calcul}. The mask incompleteness is not rigorously corrected (cf. App. \ref{App:pair_count}), except if $\eta_{\rm mask}$ is scale-independent.\footnote{For scales that are neither too small nor too large, it should be a valid hypothesis, at least to first order.} 
By default, we consider the first estimator (Eq. \ref{eq:weightproc1}), since understanding the weighting is straightforward, and the mask is better corrected, but we will test and compare the two in the results section.\footnote{
Of course, in practice, all the previous integrals are discrete, with 
\begin{equation}
    \int \drp\longrightarrow \sum_i \Delta \rp^i\,.
\end{equation}}

\subsubsection{Covariance matrix}
We used the jackknife covariance matrix as our fiducial covariance.
For a measured data vector $w$, the covariance matrix is
\begin{equation}
    \mathcal{C}_w=\frac{N_{\rm Jkk}-1}{N_{\rm Jkk}}\sum_{k=1}^{N_{\rm Jkk}} (w^k-\Tilde{w})(w^k-\Tilde{w})^\top\,,
\end{equation}
where $N_{\rm Jkk}$ is the number of jackknife regions, $w^k$ is the data vector excluding the data from the $k^{\rm th}$ region, and $\Tilde{w}$ is its mean value. Here, $w$ can be, for example, the vector $w_{\rm sp}(\rp)$, the ratio of vectors $w_{\rm sp}(\rp)/\sqrt{w_{\rm ss}(\rp)}$, the scalars $\overline{w}_{\rm sp}$, or the ratio of scalars $\overline{w}_{\rm sp}/\sqrt{\overline{w}_{\rm ss}}$. 
In this article, we use a sufficiently large number of regions (200), with a small enough number of points (around a dozen), so that Hartlap \citep{Hartlap} or Percival \citep{Percival_factor} corrections to the inverse of the covariance are at the percent level and can be safely neglected.

\subsection{\label{sc:metho_Fitting} Extracting the unknown distribution moments}

\subsubsection{clustering-redshifts requirements for cosmic shear} \label{sc:requirement}

The success of the \Euclid mission requires achieving a particular precision in the characterisation of $n(z)$ of the tomographic bins $i$.
The requirements usually focus on the mean redshift, expressed as
\begin{equation}
\langle z \rangle_{i} = \int \dz \; z\, n_{\rm p_i}(z)\,,\label{eq:def_meanz}
\end{equation}
as illustrated in works such as \cite{Gatti_DESY1}, \cite{van_den_Busch_2020}, and \cite{Gatti_Giulia_DESY3}. The desired precision for the \Euclid mission requires knowledge of the redshift standard deviation, however, which is denoted as
\begin{equation}
\sigma_{n_{i}}= \sqrt{\int \dz \; (z-\langle z\rangle)^2\, n_{\rm p_i}(z)}\,,
\end{equation}
as outlined in \cite{propagating_photo_z_uncertainty}. The  criteria are as follows: the mean redshift bias \citep{Laureijs11,propagating_photo_z_uncertainty} and the redshift standard deviation \citep{propagating_photo_z_uncertainty} for every bin should be measured with precision
\begin{align}
&\sigma(\langle z\rangle_i)\leq 0.002\, (1+\langle z\rangle_i),\label{eq:requ_z}\\
&\sigma(\sigma_{n_i})\leq 0.1\, \sigma_{n_i}\,.\label{eq:requ_s}
\end{align}

\subsubsection{Discretisation of a distribution}\label{sec:discrete_nzi}

\begin{figure}
    \centering
    \includegraphics[width=1\linewidth]{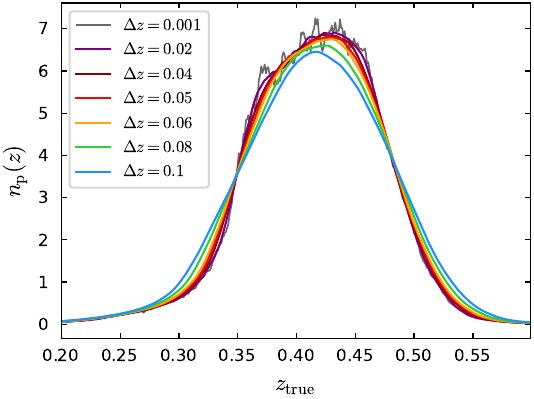}
    \caption{ Galaxy distribution observed through different $\Delta z$ averaging as Eq. \eqref{eq:nz_avg}. The redshift convolution modifies the shape of the distribution for larger slicing compared to the real slicing as estimated with a very thin slicing (grey). For this case, all slicings $\Delta z \leq 0.06$ produce a good estimate.}
    \label{fig:nz_binning}
\end{figure}

The last step of a clustering-redshifts pipeline is to derive the $n_{\rm p}(z)$ moments from the discrete measurements $n_{\rm p}(z_i)$. 
The mean redshift of a bin is evaluated with
\begin{align}
    \l z\r 
    &\approx \sum_j \Delta z\, z_j \,n^{\rm meas}_{\rm p}(z_j)\label{eq:mean_z}\,.
\end{align}
We are not directly measuring $n_{\rm p}(z_j)$ but rather its averaged value over ranges $z_j\pm \Delta z/2$,
\begin{equation}
    n_{\rm p}^{\rm meas}(z_j)\approx\frac{1}{\Delta z}\int_{-\Delta z/2}^{\Delta z/2} \dz\; n_{\rm p}(z_j+z)\,. \label{eq:nz_avg}
\end{equation}
As a consequence, Eq. \eqref{eq:mean_z} is an accurate approximation of the exact integral form as long as $\Delta z$ is small compared to the tomographic bin width. As an example, the numerical difference between  Eq. \eqref{eq:mean_z} and Eq. \eqref{eq:def_meanz} is ${\bigO{10^{-4}}}$ for our \Euclid tomographic bins (cf. Sect. \ref{sc:data_euclid_photo}), and $\Delta z=0.02$--$0.5$. Nonetheless, the recovered shape of the distribution is affected by Eq. \eqref{eq:nz_avg} and for large $\Delta z$ the measured distribution would be flattened. We illustrate this in Fig. \ref{fig:nz_binning}, plotting a galaxy distribution (which in practice corresponds to a very thin slicing), and the same distribution with a $\Delta z$ averaging as Eq. \eqref{eq:nz_avg}. We see that the flattening associated with the bin increases. These effects will also be tested in Sect. \ref{sc:Result}. 

\subsubsection{Overall strategy}

The uncertainty on the mean-$z$ directly inferred with Eq. \eqref{eq:mean_z} would not fulfil the \Euclid requirement Eq. \eqref{eq:requ_z}. In doing this, we would be very conservative, allowing for any tomographic bin shape, whereas in reality, we know their shape a priori. Thus, we can define some priors on the $n_{\rm p}(z)$ properties to decrease the uncertainty. We describe two possibilities: 
\begin{enumerate}
    \item we assume a prior on the shape of the distribution, through a $n(z)$ model;
    \item we use Gaussian process formalism, to generate coherent realisations, assuming a kernel, which is related to how ‘‘fast’’ the distribution is evolving in redshift.  
\end{enumerate}
Thus in one case, we put a prior on the shape of the tomographic bin, and in the other, on the redshift variation. 
\begin{figure*}
    \centering
    \includegraphics[width=1\linewidth]{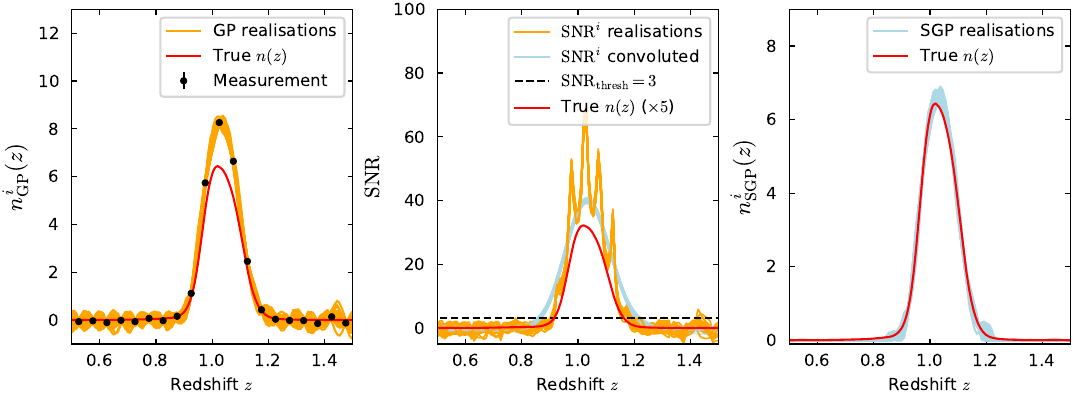}
    \caption{\textit{Left}:  $n(z)$ distribution measurement, with 50 associated GP realisations (orange), and the true distribution (red). \textit{Middle}: SNR of the GP realisation (orange), and their convolution (blue). The SNR threshold (=3) is reported with the dashed line. \textit{Right}: Associated SGP realisations, renormalised to unity (blue), compared to the true distribution (red). }
    \label{fig:SNR_GP}
\end{figure*}
\subsubsection{Shifted-stretched model }
For this approach, we evaluated a given model $n^{\rm model}(z)$ with some parameters $\{p_1,p_2\ldots \}$, by minimising the likelihood,
\begin{align}
    \ln \mathcal{L}&(\{p_1,p_2\ldots \}\vert n, \mathcal{C}_n)\nonumber\\
&=-\frac{1}{2}  \left[ n(z_i)-n^{\rm model}(z_i)\right]^\top \mathcal{C}_n^{-1}\left[ n(z_i)-n^{\rm model}(z_i)\right]\,,
\end{align}
where $\{n_{\rm p}(z_i)\}_i$, and the associated covariance $\mathcal{C}_n$ are the measured quantities. 
Then, using a Markov chain Monte Carlo,\footnote{We use the python package \texttt{emcee} \citep{emcee}.} the mean redshift and standard deviation were extracted from one thousand realisations of $\{n^{\rm model}\}$. 

For the $n^{\rm model}$, we use the shifted-stretched model (SSM), which is commonly used for clustering redshifts; given a particular distribution $n(z)$ with a mean redshift $\langle z\rangle$, two free parameters are used describing a redshift shift $\delta z$ and a stretching $s$,
\begin{equation}
    n^{\rm model}(z \vert s,\delta z)=\frac{1}{s}n\left(\frac{z-\langle z\rangle -\delta z}{s}+\langle z\rangle \right)\,.
\end{equation}
This model is particularly convenient because for a given $\delta z,\, s$, the bias in mean redshift and standard deviation with respect to the model are $\delta z$ and $s$. In our case, as the model is the true redshift distribution, we have 
\begin{align}
    \delta z=&\l z\r -\l z\r_{\rm true},\,\\
    s=&\sigma_n/\sigma_n^{\rm true}\,.
\end{align}
As we use the true redshift distribution of the bin as a model (simulation only), the performance may be overestimated compared to a realistic scenario, where the model is taken from a Self-Organised-Map estimate \citep[e.g.][]{KIDS_redshift_dis}, or a candidate from a simulation \citep[e.g.][]{Gatti_DESY1,Gatti_Giulia_DESY3,Cawton2022}. That is why we also consider a more direct method. 
\label{sec:SSM}

\subsubsection{Suppressed Gaussian process}\label{sec:SGP}

We introduce the suppressed-Gaussian-process model (SGP) developed by \cite{Naidoo23}; GP reconstruction from clustering-redshifts was also performed in \cite{GP_Johnson}. The main benefit of this approach is that it is non-parametric. 
The procedure is as follows:
\begin{enumerate}
    \item we generate realisations of Gaussian Process (GP) where the multivariate normal distributions are the measurements, and model the correlation between the points through a specific kernel;\footnote{We use the python package \texttt{Georges} \citep{George}.}
    \item we apply to the realisations a suppression function that damps signals in regions where the measurements are consistent with zero; 
    \item the realisations are renormalised to unity;
    \item the moments of the true distribution and their uncertainties are estimated from this large sample of normalised SGP realisations.
\end{enumerate}

For the GP covariance, we chose a 3/2 Matern kernel function, and assume the length-scale $l$ to be  $l=\Delta z/2$ where  $\Delta z$ corresponds to the redshift slicing.\footnote{The length-scale represents the characteristic distance over which the function values are correlated.} To validate these choices, we tested different kernels and length-scale, by comparing generated GP from discrete $n(z_i)$ and covariance to true $n(z)$; see, for example, the right panel of Fig. \ref{fig:SNR_GP}.

For the second step, given a GP realisation $i$, the associated SPG is 
\begin{equation}
    n_{\rm SGP}^i (z)=n_{\rm GP}^i(z)\; S(x_i,k)\,,
\end{equation}
where $S$ is a suppression factor that removes the noisy part of the distribution. We chose the parametrisation 
\begin{equation}
    S(x,k)=\begin{cases} 
0 & \text{if } x<0  \\
1-(1-x)^k &\text{if } 0<x<1\\
1 & \text{if } x>1\,.
\end{cases}\label{eq:S_x}
\end{equation}
 In the above, the $x$ variable is a smoothing of the continuous signal-to-noise ratio (SNR)
\begin{equation}
    x_i=\frac{1}{{\rm SNR}_{\rm thresh}}\mathcal{G}\ast \frac{n_{\rm GP}^i(z)}{\sigma_i(z)}\,.
\end{equation}
Here $\mathcal{G}$ is a Gaussian with a standard deviation $\Delta z$, $\ast$ represents the convolution operation, which smooths the SNR between data points (and avoids non-physical cuts), and ${\rm SNR}_{\rm thresh}$ is the SNR threshold below which the suppressed factor is applied. Thus, if ${\rm SNR}<{\rm SNR}_{\rm thresh}$ over a redshift range larger than $\Delta z$, the amplitude of the realisation $n_{\rm GP}^i(z) $ would be reduced by a factor $k/3\times {\rm SNR}$.
We fix $k=0.3$ in Eq. \eqref{eq:S_x}, so that the suppressed factor rapidly drops for $x<1$, but we ensured that the impact of this parameter was small.  

In Fig. \ref{fig:SNR_GP} we illustrate the whole procedure. We show for a measurement $\{n_{\rm p}(z_i)\}_i,\, \mathcal{C}_n$ (not normalised to unity), GP realisations, their SNR $n_{\rm GP}/\sigma(z)$, and the associated normalised SGP realisations compared with the true distribution. It is clear that going from the left to the right panel, the overall procedure greatly improves the realisations. Nonetheless, the procedure could be optimised further in the future, since we did not rigorously explore all of the SGP parametrisation.

\section{\label{sc:Result}Results}
\begin{figure*}
    \centering
    \includegraphics[width=0.8\linewidth]{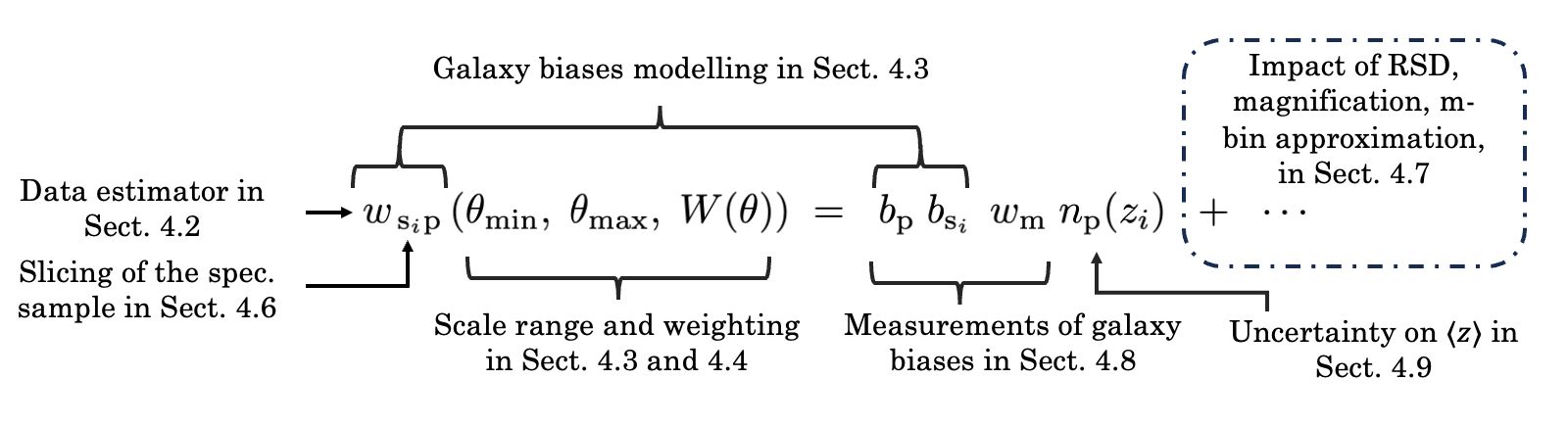}
    \caption{Modelling the cross-correlation between spectroscopic samples in redshift slices and photometric bins, and reference to the corresponding tests we performed to optimise the pipeline. }
    \label{fig:investigation}
\end{figure*}
In this section, we present the clustering-redshifts pipeline optimisation. 
 In Fig. \ref{fig:investigation} we summarise the different ingredients entering the modelling of clustering redshifts, and the corresponding validation tests presented in this section. In Sect. \ref{sub:Res_strat}, we describe the simulated data used. 
In Sect. \ref{sub:Res_estimator}, we compare the performance of the clustering estimator. In Sect. \ref{sub:Res_biashyp}, we evaluate the hypothesis that the galaxy biases of cross-correlations and auto-correlations are the same at small scales. In Sect. \ref{sub:Res_scales}, we determine the optimal scale range, and in Sect. \ref{sub:Res_weight}, the optimal scale weighting. In Sect. \ref{sub:Res_bin_size}, we test different spectroscopic slicing $\Delta z$ for measurements affected by systematics, with the effects of these systematics tested individually in Sect. \ref{sub:Res_indep}. In Sect. \ref{sub:Res_bias_corr}, we compare different galaxy bias correction methods. Finally, in Sect. \ref{sub:tomobin}, we investigate the impact of the tomographic bin definition on the redshift constraints.

\subsection{\label{sub:Res_strat} Investigation strategy}

Our investigation strategy consists of varying a set of analysis parameters to check the impact on clustering redshifts. We chose to use three tomographic bins covering the full redshift range of interest and corresponding to different types of galaxies.

The three tomographic bins: ‘‘low-$z$’’, ‘‘mid-$z$’’, and ‘‘high-$z$’’, were generated within the Flagship simulation with the photo-$z$ cuts:
\begin{align}
    &0.35<z_{\rm p}<0.47\,,\\
    &0.86<z_{\rm p}<1\,,\\
    &1.30<z_{\rm p}<1.6\,.
\end{align}
They corresponds to bin 2, 7, and 10 in Fig. \ref{fig:TB_spec_samples}.
For reference samples, we used BOSS LOWZ and CMASS, DESI ELG, and \Euclid NISP-S like samples. We report the redshift distribution of the tomographic bins and spectroscopic samples in Fig. \ref{fig:TB_spec_samples}.

To characterise the uncertainty from sample variance in our analysis, we split each tomographic bin into five independent Flagship2 sky patches\footnote{We applied cuts on the RA coordinates, with a width of $\Delta {\rm RA}=17.5$ deg.}  of $1000$ deg$^2$. As the simulation is limited to $5000$ deg$^2$, this is the maximum number of large independent patches that we can use.
When investigating the optimal spectroscopic slicing we do not want our results to be biased because the points luckily cover the optimal part of
the distribution. Thus, for every patch $j=1,\dots 5$, we varied the redshift slicing of the spectroscopic samples $z_{i};\, \Delta z$, by shifting the slice centres by $j\times \Delta z/5$. 
To summarise, we tested over five independent $1000$ deg$^2$ sky patches for three tomographic bins, each cross-correlated with spectroscopic samples distributed into different slices for each patch (but with the same $\Delta z$).

\subsection{\label{sub:Res_estimator} Choice of the estimator weighting scheme}

\begin{figure}
    \centering
    \includegraphics[width=1\linewidth]{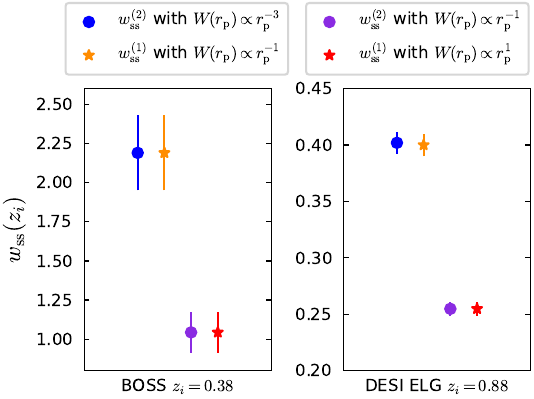}
    \caption{Auto-correlations of two spectroscopic samples for a single redshift bin ($\Delta z=0.05$). The yellow and red points are obtained using the first estimator $w_{\rm ss}^{(1)}$ with $\gamma\in \{-1,\,+1\}$.  The blue and purple points are obtained using the second one $w_{\rm ss}^{(2)}$ with  $\gamma\in \{-3,\,-1\}$. The consistency of the points with different weights illustrates the effective weighting due to the estimator definition.}
    \label{fig:estim_ls}
\end{figure}

In this section, we compare the two different estimators of the angular correlation function, described in Sect. \ref{sec:estimator_average}. 
We measured the auto-correlations of spectroscopic samples for a $1000$ deg$^2$ sky patch, a single redshift bin of width $\Delta z=0.05$, and the two estimators:
\begin{itemize}
    \item $w_{\rm ss}^{(1)}$ defined by Eq. \eqref{eq:weightproc1}, where weighting is applied to the ratio of pair counts;
    \item $w_{\rm ss}^{(2)}$ defined by Eq. \eqref{eq:weightproc2}, where weighting is applied directly the pair counts.
\end{itemize}
Correlations were evaluated for the scale range 0.5-- 5 Mpc, and with different weighting functions $W\propto \rp^{\gamma}$. The results are reported in Fig. \ref{fig:estim_ls}. We used logarithmic binning, 
and expect the measurements to be weighted by an effective function $W_{\rm eff}\propto \rp ^{\gamma+2} $ (cf. Eq. \ref{eq:Weff}), which is confirmed by the matched amplitudes of blue-yellow and purple-red points. Importantly, we do not observe significant differences in the associated uncertainties between the two estimators with the corresponding weights. Given that the weighting variation does not offer any practical advantage, and leads to unnecessary complexity in the weighting process (cf. App. \ref{App:pair_count}), we adopted the first estimator, Eq. \eqref{eq:weightproc1}, as our fiducial choice for this analysis, and we will evaluate the optimal value of $\gamma$.

\subsection{\label{sub:Res_biashyp} Testing the bias-cancellation hypothesis}
\begin{figure}
    \centering
    \includegraphics[width=1\linewidth]{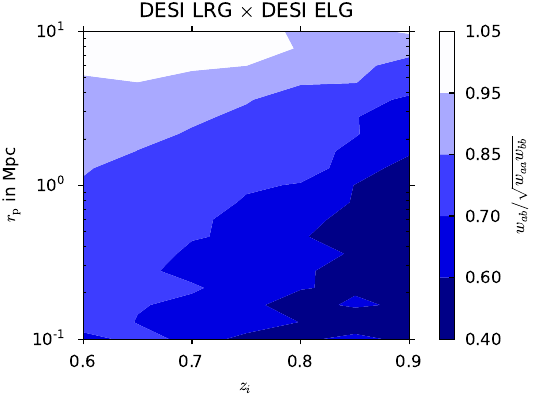}
    \caption{Variation in the Pearson coefficients $r_{ab}$ for Flagship LRGs with ELGs. This coefficient informs on the small-scale under-correlation of blue and red galaxies. }
    \label{fig:LRGxELG}
\end{figure}

\begin{figure*}
    \centering
    \includegraphics[width=1\linewidth]{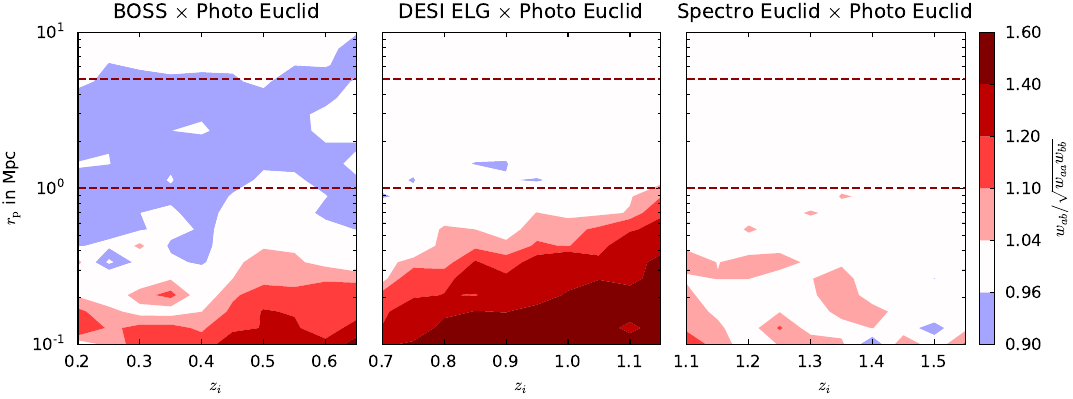}
    \caption{Pearson coefficient for our three simulated bins: low-$z$, mid-$z$, and high-$z$. This term is assumed to be one for clustering-redshifts calibration. We see different behaviours, $r=1$, $r<1$, and $r>1$ for different scales and redshifts, except for the high-$z$ bin, for which $r=1$ (almost) everywhere. }
    \label{fig:ratio_bias_3bins}
\end{figure*}

The scales used for clustering-redshifts calibration are typically very small [e.g. (0.1--5) Mpc] in order to minimise correlations between the calibration and the cosmological analysis, and to maximise the SNR. The associated nonlinear effects are often overlooked, however. Most authors assume that for two galaxy samples located within the same redshift bin, the two-point angular correlation function follows the relation: 
\begin{equation}
    w_{ab}(\rp)\approx\sqrt{w_{aa}(\rp)\; w_{bb}(\rp)}\,.
\end{equation}
If we were using scalar product, instead of scale correlation function, this would be equivalent to the Cauchy--Schwarz equality, which implies that the density contrasts $\Delta_a$ and $\Delta_b$ are linearly dependent, and the Pearson correlation coefficient $r_{ab}$ is 1, with
\begin{equation}
r_{ab}:=\frac{\l \Delta_a\,\Delta_b\r_{\rp}}{\sqrt{ \l \Delta_a\,\Delta_a\r_{\rp}\; \l \Delta_b\,\Delta_b\r_{\rp}}}=1 \label{eq:dab_daa_dbb} \, . 
\end{equation}
At large scales ($>10 $ Mpc) galaxy samples act as linear tracers of the underlying dark matter density field. In this case, $\delta_a=b_b/b_a\times  \delta_b$, which means that galaxy field are linearly dependant. At smaller scales, the linearity between the dark matter field and galaxy statistics begins to break down. For a more complex biasing between matter field and galaxy field, however, we would still expect $\vert r_{ab}\vert\leq 1$.\footnote{As shown in App. \ref{app:bias_next_order} for quadratic non-linearities (second-order effects), we expect $r_{ab}\approx 1$, with the correction arising from third-order terms. Pushing further the perturbative development, we can show that indeed $r_{ab}\leq 1$, analogously to the Cauchy--Schwarz inequality.} 

At even smaller scales (within the one-halo regime), the modelling of galaxy fields as random variables of the dark matter field breaks down \citep[e.g. in][clustering properties between galaxy subpopulations was investigated]{2008_stock}, potentially impacting cosmological inference \citep[e.g. ][in the context of weak lensing]{lensing_is_low}. We illustrate this issue with an extreme example: one sample consists only of central galaxies, whilst the other consists only of satellite galaxies. If we assume there is at most one satellite galaxy per halo, the auto-correlation of central and satellite galaxies would be very small on scales smaller than the halo size, but still non-zero as we are not evaluating 3d clustering. The positive cross-correlation between the two would be strong, however. This would result in $r_{ab}\gg 1$, which is not possible for random variables. The reason for this discrepancy is that the statistics of the satellite distribution depend on those of the central galaxy distribution. In typical halo occupation distribution models, the probability of hosting a satellite galaxy is conditional on the presence of a central galaxy. Alternatively, if two galaxy samples populate different haloes, the cross-correlation would be very small on scales smaller than the halo size, while the auto-correlations would remain significant, leading to $r_{ab} \ll 1$. 

 The scale range mentioned earlier (referred to as ‘‘small’’ and ‘‘smaller’’), and the behaviour of the $r_{ab}$ coefficients are not clearly known. We decided to improve this aspect for our \Euclid calibration method.  
 
The small-scale effects that we have described are not a hypothetical framework, as this type of behaviour has already been observed in spectroscopic surveys.
 For instance, the first DESI data provides clear evidence that  LRGs and ELGs exhibit low cross-correlation signals at small scales \citep[a phenomenon so-called ‘‘conformity’’; ][]{ROCHER_conform, DESI_ELG_LRG,conformity_DESI}. This occurs because these galaxies reside in different haloes, with their properties influenced by the history of the halo.
 
Given that the under-correlation between red and blue galaxies is well established, we aimed to test whether the Flagship simulation could reproduce this small-scale effect. Our strategy was to split the two samples into redshift slices ($\Delta z=0.05$) using true-redshift, and to measure the ratio with angular correlation,
\begin{align}
    r_{ab}(\rp,\,z_i)&=\frac{w_{a_ib_i}(\rp)}{\sqrt{w_{a_ia_i}(\rp)\,w_{b_ib_i}(\rp)}}\,,
\end{align}
which is expected to be 1 under the linear approximation. 

In Fig. \ref{fig:LRGxELG} we show $r_{ab}(\rp, \,z)$ for the Flagship LRG and ELG-like samples. We recover the under-correlation at small scales, through $r_{ab}(\rp,\,z)<1$. This ratio not only decreases with scales, but also with redshift. Since the recovered $n(z)$ is degenerate with $r$,  an evolving $r$ with redshift would introduce a bias in the reconstructed $n(z)$. This confirms the presence of one-halo behaviour in the Flagship simulation. However, since these behaviours were only recently identified and are still being actively studied, it is likely that their implementation in the Flagship simulation is incomplete or not fully accurate. A thorough study of this aspect of the simulation and one-halo physics is beyond the scope of this article.

In Fig. \ref{fig:ratio_bias_3bins} we report $r_{ab}(\rp, \, z)$ for the three simulated photometric $\times $ spectroscopic bins (defined in Sect. \ref{sub:Res_strat}). We observe significant scale and redshift evolution of the ratio, at least for the BOSS and DESI correlations, at scales smaller than 1 Mpc. Notably, we find $r > 1$, which we attribute to the physics of galaxy-halo connection and galaxy formation rather than non-linearities in the matter field (cf. the beginning of this subsection). Proving whether this is due to limitations in the HOD modelling or is a reliable prediction of the simulation is beyond the scope of our work. There is also a weak under-correlation for the low-$z$ bin (and the mid-$z$) at scales (1--8) Mpc, but it does not appear to evolve with redshift, and is therefore not expected to introduce a bias into the $n(z)$. This may correspond to the correction for non-linearities described above ($r\lesssim 1$). For the high-$z$ bin, we do not measure deviation from unity at any scale. 

Given these results, we are confident that using projected scales larger than 1.5 Mpc would avoid this one-halo effect.\footnote{We note that we are using angular correlation, so most of the 3-D separation between galaxies is much larger than this 1.5 Mpc. } As simulations improve in their modelling of galaxy halo properties, this test will need to be repeated.
The ratio amplitude, scale, and redshift evolution are different depending on the tracers: under-correlation for DESI LRG and ELG in Fig. \ref{fig:LRGxELG}, and over-correlation for \Euclid photo and DESI ELG in Fig. \ref{fig:ratio_bias_3bins}. We recommend to use only one type of spectroscopic tracer for the sample s when evaluating cross-correlations between photometric and spectroscopic samples at a specific redshift: $\wsp(z_i)$, avoiding mixing ELGs and LRGs for example.  If different spectroscopic tracers $\{{\rm s}_j\}_j$ are available for the same redshift $z_i$, we can then predict different $n_{\rm p}(z_i\,\vert s_j)$ and combine them \citep[e.g.  ][]{KIDS_redshift_dis}.  Small-scale effect could lead to incompatibility as we show in this section,  so that this procedure is more likely to detect systematics, and  more robust.

\subsection{\label{sub:Res_scales} Choice of the scale range}

\begin{figure*}
    \centering
    \includegraphics[width=1\linewidth]{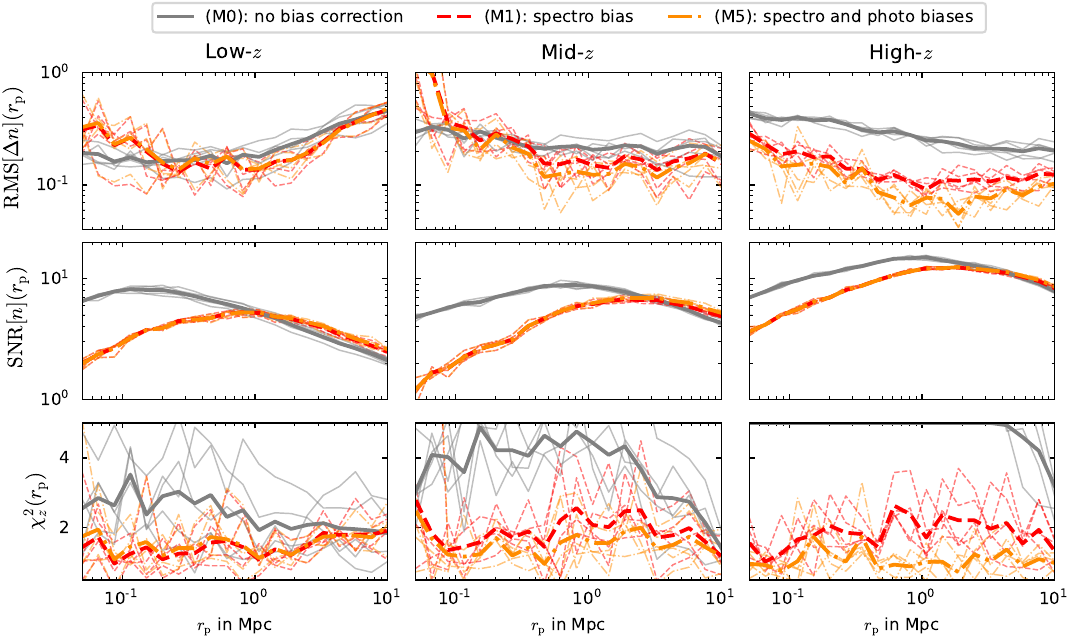}
    \caption{RMS of $\Delta n$, SNR of $n$, and reduced $\chi^2$ of $\Delta n$ (top to bottom) as a function of the projected scale $\rp$ for the three redshift bins (left to right) for three bias corrections (colours). The fine lines show five independent realisations of $1000$ deg$^2$, and the thick lines show their means.}
    \label{fig:scales}
\end{figure*}
In this section, we investigate the best scale range that minimises the deviation
\begin{equation}
    \Delta n\,(z_i\vert\, \rp)=n_{\rm p}(z_i\vert \,\rp)-n_{\rm true}(z_i)\,,
\end{equation}
while maximising the SNR. We explored a wide range of scales, including those previously discarded in Sect. \ref{sub:Res_biashyp}.
Unlike the estimator of $w$ in Eq. \eqref{eq:weightproc1}, which averages over scales, our estimator was evaluated for a single scale\footnote{In practice, it is evaluated for a small scale bin.} as in  Eq. \eqref{eq:npz=wsp/bpbs}. Consequently, we refer to the distribution measured as $n_{\rm p}\,(z\,\vert \,\rp)$. We always normalised the measured distributions to unity. 
 We employed three methods for correcting the bias, covering all the bias correction cases: no correction (i.e. M0),  spectroscopic correction (i.e. M1), and full correction (i.e. M5).\footnote{We expect the results with M2-3-4 to be similar to that of M5.} This approach allows us to assess how bias correction influences the optimal scale range.
 
We used three statistical indicators, as a function of the projected scale
\begin{itemize}
    \item the root mean square 
\begin{equation}
    {\rm RMS}[\Delta n]\,(\rp)=\sqrt{\frac{1}{N_z}\sum_{z_i} \Delta n\,(z_i\vert\, \rp) ^2 }\,;\label{eq:RMS}
\end{equation} 
\item the (mean) signal-to-noise-ratio 
\begin{equation}
    {\rm SNR}[n] (\rp)=\frac{1}{N_z} \sum_{z_i} \frac{\vert n_{\rm p}(z_i)\vert}{\sigma_n (z_i)}\,;\label{eq:SNR}
\end{equation}
\item the reduced $\chi^2$ 
\begin{equation}
    \chi_z^2(\rp)=\frac{1}{N_z} \left[\Delta n\,(z_i\vert\, \rp\right]^\top_{z_i}\,\;\mathcal{C}_{n\vert \rp)}\;\left[\Delta n\,(z_i\vert\, \rp)\right]_{z_i}\,,\label{eq:chi2_red}
\end{equation}
\end{itemize}
with $N_z$ the number of points (equivalently the number of redshift slices of the spectroscopic sample).
These indicators, evaluated at scales ranging from (0.05--10) Mpc for the different bins and bias corrections, are reported in Fig. \ref{fig:scales}. The thin lines represent individual realisations of $1000$ deg$^2$ patches, while the thicker lines indicate their means.

We first discuss the impact of the bias correction. 
Whilst the no-bias correction model M0 shows a higher SNR for scales lower than 1 Mpc, its corresponding $\chi_z^2$ value is very high, indicating that it is not accurate. Therefore, we focus exclusively on M1 and M5 for the remainder of this section. 
Both methods perform similarly for the SNR, but for mid-$z$ and high-$z$, M5 provides a $\chi^2_z$ closer to unity.

For the three tomographic bins, the RMS is increasing for smaller scales. This increase is associated with a lower SNR, and the $\chi_z^2$ remains relatively constant. This suggests an overall consistency, and we can choose the scale range based on the SNR only.  For the low-$z$ bin, it peaks at 1 Mpc and then decreases. For the other two bins, it appears to peak around 2 Mpc, with a small decrease for larger scales. 

In conclusion,  the scale range (1.5--5) Mpc appears to be well suited for clustering redshifts; it is nearly optimal regarding the SNR;
the $\chi_z^2$ is almost minimal for the whole range;
 it avoids the $\rp<1.5$ Mpc one-halo regime (excluded in Sect. \ref{sub:Res_biashyp});
these scales are not used for cosmological analysis. 

This scale range is the same as in \cite{Gatti_Giulia_DESY3} and \cite{DESY3_MAGLIM_z}, but significantly larger than in other works such as \cite{van_den_Busch_2020}, \cite{Cawton2022}, and \cite{Naidoo23}, which used scales of 0.1–1 Mpc$ \,h^{-1}$ or 0.5–1.5 Mpc$\,h^{-1}$. Early clustering-redshifts studies even advocated for using smaller scales, down to the kpc range \citep{Menard2013, Schmidt2013}.
We note that we have chosen scales that avoid the one-halo term and the range of scales used for cosmology. 
One way to be conservative on the small-scales bias effects would be to use only large scales, overlapping those used in the $3\times2$ point analysis. In previous calibrations such as HSC, KIDS, and DES, scale choice was largely driven by the lack of spectroscopic sample \citep{KIDS_redshift_dis}, and the range was chosen to be at the peak of the SNR \citep{Gatti_DESY1}. For \Euclid, this will no longer be the case, and from Fig. \ref{fig:scales} it is clear that larger scales can be used for calibration. However, it requires the generation of a more complex covariance to accurately propagate the correlation. This was proved to be informative through a forecasting work in \cite{johnston20246x2ptforecastinggainsjoint}, but never implemented in practice with simulated or real data, and is beyond the scope of this article.

\subsection{\label{sub:Res_weight} Choice of the scale weighting}

\begin{figure*}
    \centering
    \includegraphics[width=0.9\linewidth]{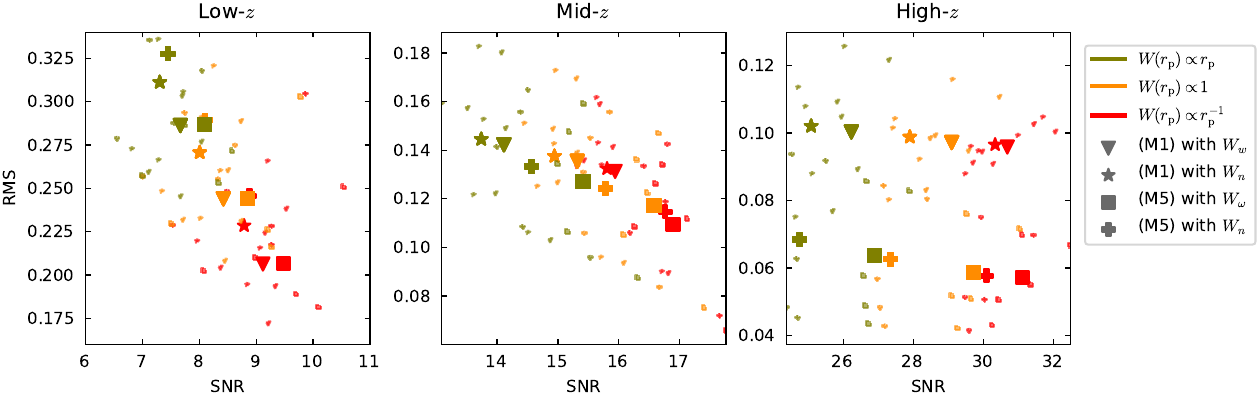}
    \caption{SNR and RMS for the three tomographic bins (left to right), with three power laws as weighting functions (colours), two weighting schemes, and two bias-correction methods (M1 and M5).}
    \label{fig:weight}
\end{figure*}

\begin{figure*}
    \centering
    \includegraphics[width=0.9\linewidth]{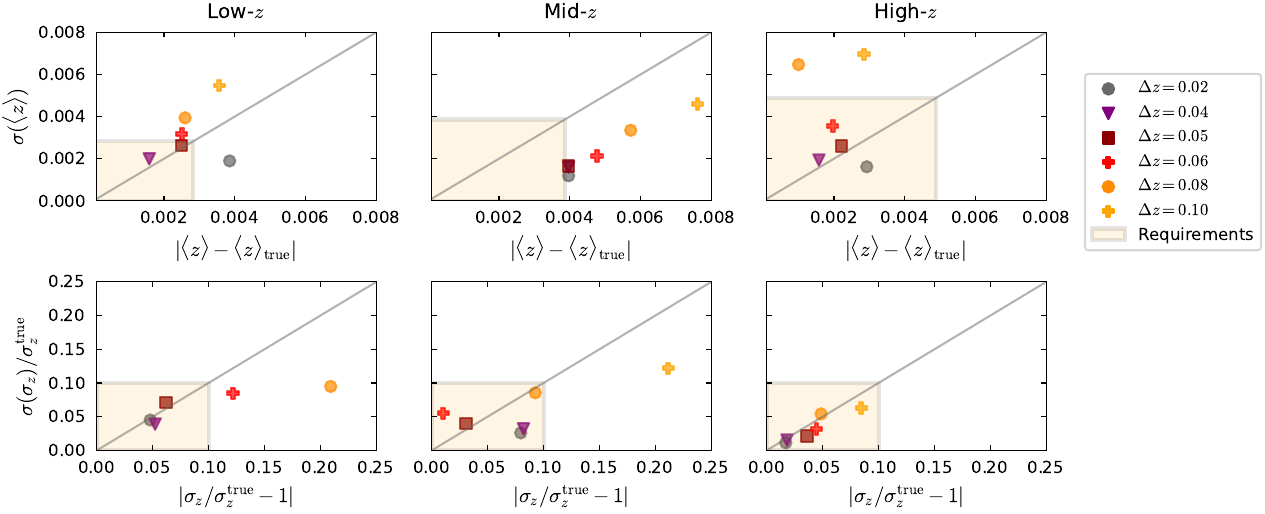}
    \caption{Mean redshift and shape reconstruction with SGP, for the three tomographic bins. The different markers and colours are associated with different slicing $\Delta z$. The sandy region corresponds to the \Euclid requirements, and the diagonal line shows a coherent deviation of 1$\sigma$.}
    \label{fig:bin_size}
\end{figure*}
 In this section, we varied the weighting $W(\rp)$ in Eq. \ref{eq:weight_w} and evaluated the RMS and SNR defined in Eqs. \eqref{eq:RMS} and \eqref{eq:SNR}. The scale range was fixed to (1.5--5) Mpc following the last section's conclusion. We tested three power laws
\begin{equation}
    W(\rp)=\frac{\rp^\gamma}{\int_{\rpmin}^{\rpmax}\drp\; \rp ^{\gamma}}\,\text{ with  } \gamma \in \{-1,0,1\}\,,
\end{equation}
 and two different weighting schemes:
\begin{itemize}
    \item we scale-average the different correlation functions and compute their ratio, as in Eq. \eqref{eq:weight_w},
    \begin{equation}
        \Tilde{w}_{ab}(z_i)=\int_{\rpmin}^{\rpmax}\drp\; W_w (\rp) \,w_{ab}(\rp,\,z_i)\,;
    \end{equation}
    \item we scale-average the estimated $n(z_i\vert\, \rp)$ as in  Eq. \eqref{eq:weight_n},
    \begin{equation}
        \Tilde{n}(z_i)=\int_{\rpmin}^{\rpmax}\drp\; W_n (\rp)\,n(z_i\vert\, \rp)\,.
    \end{equation}
\end{itemize}
In Fig. \ref{fig:weight} we show the comparison between the SNR and the RMS for the three tomographic bins, with three power laws, two weighting schemes, and two bias correction methods (M1 and M5). The small markers are the realisations for the five individual $1000$ deg$^2$ patches, while the larger markers indicate their means.
The region of interest covers the lower right part of the plots, as it corresponds to minimum deviation and high SNR. Notably, the RMS/SNR plot displays a linear trend, indicating that points with a smaller deviation are associated with a better SNR, which is consistent with the expectations for an unbiased measurement. For all three redshift bins, the optimal weighting is the inverse scale weighting (darker red marker), a similar result to \cite{Schmidt2013} and \cite{Menard2013}. The reason is that SNR is higher, while RMS is lower, for scales around 1.5 Mpc than 5 Mpc (cf. Fig. \ref{fig:scales}). $W_w$ and $W_n$ are similar for the RMS, but the first performs slightly better for the SNR.  These findings hold for M1 and M5. Additionally, we confirm that correcting the photometric bias for the first redshift bin is not strictly necessary in terms of RMS, but it does lead to a better SNR.

Thus, the best weighting scheme appears to be the inverse scale weighting applied to correlation functions, combined with a full bias correction approach.

\subsection{\label{sub:Res_bin_size} Finding the optimal spectroscopic slicing}

In the previous sections, we did not include systematics (e.g. magnification) in the measurement because we were questioning the optimal statistical choices on the two-point clustering itself, hence the use of statistical quantities such as RMS and SNR. In the rest of the article, we consider realistic measurements, including systematics, and evaluate their impact on the moments of the redshift distribution we aim to calibrate.  

In this section, we varied the slicing of the spectroscopic sample $\Delta z$, cf. Sect. \ref{sec:real_spec_bins}. On the one hand, narrow slices are more sensitive to corrections associated with bin modelling (cf. Sect. \ref{sec:comparison_corr}), magnification (cf. Sect. \ref{sec:magn}), or RSD (cf. Sect. \ref{sec:rsd}). On the other hand, if the slices are too wide, there may not be enough points covering the distribution to properly measure the $n(z)$, the shape would be flattened (cf. Sect. \ref{sec:discrete_nzi} and Fig. \ref{fig:nz_binning}), and the assumption that galaxy biases and redshift distributions are constant may break (cf. Sect. \ref{sec:real_spec_bins}). Therefore, we aim to find a compromise between narrow slices affected by systematic errors and large slices that are not statistically optimal, and for which the modelling might not be precise enough. The impacts of the systematic errors are tested one by one in the next Sect. \ref{sub:Res_indep}.

We introduced the systematics using magnified sky coordinates and fluxes for the photometric and spectroscopic samples, and spectroscopic redshift (including peculiar velocity) for the spectroscopic sample. We used M5 as the bias correction method. For extracting the distribution moments from the $\{n(z_i)\}$ we employed the SGP $n(z)$ reconstruction introduced in Sect. \ref{sec:SGP}. We did so because, even with few points, fitting the true distribution (as with the SSM, cf. Sect. \ref{sec:SSM}) is likely to yield good yet unrealistic results, whereas the SGP provides weaker yet realistic constraints.  

We measured the mean (absolute) biases on the mean $\vert \l z\r-\l z\r_{\rm true}\vert$ and standard deviation $\vert \sigma_z/\sigma^{\rm true}_z-1\vert$ of the $n_{\rm p}(z)$ for different slicing, where the mean is evaluated over the five independent patches. We reported the mean values and their associated mean uncertainties in Fig. \ref{fig:bin_size}. We also created a similar plot using the SSM and obtain comparable results except for the uncertainties, which remain constant across the different slicing. This confirms our previous statement and justifies the choice of the SGP. 

First of all, the uncertainties ($y$ axis) on mean redshift and shape increase with the slicing $\Delta z$, and we also observe larger deviation for wider slices. Regarding the shape reconstruction (related to $\sigma_z$), narrow slices are not associated with larger deviations, except for the mid-$z$ bin. For the mean redshift, small slicing is associated with similar or larger deviations than $\Delta z\sim 0.05$, but not larger uncertainties. The optimum slicing appears to correspond to $0.04\leq \Delta z \leq 0.06$. 

Consequently, we decided to use a redshift slicing of $\Delta z=0.05$ for every bin, for the rest of the study. This is larger than the slicing used for DES  \citep{Gatti_Giulia_DESY3, DESY3_MAGLIM_z}, but the same as the previous \Euclid clustering-redshifts paper \citep{Naidoo23}. As illustrated by the sandy region in Fig. \ref{fig:bin_size}, we predict that the systematic errors for this redshift slicing will not shift the data points outside the \Euclid requirement. Since the $\Delta z=0.05$ points are predominantly on the diagonal (except for the mid-$z$ mean-$z$), this suggests that the corrections are fairly small for this slicing: the parameters are not biased more than 1$\sigma $ away from the truth. The mid-$z$ is the only case where we might fail to fulfil the requirement if we do not correct for the systematic errors. We investigated the origin of this deviation in detail in the next section. 

\subsection{\label{sub:Res_indep} Evaluating the (mis-)modelling of the clustering, magnification, and RSD}

\begin{figure*}
    \centering
    \includegraphics[width=0.9\linewidth]{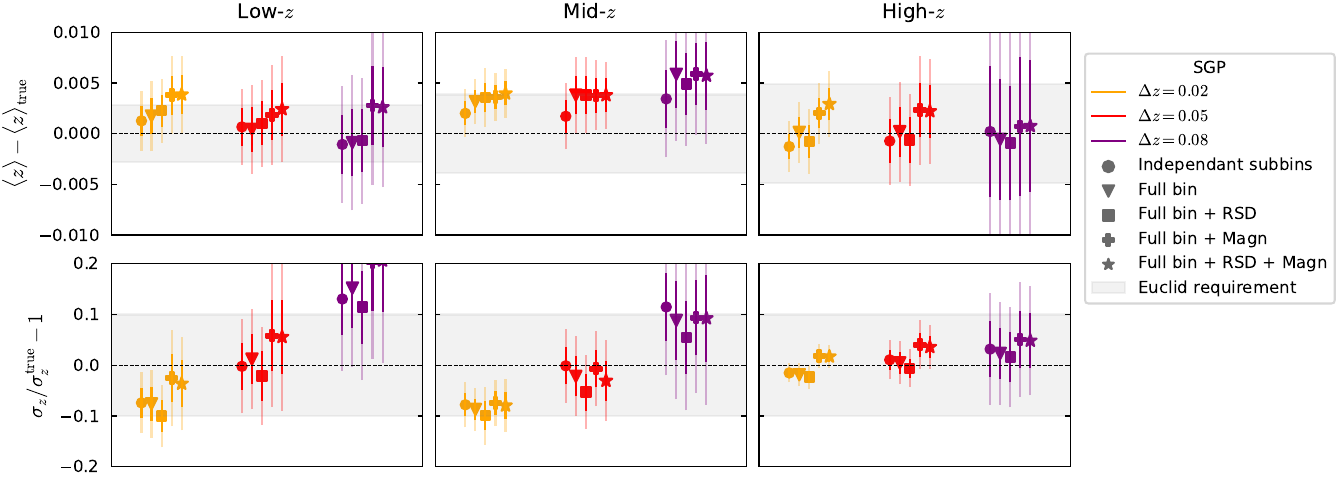}
    \caption{Deviations on the mean-$z$ (\textit{top}) and standard deviation (\textit{bottom}), with SGP fitting, for the three tomographic bins (\textit{left, middle, and right}), with three $\Delta z$ slicings  (colours) and different implemented systematic effects  (marker). The shadowed bands correspond to the \Euclid requirement on both parameters. The dark and light error bars indicate $68.3\%$ and $95.5\%$ uncertainties}
    \label{fig:deviation_systematics_SGP}
\end{figure*}

\begin{figure*}
    \centering
    \includegraphics[width=0.9\linewidth]{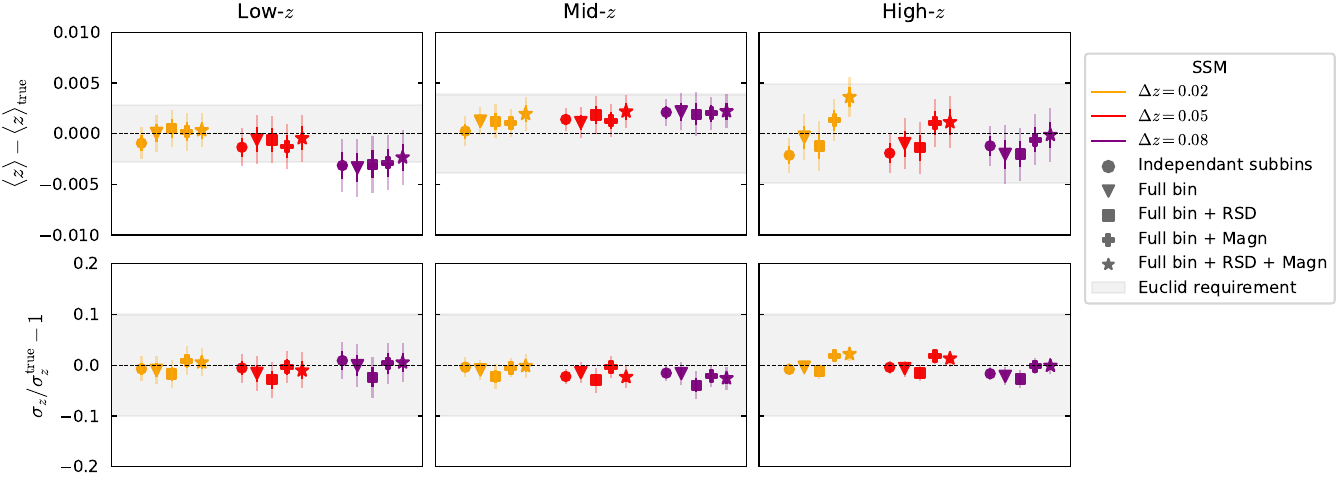}
    \caption{Same as Fig. \ref{fig:deviation_systematics_SGP} for the two parameters of the SSM model, which are equivalent to $\l z\r -\l z\r_{\rm true}$ and $\sigma_z/\sigma_z{\rm true}-1$.}
    \label{fig:deviation_systematics_shift}
\end{figure*}

For clustering redshifts, we measure the cross-correlation of a wide photometric tomographic sample with several spectroscopic slices, and we assume that only the photometric galaxies within the redshift range of the spec-$z$ cuts contribute to a non-zero correlation. There can be three corrections:
\begin{itemize}
    \item galaxies in the nearby redshift (nearby bins) can correlate as well (we refer to this as the one-bin approximation);
    \item galaxies not covering the same redshift range can be correlated through magnification processes;
    \item spectroscopic redshifts can be different than true redshifts,  leading to correction on the $n_{\rm s}$ at the boundary of the bin. 
\end{itemize}
Two of these effects can be understood as bin border effects, and for larger slices they are expected to decrease in proportion. 
For wide spectroscopic slices, the assumption that $n(z)$ and $ b(z)$ are constant within the bin might break down. We also note that spectroscopic interlopers were not included in the \Euclid spectroscopic samples.

In the previous Sect. \ref{sub:Res_bin_size}, we demonstrated that the compromise between having enough points $n_{\rm p}(z_i)$ and limiting these effects, corresponds to slicings $\Delta z =0.04$--$0.06$, regardless of the tomographic bin. 
To assess the influence of each effect, we used the idealised bias correction M5 to avoid the impact of galaxy bias, and proceed as follows:
\begin{itemize}
    \item To test the impact of the one-bin approximation, we compared the cross-correlation  $w_{\rm sp}(z_i)$ with spectroscopic slices and the tomographic bin for two cases. In the first one, 
    referred to as the ‘‘full bin’’, the photometric sample is the fiducial one used in this paper.
    For the second case, referred to as the ‘‘independent bins’’, we used idealised tomographic bins. To generate these idealised tomographic bins, for every $z_i$, we keep only the photometric galaxies within the same redshift range as the spectroscopic  galaxies (using true-$z$), and then complete this sample with randoms to maintain the total number of galaxies as in the full tomographic bin. This ensures that only the galaxies within the redshift slice of the spectroscopic sample contribute to the correlation, for each spectroscopic slice.
    \item To test the magnification effect, we compared the results using the magnified variables for position and fluxes (over which we apply the samples cuts), or the unmagnified ones.   
     \item To test the  RSD effect, for the spectroscopic samples we used either the Flagship true redshifts, or the observed ones, 
      the latter corresponding to a spectroscopic-like redshift estimate (including velocities).
      \item We also tested all these effects together, referred to as ‘‘Total’’.
\end{itemize}

In Fig. \ref{fig:deviation_systematics_SGP}  we show the deviations associated with the different systematic effects for different slicing, for the SGP. The same test for SSM methods is presented in Fig. \ref{fig:deviation_systematics_shift}.
The corresponding values for the two procedures are listed in Table \ref{tab:bin_dev_0.05} for $\Delta z=0.05$, in App. \ref{app:bin_size_sys}. 
For RSD and magnification, we have to consider the full tomographic bin, which by default comes with the one-bin approximation. Therefore, the full bin assumption has to be compared with ‘‘independent bins’’, and the RSD or magnification has to be compared with the full bin.

For the SSM and SGP, with $\Delta z=0.05$, the constraints on the redshift standard deviation $\sigma$ are within the requirements independently from the effects. Only a large slicing of $\Delta z=0.08$ leads to some measurements out of the requirements because not enough  points cover the distribution, and with larger slicing, the distribution appears flattened ($\sigma_z/\sigma_z^{\rm true}>1$) as shown in Fig. \ref{fig:nz_binning}. 

Regarding the mean redshift bias, all the ‘‘independent bin’’ measurements are consistent with no deviation at the 1 to 2$\sigma$ level with the SGP and SSM. This suggests that our choice of scale and weighting is sufficient to limit the small-scale effects (cf. Sect. \ref{sub:Res_scales}). 
 In general, for the three tomographic bins, all the systematics lead to some deviations compared to the independent bins, but distinguishing these variations from statistical fluctuations is complicated. 
It is clear that the SGP procedure is more sensitive to these effects. The reason for this is that with SSM we compare the measurements with a model free from these effects. As this model is, in our case, the truth, its results may be optimistic. Nevertheless, with a SOM estimate as a model, we might expect a similar low sensitivity to these effects. 
For smaller bin sizes, the error bars are smaller, but the deviation due to systematic effects is at least as large as for other sizes. Wider slices are associated with greater uncertainty, and the balance between accuracy and precision is achieved for $\Delta z = 0.05$. 
For this bin size, the systematic effects do not produce deviations that exceed the requirements.
However, for the mid-$z$ bin, with SGP these systematic effects result in a deviation right at the limit of the budget on the mean-$z$, as the bias is $0.0038\pm 0.0017$ (the requirements being $0.0038$). For the optimistic case (SSM), the deviation remains within the requirements: the bias in mean redshift being $0.0022 \pm 0.008$. The effect causing this large deviation for SGP seems to be the one-bin approximation. 
\begin{figure*}
    \centering
    \includegraphics[width=0.9\linewidth]{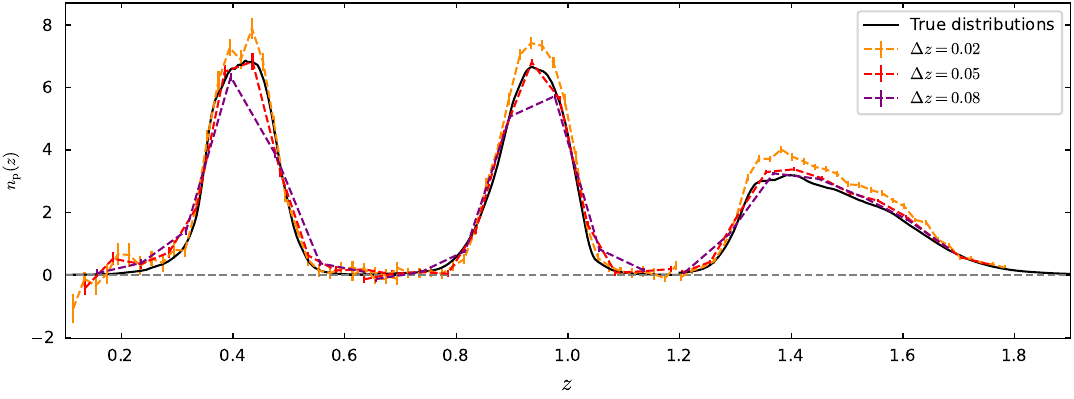}
    \caption{True-redshift distributions of tomographic bins in black, and the clustering-redshifts measurements with M5 for three spectroscopic slicings: $\Delta z=0.02,\,0.05,\,0.08$  (orange, red, and purple). 
    }
    \label{fig:nz_diff_bin_eta}
\end{figure*}
 In Fig. \ref{fig:nz_diff_bin_eta}, we report the true redshift distribution of three bins are shown alongside clustering-redshifts measurements for three slicing values: $\Delta z = 0.02,\,0.05,\,\text{and }\,0.08$, excluding the effects of RSD and magnification. We observe increasing deviations with finer slicing, whilst the shape reconstruction becomes less precise for coarser slicing (cf. Fig. \ref{fig:nz_binning}).  The small slicing deviations are higher for the mid-$z$ and high-$z$ bin, as one can expect since $\eta_\pm$ increase with redshift. We also see that large slices lead to worst shape estimate for the low-$z$ bin, as its width is narrower. 

In App. \ref{app:m-bin} we explore an alternative modelling including correlation with neighbours slices ( cf. Sect. \ref{sec:comparison_corr}). In short,  despite our effort we did not manage to improve the accuracy of the modelling with extra-terms.

\subsection{\label{sub:Res_bias_corr} Comparison of the different bias-correction methods}
In this section, we explore the different possibilities for breaking the degeneracy between $n_{\rm p}(z)$ and $b_{\rm p}\, b_{\rm s}$ from $w_{\rm sp}$ (cf. Eq. \ref{eq:eq_wsp_n_b}). In the first subsection, we focus on measuring the photometric bias alone, since this is the challenging aspect. 

\begin{figure}
    \centering
    \includegraphics[width=0.9\linewidth]{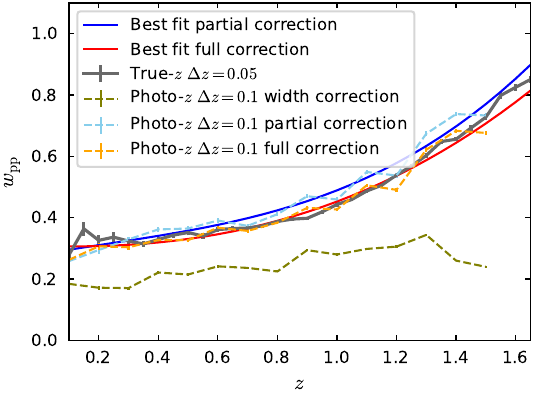}
    \caption{ True $w_{\rm pp}$ for $\Delta z=0.05$ in grey and measurement with corrections factors including M3 as dashed green, blue, and orange lines for photo-$z$ binning $\Delta z=0.1$. The best fits from degree-3 polynomials are reported with solid lines. The $w_{\rm pp}$ used for  M3 is the best-fit model of the full correction (solid red line).  }
    \label{fig:M3bias}
\end{figure}

\begin{figure}
    \centering
    \includegraphics[width=0.9\linewidth]{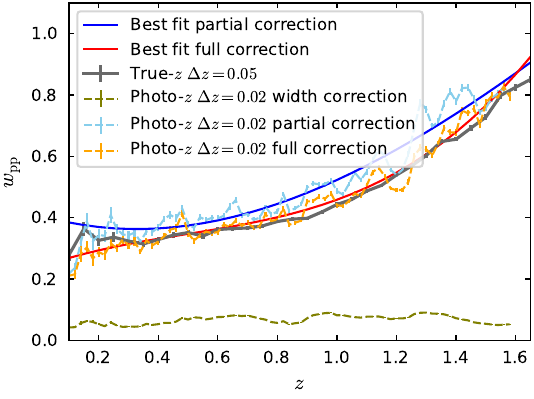}
    \caption{Same as Fig. \ref{fig:M3bias}, but for M4, with a smaller photo-z binning $\Delta z=0.02$. The $w_{\rm pp}$ used for  M4 is the best-fit model of the full correction (solid red line). }
    \label{fig:M4_bias}
\end{figure}

\begin{figure*}
    \centering
    \includegraphics[width=0.9\linewidth]{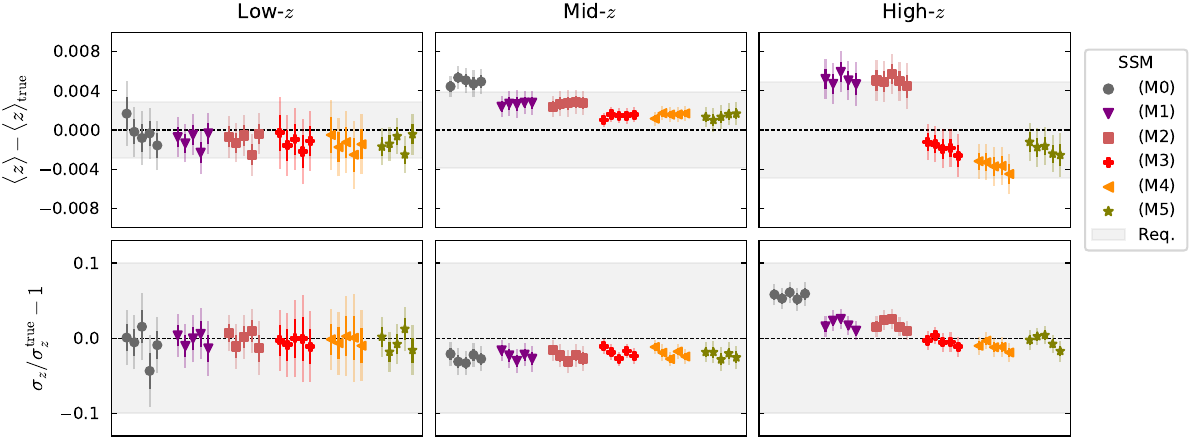}
    \caption{Deviations with the SSM fitting on the mean redshift (upper panel) and standard deviation (lower panel) for the six methods M0--M5 in colour, for five sky patches (same colour). The requirements are represented by the shaded area. The dark and light error bars indicate $68.3\%$ and $95.5\%$ uncertainties. }
    \label{fig:bias_correction_shift}
\end{figure*}

\begin{figure*}
    \centering
    \includegraphics[width=0.9\linewidth]{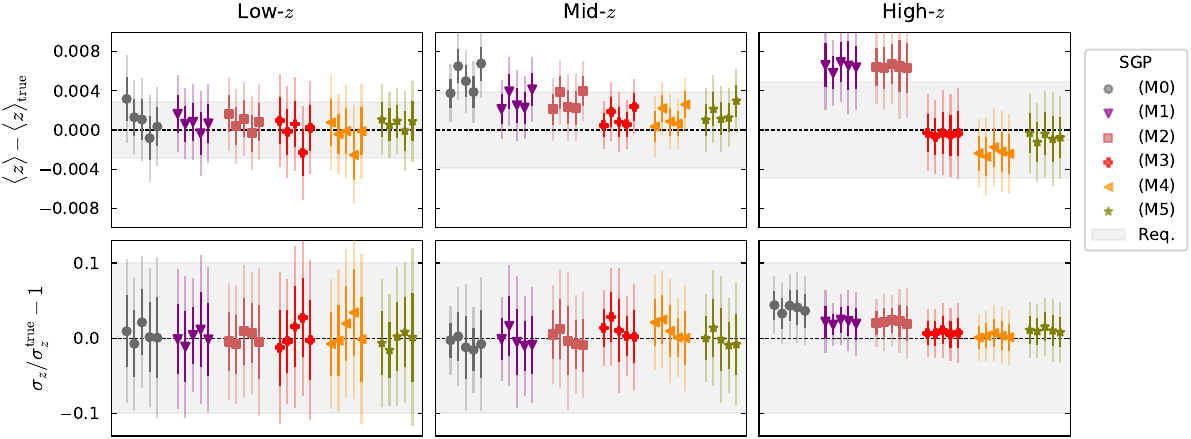}
    \caption{Same as Fig. \ref{fig:bias_correction_shift} with the SGP fitting. }
    \label{fig:bias_correction_GP}
\end{figure*}

\subsubsection{Measuring the photometric galaxy bias evolution}
\label{sec:photo_bias}

In Sect. \ref{sec:moredetails_M3M4}, we proposed to improve existing photometric bias corrections \citep{van_den_Busch_2020,Cawton2022} with two methods to measure $b_{\rm p}(z)$, applicable to real data, specifically for photometric galaxies with $z_{\rm photo}<1.6$ (i.e. fully covered by spectroscopic samples). 
 The larger binning method M3  is applied in Fig.  \ref{fig:M3bias} and the smaller binning method M4  is applied in Fig. \ref{fig:M4_bias}. They are represented by orange dashed lines.
For both figures we also include the results using either the partial correction (blue dashed) or width correction (green dashed), for comparisons with previous analyses \citep{van_den_Busch_2020,Cawton2022,DESY3_MAGLIM_z}. 
Additionally, we present the best fit degree-3 polynomial for the partial and the full corrections (solid lines) from a standard $\chi^2$ minimisation. The grey line shows the measurement using true-$z$, which we want to recover after the corrections. 

We find that assuming the photo-$z$ as the true-$z$ (width-only correction), neither the amplitude nor the redshift evolution is well measured. For the partial correction, whilst the redshift evolution appears well-recovered, differences in amplitude persist.
Finally, our full correction method significantly outperforms the others: in both figures, we recover the amplitude and redshift evolution of the bias more accurately. Notably, the bias evolves weakly for $z < 1$, suggesting that correcting the photometric bias at low redshift may not be necessary.

Since the partial and full corrections both give a good trend for the redshift evolution of the galaxy bias, one could argue their performance are identical in the context of clustering redshifts, as the measured $n_{\rm p}(z)$ can always be latterly renormalised. With the correct amplitude for the galaxy bias, evaluating whether the $n_{\rm p}(z)$ measured is directly normalised to unity can provide valuable information. 
Deviations from unity could indicate
\begin{itemize}
    \item a contamination from systematics, such as the small scale effect illustrated in Sect. \ref{sub:Res_biashyp};
    \item the  photometric sample is not fully covered by spec (and what is the fraction);
    \item a problem with the spectroscopic redshifts, such as interlopers.
\end{itemize}
In that sense, the full correction performs significantly better than the partial one. The performance of our bias measurement may depend on the specific photo-$z$ code and the quality of the available photometry, and data implementation will have to address this question.

\subsubsection{Comparison of the methods}

In this section, we compare the different bias correction methods introduced in Sect. \ref{sec:correcting_bias}, including M3 and M4. We aim to evaluate the bias correction performance, thus we work in an idealised framework with independent bins, no RSD, and no magnification (i.e. the ‘‘independent bin’’ test of Sect. \ref{sub:Res_indep}). We employ a redshift slicing $\Delta z=0.05$. 
We report in colours the deviations on the mean redshift and standard deviation for the six methods M0--M5, in Figs. \ref{fig:bias_correction_shift} (for SSM) and \ref{fig:bias_correction_GP} (for SGP). For each method, we have five realisations by varying the sky patch.  We also give the values for the biases on mean-$z$ and standard deviation in Table \ref{tab:bin_dev_0.05}, in App. \ref{app:bin_size_sys}.

We can see that at low-$z$ it is not necessary to correct for the photometric bias, and spectroscopic correction M1 produces good estimates. At mid-$z$ it is necessary to correct at least for the spectroscopic bias, and ideally for both with M3 or M4. At high-$z$, it is essential to correct both biases. We observe that M3 performs significantly better than M4 for SSM and SGP; indeed, the redshift evolution is better captured in Fig. \ref{fig:M3bias} than in Fig. \ref{fig:M4_bias}. This difference may be attributed to the fact that smaller slices are more sensitive to systematic effects, as discussed in Sect. \ref{sub:Res_bin_size}.
Overall, M3 performs similarly to the idealistic M5 and is consistently the best bias correction method, exhibiting performance akin to M1 for low-$z$ (though with slightly larger error bars).

In conclusion, using M3 for every bin yields unbiased estimates of the mean redshift and standard deviation across all bins, with uncertainties that fall within the requirements for both methods. The values presented in these two figures are reported in  Table \ref{tab:dev_M05}.

\subsection{Tomographic bin uncertainties}\label{sub:tomobin}

We defined the tomographic bins as equi-populated, based on the assumption that the uncertainty scales with the total number of galaxies in each bin. For the 10 spectroscopic samples, we used 1-DESI-BGs, 2-BOSS, 3-DESI-LRGs, 4-DESI-LRGs, 5-DESI-LRGs, 6-DESI-ELGs, 7-DESI-ELGs, 8-\Euclid-NISP, 9-\Euclid-NISP, 10-\Euclid-NISP.

\begin{figure}
    \centering
    \includegraphics[width=1\linewidth]{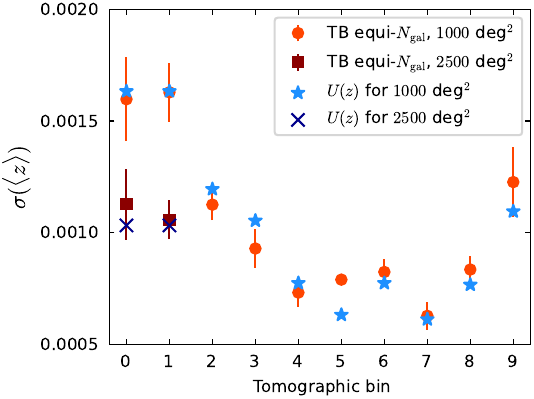}
    \caption{Uncertainties on mean-$z$ with SSM fitting for five sky realisations in red. We also plot the best-fit uncertainty model in blue.}
    \label{fig:Error_volume}
\end{figure}

In Fig. \ref{fig:Error_volume} we plot in red the uncertainty on the mean redshift $\sigma ({\l z \r})$ for every tomographic sample, and its standard deviation $\sigma(\sigma(\l z\r))$, evaluated for the five $1000$ deg$^2$ sky-patches with jackknife covariance, with SSM fitting. Interestingly, we find that the uncertainty is not constant. It remains the same for the first two samples, then decreases across the next six samples, before increasing again for the final ones. We also report for the first two samples, the case of $2500$ deg$^2$ sky patches in dark red, and observe a mean-$z$ uncertainty reduction of 40$\%$.

\cite{clust_z_mcQuinn_white} predicted that "the linear bias times number of objects in a redshift bin generally can be constrained with cross-correlations to fractional error $\approx \sqrt{10^{2} N_{\Delta z}({{\rm d} N_{\rm s}/{\rm d} z})^{-1}} $", where $N_{\Delta z}$ is the number of spectroscopic redshift slices, and $N_{\rm s}$ the number of spectroscopic galaxies.\footnote{ We have slightly adjusted the convention to remain consistent with the rest of our article.} This expression describes how well the redshift distribution can be measured, whereas we are primarily interested in its moments. We assume that, at first order, the scaling should behave similarly, but with a different coefficient, which is complicated to predict because of the complexity of the mean-$z$ fitting procedure. 
As a consequence, the uncertainty should depend on the number of spectroscopic galaxies, rather than on the volume or the number of photometric galaxies. However, for the low-$z$ bins, the spectroscopic density is sufficiently high, especially with DESI-BG-like sample, that we might enter a cosmic variance (CV) regime. In this case, uncertainties are no longer sample-dominated (i.e. the shot noise $1/n_{\rm s}$ is subdominant). 
Thus, we introduce a maximum spectroscopic density to reach the cosmic variance limit $n_{\rm cv}$, which in principle depends on $z$, but that we assume to be constant. 
Our uncertainty model is
\begin{equation}
    U(z)=A \;\sqrt{\frac{N_{\Delta z}}{V_{\rm TB}\,\min\, (n_{\rm spec}, \, n_{\rm cv})}}\,,\label{eq:Uz}
\end{equation}
where $A$ is a constant independent of the tomographic bin, and $V_{\rm TB}$ is the volume of the tomographic bin.

The equation above also indicates that the uncertainty scales with the inverse of the square root of the tomographic volume. This explains why in Fig.8 of \cite{Naidoo23}, the uncertainty scales as a power law of the survey area $A^{\gamma}$, and the best-fit power law is close to  $\gamma=-1/2$. 

We constrained the $A$ and $n_{\rm cv}$ coefficients with the set of tomographic bin measurements using a $\chi^2$ procedure, and report in Fig. \ref{fig:Error_volume} the best-fit model in blue for $1000$ and $2500$ deg$^2$. We observe a good agreement, particularly given that we neglected complications arising from galaxy biases and the mean-redshift fitting procedure. The uncertainty is largest for the low-$z$ tomographic bins where the survey volume is much smaller and precision is limited by the CV regime. For the high-$z$ samples, the uncertainty increases due to the reduced density of spectra, which is not sufficiently offset by the larger volume. The best-fit $n_{\rm cv}$ roughly corresponds to the DESI-BGS density divided by a factor of four (and the BOSS density by a factor of 1.8), and it only plays a role for the first two bins.

As a result, to achieve consistent uncertainty across bins with clustering-$z$, the tomographic bin definition would need to be adjusted according to an uncertainty model like Eq. \eqref{eq:Uz}. However, the tomographic bin definition is primarily influenced by cosmological constraining power \citep[e.g. ][]{Pocino-EP12}, and our results alone do not provide sufficient grounds to recommend a change in binning strategy. However, the precision of the mean redshift is one of the key limiting factors in cosmic shear analyses \citep[see, e.g.,][]{propagating_photo_z_uncertainty}, and the impact of bin definitions on the accuracy of redshift calibration is often overlooked.

\section{\label{sc:forecast} Application to the ten \Euclid tomographic bins}

\begin{figure*}
    \centering
    \includegraphics[width=0.9\linewidth]{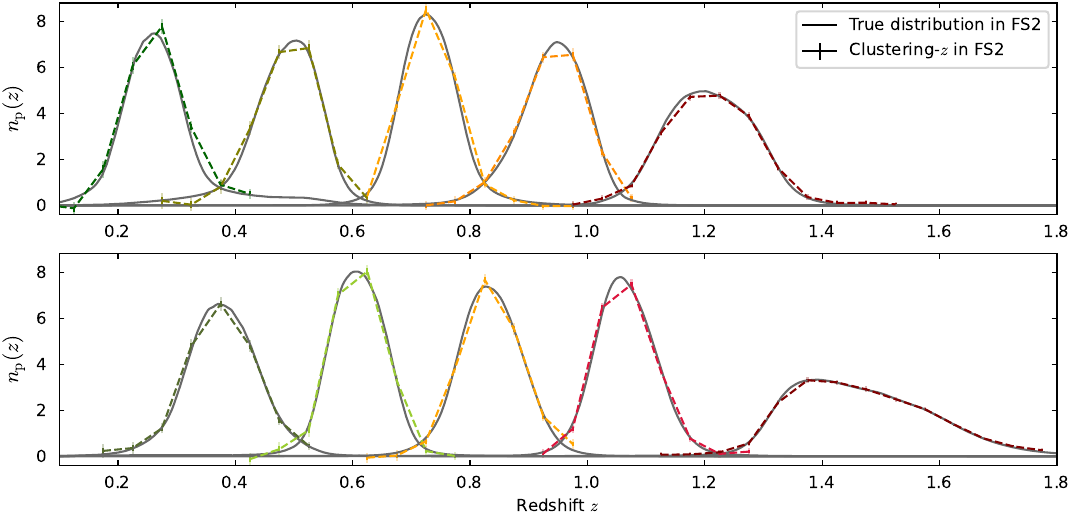}
    \caption{True-redshift distribution of the simulated Flagship2 ten tomographic bins with $z_{\rm p}<1.6$ (solid lines) and their clustering-redshifts measurements with our pipeline, including all systematics and realistic bias correction M3. The first two bin measurements are realised with $2500$ deg$^2$ sky patches, and the other eight are realised with $1000$ deg$^2$ sky patches. The dark and light error bars indicate $68.3\%$ and $95.5\%$ uncertainties, but are barely visible by eye. }
    \label{fig:10bind_measurment}
\end{figure*}

\begin{figure*}
    \centering
    \includegraphics[width=0.85\linewidth]{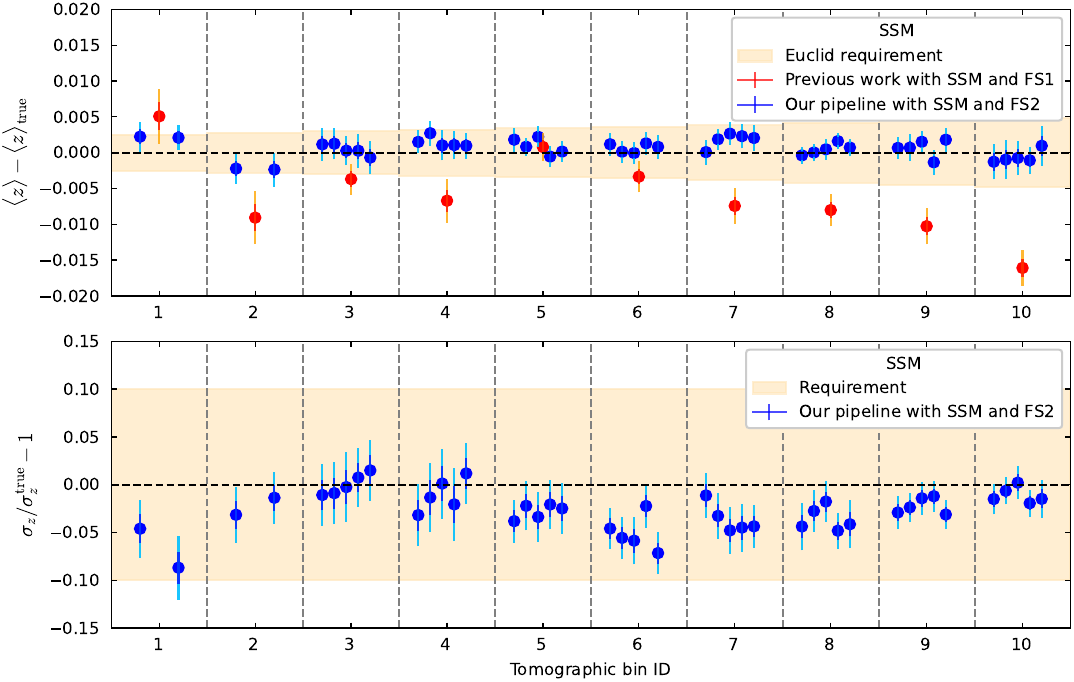}
    \caption{Deviations in the mean redshift (\textit{top}) and the redshift standard deviation (\textit{bottom}) for the ten Flagship2 tomographic bins relative to the true-redshift distributions using the SSM method. The deviations are shown for each bin and sky patch. We consider patches of $2500$ deg$^{2}$ for the first two bins, and $1000$ deg$^2$ for the others. The \Euclid requirements on the uncertainty are highlighted by the shaded area centred around zero. The uncertainties are consistent across all the bins for the mean redshift and we achieve a clear improvement from previous work shown in red. For the redshift standard deviations, the uncertainties are slightly underestimated, but this is within an acceptable range given the requirements.}
    \label{fig:mean_z_SSM}
\end{figure*}

\begin{figure*}
    \centering
    \includegraphics[width=0.85\linewidth]{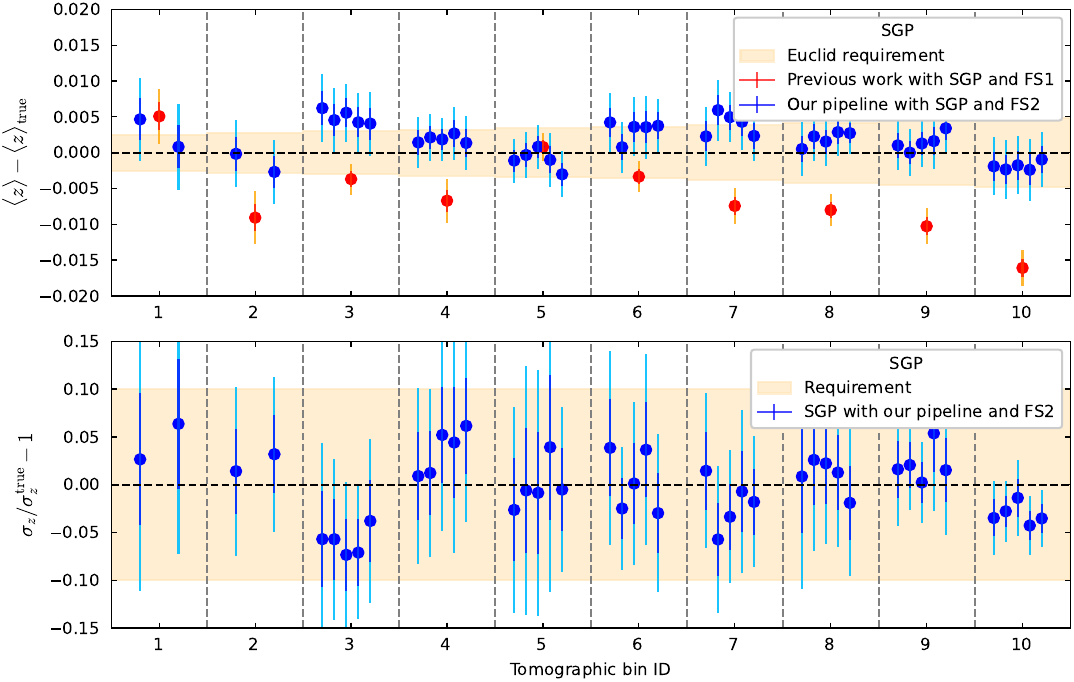}
    \caption{Deviations in the mean redshift (\textit{top}) and the redshift standard deviation (\textit{bottom}) using the SGP method for the same bin configuration as in Fig. \ref{fig:mean_z_SSM}. The \Euclid requirements on the uncertainty are highlighted by the shaded area centred around zero. For some bins (particularly bins 1, 3, and 7), we observe slightly larger deviations and uncertainties than permitted on the mean redshift. However, a clear improvement compared to previous work (shown in red) is still evident. The redshift standard deviations are unbiased and fall within the required range (shaded sandy area). }
    \label{fig:mean_z_SGP}
\end{figure*}

In this section, we applied our pipeline to calibrate the ten tomographic bins described in Sect. \ref{sc:data_euclid_photo}. For the spectroscopic samples, following our recommendation (see Sect. \ref{sub:Res_biashyp}) we only correlated a tomographic bin with one spectroscopic tracer at a time. We used the following tracers for the different bins: 1-DESI-BGs, 2-BOSS, 3-DESI-LRGs, 4-DESI-LRGs, 5-DESI-LRGs, 6-DESI-ELGs, 7-DESI-ELGs, 8-\Euclid-NISP, 9-\Euclid-NISP, 10-\Euclid-NISP.
For simplicity, we did not consider the joint constraint of several spectroscopic tracers covering the same bin, which would provide additional information. For each bin, we produced several independent realisations by selecting distinct patches in Flagship 2. For the first two tomographic bins, we generated two  $2500$ deg$^{-2}$ patches, while for the remaining eight bins, we used five $1000$ deg$^{-2}$ patches. This approach ensures similar uncertainties across bins, as explained in Sect. \ref{sub:tomobin}.

The true tomographic bin distribution, and the clustering-redshifts measurements are reported in Fig. \ref{fig:10bind_measurment}. The measurements are visually indistinguishable from the truth. To quantify how good they are, we show the (optimistic) SSM calibration of the mean-$z$ and standard deviations in Fig. \ref{fig:mean_z_SSM}. The results with the conservative SGP calibration are shown in Fig. \ref{fig:mean_z_SGP}. For comparison, we also plot the previous results of \cite{Naidoo23} for the mean redshift (with Flagship1 samples). Our results using the SSM method are fully successful; for every bin and patch realisation, we meet the \Euclid requirements for the mean redshift and shape, unlike prior results. However, as expected, the "conservative" SGP method yields larger uncertainties and deviations. In particular, for some bins (especially bins 1, 3, and 7), we observe slightly larger deviations and uncertainties than allowed by the mean redshift requirement.
We tested the effect of increasing the patch sizes by a factor of two and found that the mean-$z$ bias was reduced on average by $40\%$, which was sufficient to meet the requirement for all bins. 

Part of the improvement to \cite{Naidoo23} is due to the use of a larger cosmic volume with Flagship2 instead of Flagship1. Nevertheless, with our pipeline we also got significantly lower mean redshift biases than \cite{Naidoo23}. The main differences between our pipeline and the previous one are our estimator (cf. Sect. \ref{sub:Res_estimator}), the scale range (cf. Sect. \ref{sub:Res_scales}), the measurements being evaluated for each spectroscopic tracer (cf. Sect. \ref{sub:Res_biashyp}), the bias correction method (cf. Sect. \ref{sub:Res_bias_corr}), and some improvement in the implementation of SSM and SGP (cf. Sect. \ref{sc:metho_Fitting} ). We note that we did include the effect of magnification contrary to \cite{Naidoo23}.
In conclusion, we predict that clustering-redshifts will be able to satisfy the \Euclid requirements.

\section{Conclusion}\label{sub:Res_summary}
We optimised and evaluated the performance of the clustering-redshifts method for calibrating the redshift distributions of \Euclid tomographic bins.

Using the Flagship2 simulation, we generated realistic spectroscopic and photometric mock samples based on BOSS, DESI, and \Euclid. The photometric sample was divided into ten equally populated tomographic bins, with a maximum photo-$z$ of 1.6. We used up to five $1000$ deg$^2$ patches to fully exploit the volume of the simulation.

We measured the angular auto- and cross-correlations of these samples under various methodological choices and investigated alternative approximations to model the clustering. 

From these clustering measurements, we calibrated the redshift distributions via two distinct methods. One method involved fitting the data points using a model $n(z\vert \, p_1,\dots)$ with free parameters. In practice, a complementary calibration, such as the self-organising map can be used to define the model \citep[e.g., ][]{KIDS_redshift_dis,Gatti_Giulia_DESY3}. The other method relies on Gaussian processes to directly fit the data without assuming a specific model. With the latter, the SOM and clustering-redshifts calibrations can both be run independently, or be combined as in the case of DES \citep{3sdir,Gatti_Giulia_DESY3}.   

In Sect. \ref{sc:Result} we investigated and concluded the following about the clustering-redshifts technique and angular clustering. 
\begin{itemize}
    \item In Sect. \ref{sub:Res_estimator} we argued that the integrated pair-count estimator $w^{(1)}$ is a better choice than $w^{(2)}$ for measuring the integrated clustering (i.e. integrating the ratio $\rm DD/RR$ instead of taking the ratio of the integrated $\rm DD$ and $\rm RR$).
    \item In Sect. \ref{sub:Res_biashyp} we explained a potential issue with very small scales that arises from the one-halo properties of the clustering. We concluded that scales should always be larger than 1.5 Mpc to limit this problem. We also recommended separating the clustering-redshifts measurements for the different spectroscopic tracers. 
    \item In Sects. \ref{sub:Res_scales} and \ref{sub:Res_weight}, we ran an optimisation of the scale range choice and weighting. With the Flagship simulation, we achieved a good compromise with $1.5<\rp<5$ Mpc with an inverse scale weighting.
    \item In Sect. \ref{sub:Res_bin_size} we showed that the optimal spectroscopic slicing given our tomographic bins and the different systematics was $\Delta z\approx 0.05$.
    \item In Sect. \ref{sub:Res_indep} we showed that approximating $\int \dz\, n^2(z)\cdots$ (Limber-full-$z$) with some $n^2(z_i)\int \dz\cdots$ expressions (Limber-1-bin) is the main systematic effect and that magnification has some limited impact on the $n_{\rm p}(z)$ inference. 
    \item In Sect. \ref{sub:Res_bias_corr} we compared different bias corrections and introduced a way to measure the photometric galaxy bias by splitting the TB into smaller bins and interpolating these effective biases over redshift (i.e. M3). The results of this method were comparable to the ideal case M5, in which photometric bias is measured using true redshifts. It should be used as the fiducial correction scheme.
    \item Finally, we investigated in Sect. \ref{sub:tomobin} the dependence of the mean-$z$ error bars on the clustering-redshifts parameters, and confirmed that they are mainly driven by the number of spectroscopic galaxies that is available in the footprint. 
\end{itemize}

In Sect. \ref{sc:forecast} we applied the optimised pipeline to calibrate the \Euclid simulated tomographic bins. The calibration performance using the SSM model met or exceeded the required accuracy, while the SGP method yielded comparable results, but they were slightly below the requirements in certain cases. It is important to note that these results were derived from limited sky patches with correction methods that can still be improved in future work. These results represent a significant improvement over the previous \Euclid clustering-redshifts study \citep{Naidoo23} overall and demonstrated that the clustering-redshifts technique is robust.

Additionally, as discussed in Sect. \ref{sc:Data}, we assumed that the ${\rm H}\,\alpha$ sample had a purity of $100\%$. In reality, however, when we bin the sample into small redshift intervals, we expect that some of these galaxies may be interlopers, that is, they might be located at different redshifts due to noise or a misidentification of a spectral lines.  
This situation might significantly impact clustering-redshifts measurements because photometric galaxies at a similar redshifts as the interlopers would contribute to the clustering signal, which would bias the inferred $n(z)$. It is therefore crucial to determine the fraction of these interlopers and their redshift distribution to correct for this effect. The clustering-redshifts technique can indeed be used to constrain the properties of interlopers by cross-correlating the NISP sample with other spectroscopic data or with photometric data with priors on their $n(z)$.

As already mentioned in Sect. \ref{sub:Res_biashyp}, the galaxy-halo connection implemented in the Flagship simulation might not be accurate, and our statistical estimator might be optimistic. It might therefore still be beneficial to use larger scales even though they are not statistically optimal. 
These larger scales will be used for the $3 \times 2$ analysis, which necessitates the generation of a more complex covariance to accurately propagate the correlation with the calibration \citep[e.g. the work of][]{johnston20246x2ptforecastinggainsjoint}. Although this approach is more complicated, the advantage is that the distribution would be free of small-scale potential bias,  which may prove worthwhile.

Finally,  we restricted the calibration to galaxies with a maximum photo-$z$ of 1.6 because we lack dense spectroscopic samples above redshift 1.8, which sets a strict upper limit. Clustering-$z$ for HSC, KIDS, and DES was even more limited than \Euclid, with a maximum redshift of $z\sim 1$. 
In the far future numerous spectroscopic samples from Lyman-break galaxies are expected to be observed by spectroscopic experiments such as DESI-II, MUST, and WST (if funded). In a much shorter time frame, QSOs might be used one could utilise QSOs to cover the redshift range $1.5<z<3$. These constraints alone would most likely not fulfil the \Euclid requirements because  QSOs have a low-density, however, but they would still provide valuable information.

\begin{acknowledgements}
The authors thank Alex Alarcon, Carles Sánchez, Elisa Chisari, and Santi Ávila  for enlightening discussions, as well as Krishna Naidoo for his help, code, and results sharing. The authors also acknowledge Mathias Urbano for his code support.
This work was supported by the  MICINN projects PID2019-111317GB-C32, PID2021-123012NB-C41, PID2022-141079NB-C32, CNS2023-144328
(Carteu) as well as predoctoral program AGAUR-FI ajuts (2024 FI-1 00692) Joan Oró. IFAE is partially funded by the CERCA program of the Generalitat de Catalunya.
This work has made use of CosmoHub, developed by PIC (maintained by IFAE and CIEMAT) in collaboration with ICE-CSIC. It received funding from the Spanish government (grant EQC2021-007479-P funded by MCIN/AEI/10.13039/501100011033), the EU NextGeneration/PRTR (PRTR-C17.I1), and the Generalitat de Catalunya.
\AckEC
\end{acknowledgements}

%
%

\bibliography{clust_z}

@ARTICLE{WZ_DES_dassignies,
       author = {{d'Assignies}, W. and {Bernstein}, G.~M. and {Yin}, B. and {Giannini}, G. and {Alarcon}, A. and {Manera}, M. and {To}, C. and {Yamamoto}, M. and {Weaverdyck}, N. and {Cawthon}, R. and {Gatti}, M. and {Amon}, A. and {Anbajagane}, D. and {Avila}, S. and {Becker}, M.~R. and {Bechtol}, K. and {Chang}, C. and {Crocce}, M. and {De Vicente}, J. and {Dodelson}, S. and {Fang}, J. and {Fert{\'e}}, A. and {Gruen}, D. and {Legnani}, E. and {Porredon}, A. and {Prat}, J. and {Rodriguez-Monroy}, M. and {S{\'a}nchez}, C. and {Schutt}, T. and {Sevilla-Noarbe}, I. and {Sanchez Cid}, D. and {Troxel}, M.~A. and {Abbott}, T.~M.~C. and {Aguena}, M. and {Alves}, O. and {Bacon}, D. and {Bocquet}, S. and {Brooks}, D. and {Camilleri}, R. and {Carnero Rosell}, A. and {Carrasco Kind}, M. and {Carretero}, J. and {Castander}, F.~J. and {da Costa}, L.~N. and {da Silva Pereira}, M.~E. and {Davis}, T.~M. and {Desai}, S. and {Doel}, P. and {Doux}, C. and {Drlica-Wagner}, A. and {Eifler}, T. and {Elvin-Poole}, J. and {Everett}, S. and {Flaugher}, B. and {Fosalba}, P. and {Frieman}, J. and {Garcia-Bellido}, J. and {Gaztanaga}, E. and {Giles}, P. and {Gutierrez}, G. and {Hinton}, S.~R. and {Hollowood}, D.~L. and {Honscheid}, K. and {Huterer}, D. and {Jain}, B. and {James}, D.~J. and {Kuehn}, K. and {Lahav}, O. and {Lee}, S. and {Marshall}, J.~L. and {Mena-Fernandez}, J. and {Menanteau}, F. and {Miquel}, R. and {Muir}, J. and {Myles}, J. and {Ogando}, R.~L.~C. and {Palmese}, A. and {Paterno}, M. and {Petravick}, P. and {Plazas Malagon}, A.~A. and {Raveri}, M. and {Roodman}, A. and {Samuroff}, S. and {Sanchez}, E. and {Sheldon}, E. and {Shin}, T. and {Smith}, M. and {Suchyta}, E. and {Swanson}, M.~E.~C. and {Tarle}, G. and {Thomas}, D. and {Vikram}, V. and {Walker}, A.~R.},
        title = "{Dark Energy Survey Year 6 Results: Clustering-redshifts and importance sampling of Self-Organised-Maps $n(z)$ realizations for $3\times2$pt samples}",
      journal = {arXiv e-prints},
         year = 2025,
        month = oct,
          eid = {arXiv:2510.23565},
        pages = {arXiv:2510.23565},
          doi = {10.48550/arXiv.2510.23565},
archivePrefix = {arXiv},
       eprint = {2510.23565},
 primaryClass = {astro-ph.CO},
       adsurl = {https://ui.adsabs.harvard.edu/abs/2025arXiv251023565D}
}

@ARTICLE{Roster25,
       author = {{Roster}, W. and {Wright}, A.~H. and {Hildebrandt}, H. and others},
       title = "{Euclid: Photometric redshift calibration with self-organising maps}",
      journal = {A\&A, submitted},
         year = 2025,
        month = aug,
          eid = {arXiv:2508.02779},
        pages = {arXiv:2508.02779},
          doi = {10.48550/arXiv.2508.02779},
archivePrefix = {arXiv},
       eprint = {2508.02779},
 primaryClass = {astro-ph.CO},
       adsurl = {https://ui.adsabs.harvard.edu/abs/2025arXiv250802779R}
}

@ARTICLE{EP-Risso,
       author = {{Euclid Collaboration: Risso}, I. and {Veropalumbo}, A. and {Branchini}, E. and others},
        title = "{Euclid preparation. The impact of redshift interlopers on the two-point correlation function analysis}",
      journal = {A\&A, submitted},
         year = 2025,
        month = may,
          eid = {arXiv:2505.04688},
        pages = {arXiv:2505.04688},
          doi = {10.48550/arXiv.2505.04688},
archivePrefix = {arXiv},
       eprint = {2505.04688},
 primaryClass = {astro-ph.CO},
       adsurl = {https://ui.adsabs.harvard.edu/abs/2025arXiv250504688E}
}

@ARTICLE{johnston20246x2ptforecastinggainsjoint,
       author = {{Johnston}, Harry and {Elisa Chisari}, Nora and {Joudaki}, Shahab and {Reischke}, Robert and {St{\"o}lzner}, Benjamin and {Loureiro}, Arthur and {Mahony}, Constance and {Unruh}, Sandra and {Wright}, Angus H. and {Asgari}, Marika and {Bilicki}, Maciej and {Burger}, Pierre and {Dvornik}, Andrej and {Georgiou}, Christos and {Giblin}, Benjamin and {Heymans}, Catherine and {Hildebrandt}, Hendrik and {Joachimi}, Benjamin and {Kuijken}, Konrad and {Li}, Shun-Sheng and {Linke}, Laila and {Porth}, Lucas and {Shan}, HuanYuan and {Tr{\"o}ster}, Tilman and {van den Busch}, Jan Luca and {von Wietersheim-Kramsta}, Maximilian and {Yan}, Ziang and {Zhang}, Yun-Hao},
        title = "{6 {\texttimes} 2 pt: Forecasting gains from joint weak lensing and galaxy clustering analyses with spectroscopic-photometric galaxy cross-correlations}",
      journal = {\aap},
         year = 2025,
        month = jul,
       volume = {699},
          eid = {A127},
        pages = {A127},
          doi = {10.1051/0004-6361/202452466},
archivePrefix = {arXiv},
       eprint = {2409.17377},
 primaryClass = {astro-ph.CO},
       adsurl = {https://ui.adsabs.harvard.edu/abs/2025A&A...699A.127J}
}

@ARTICLE{scales_bias,
       author = {{Pandey}, S. and {Krause}, E. and {Jain}, B. and {MacCrann}, N. and {Blazek}, J. and {Crocce}, M. and {DeRose}, J. and {Fang}, X. and {Ferrero}, I. and {Friedrich}, O. and {Aguena}, M. and {Allam}, S. and {Annis}, J. and {Avila}, S. and {Bernstein}, G.~M. and {Brooks}, D. and {Burke}, D.~L. and {Carnero Rosell}, A. and {Carrasco Kind}, M. and {Carretero}, J. and {Costanzi}, M. and {da Costa}, L.~N. and {De Vicente}, J. and {Desai}, S. and {Elvin-Poole}, J. and {Everett}, S. and {Fosalba}, P. and {Frieman}, J. and {Garc{\'\i}a-Bellido}, J. and {Gruen}, D. and {Gruendl}, R.~A. and {Gschwend}, J. and {Gutierrez}, G. and {Honscheid}, K. and {Kuehn}, K. and {Kuropatkin}, N. and {Maia}, M.~A.~G. and {Marshall}, J.~L. and {Menanteau}, F. and {Miquel}, R. and {Palmese}, A. and {Paz-Chinch{\'o}n}, F. and {Plazas}, A.~A. and {Roodman}, A. and {Sanchez}, E. and {Scarpine}, V. and {Schubnell}, M. and {Serrano}, S. and {Sevilla-Noarbe}, I. and {Smith}, M. and {Soares-Santos}, M. and {Suchyta}, E. and {Swanson}, M.~E.~C. and {Tarle}, G. and {Weller}, J. and {DES Collaboration}},
        title = "{Perturbation theory for modeling galaxy bias: Validation with simulations of the Dark Energy Survey}",
      journal = {\prd},
         year = 2020,
        month = dec,
       volume = {102},
       number = {12},
          eid = {123522},
        pages = {123522},
          doi = {10.1103/PhysRevD.102.123522},
archivePrefix = {arXiv},
       eprint = {2008.05991},
 primaryClass = {astro-ph.CO},
       adsurl = {https://ui.adsabs.harvard.edu/abs/2020PhRvD.102l3522P}
}

@ARTICLE{2008_stock,
       author = {{Swanson}, Molly E.~C. and {Tegmark}, Max and {Blanton}, Michael and {Zehavi}, Idit},
        title = "{SDSS galaxy clustering: luminosity and colour dependence and stochasticity}",
      journal = {\mnras},
         year = 2008,
        month = apr,
       volume = {385},
       number = {3},
        pages = {1635-1655},
          doi = {10.1111/j.1365-2966.2008.12948.x},
archivePrefix = {arXiv},
       eprint = {astro-ph/0702584},
 primaryClass = {astro-ph},
       adsurl = {https://ui.adsabs.harvard.edu/abs/2008MNRAS.385.1635S}
}

@ARTICLE{GP_Johnson,
       author = {{Johnson}, Andrew and {Blake}, Chris and {Amon}, Alexandra and {Erben}, Thomas and {Glazebrook}, Karl and {Harnois-Deraps}, Joachim and {Heymans}, Catherine and {Hildebrandt}, Hendrik and {Joudaki}, Shahab and {Klaes}, Dominik and {Kuijken}, Konrad and {Lidman}, Chris and {Marin}, Felipe A. and {McFarland}, John and {Morrison}, Christopher B. and {Parkinson}, David and {Poole}, Gregory B. and {Radovich}, Mario and {Wolf}, Christian},
        title = "{2dFLenS and KiDS: determining source redshift distributions with cross-correlations}",
      journal = {\mnras},
         year = 2017,
        month = mar,
       volume = {465},
       number = {4},
        pages = {4118-4132},
          doi = {10.1093/mnras/stw3033},
archivePrefix = {arXiv},
       eprint = {1611.07578},
 primaryClass = {astro-ph.CO},
       adsurl = {https://ui.adsabs.harvard.edu/abs/2017MNRAS.465.4118J}
}

@ARTICLE{Salvato_2019,
       author = {{Salvato}, Mara and {Ilbert}, Olivier and {Hoyle}, Ben},
        title = "{The many flavours of photometric redshifts}",
      journal = {Nature Astronomy},
         year = 2019,
        month = jun,
       volume = {3},
        pages = {212-222},
          doi = {10.1038/s41550-018-0478-0},
archivePrefix = {arXiv},
       eprint = {1805.12574},
 primaryClass = {astro-ph.GA},
       adsurl = {https://ui.adsabs.harvard.edu/abs/2019NatAs...3..212S}
}

@ARTICLE{4MOST_verdier,
       author = {{Verdier}, Aurelien and {Rocher}, Antoine and {Bandi}, Behnood and {Richard}, Johan and {Roukema}, Boudewijn and {Loveday}, Jon and {Tempel}, Elmo and {Bilicki}, Maciej and {Kneib}, Jean-Paul and {Guitton}, Mathilde},
        title = "{The 4MOST-Cosmology Redshift Survey: Target Selection of Bright Galaxies and Luminous Red Galaxies}",
         year = 2025,
        month = aug,
          eid = {arXiv:2508.07311},
        pages = {arXiv:2508.07311},
          doi = {10.48550/arXiv.2508.07311},
archivePrefix = {arXiv},
       eprint = {2508.07311},
 primaryClass = {astro-ph.CO},
       adsurl = {https://ui.adsabs.harvard.edu/abs/2025arXiv250807311V}
}

@ARTICLE{kids2025_zcal,
       author = {{Wright}, Angus H. and {Hildebrandt}, Hendrik and {van den Busch}, Jan Luca and {Bilicki}, Maciej and {Heymans}, Catherine and {Joachimi}, Benjamin and {Mahony}, Constance and {Reischke}, Robert and {St{\"o}lzner}, Benjamin and {Wittje}, Anna and {Asgari}, Marika and {Chisari}, Nora Elisa and {Dvornik}, Andrej and {Georgiou}, Christos and {Giblin}, Benjamin and {Hoekstra}, Henk and {Jalan}, Priyanka and {William}, Anjitha John and {Joudaki}, Shahab and {Kuijken}, Konrad and {Lesci}, Giorgio Francesco and {Li}, Shun-Sheng and {Linke}, Laila and {Loureiro}, Arthur and {Maturi}, Matteo and {Moscardin}, Lauro and {Porth}, Lucas and {Radovich}, Mario and {Tr{\"o}ster}, Tilman and {von Wietersheim-Kramsta}, Maximilian and {Yan}, Ziang and {Yoon}, Mijin and {Zhang}, Yun-Hao},
        title = "{KiDS-Legacy: Redshift distributions and their calibration}",
         year = 2025,
        month = mar,
          eid = {arXiv:2503.19440},
        pages = {arXiv:2503.19440},
          doi = {10.48550/arXiv.2503.19440},
archivePrefix = {arXiv},
       eprint = {2503.19440},
 primaryClass = {astro-ph.CO},
       adsurl = {https://ui.adsabs.harvard.edu/abs/2025arXiv250319440W}
}

@ARTICLE{hildebrandt_magn,
       author = {{Hildebrandt}, H.},
        title = "{Observational biases in flux magnification measurements}",
      journal = {\mnras},
         year = 2016,
        month = feb,
       volume = {455},
       number = {4},
        pages = {3943-3951},
          doi = {10.1093/mnras/stv2575},
archivePrefix = {arXiv},
       eprint = {1511.01352},
 primaryClass = {astro-ph.GA},
       adsurl = {https://ui.adsabs.harvard.edu/abs/2016MNRAS.455.3943H}
}

@ARTICLE{eboss_qso,
       author = {{Palanque-Delabrouille}, N. and {Magneville}, Ch. and {Y{\`e}che}, Ch. and {P{\^a}ris}, I. and {Petitjean}, P. and {Burtin}, E. and {Dawson}, K. and {McGreer}, I. and {Myers}, A.~D. and {Rossi}, G. and {Schlegel}, D. and {Schneider}, D. and {Streblyanska}, A. and {Tinker}, J.},
        title = "{The extended Baryon Oscillation Spectroscopic Survey: Variability selection and quasar luminosity function}",
      journal = {\aap},
         year = 2016,
        month = mar,
       volume = {587},
          eid = {A41},
        pages = {A41},
          doi = {10.1051/0004-6361/201527392},
archivePrefix = {arXiv},
       eprint = {1509.05607},
 primaryClass = {astro-ph.CO},
       adsurl = {https://ui.adsabs.harvard.edu/abs/2016A&A...587A..41P}
}

@ARTICLE{desi_qso,
       author = {{Y{\`e}che}, Christophe and {Palanque-Delabrouille}, Nathalie and {Claveau}, Charles-Antoine and {Brooks}, David D. and {Chaussidon}, Edmond and {Davis}, Tamara M. and {Dawson}, Kyle S. and {Dey}, Arjun and {Duan}, Yutong and {Eftekharzadeh}, Sarah and {Eisenstein}, Daniel J. and {Gazta{\~n}aga}, Enrique and {Kehoe}, Robert and {Landriau}, Martin and {Lang}, Dustin and {Levi}, Michael E. and {Meisner}, Aaron M. and {Myers}, Adam D. and {Newman}, Jeffrey A. and {Poppett}, Claire and {Prada}, Francisco and {Raichoor}, Anand and {Schlegel}, David J. and {Schubnell}, Michael and {Staten}, Ryan and {Tarl{\'e}}, Gregory and {Zhou}, Rongpu},
        title = "{Preliminary Target Selection for the DESI Quasar (QSO) Sample}",
      journal = {Research Notes of the American Astronomical Society},
         year = 2020,
        month = oct,
       volume = {4},
       number = {10},
          eid = {179},
        pages = {179},
          doi = {10.3847/2515-5172/abc01a},
archivePrefix = {arXiv},
       eprint = {2010.11280},
 primaryClass = {astro-ph.CO},
       adsurl = {https://ui.adsabs.harvard.edu/abs/2020RNAAS...4..179Y}
}

@ARTICLE{deep-z,
       author = {{Eriksen}, M. and {Alarcon}, A. and {Cabayol}, L. and {Carretero}, J. and {Casas}, R. and {Castander}, F.~J. and {De Vicente}, J. and {Fernandez}, E. and {Garcia-Bellido}, J. and {Gaztanaga}, E. and {Hildebrandt}, H. and {Hoekstra}, H. and {Joachimi}, B. and {Miquel}, R. and {Padilla}, C. and {Sanchez}, E. and {Sevilla-Noarbe}, I. and {Tallada}, P.},
        title = "{The PAU Survey: Photometric redshifts using transfer learning from simulations}",
      journal = {\mnras},
         year = 2020,
        month = oct,
       volume = {497},
       number = {4},
        pages = {4565-4579},
          doi = {10.1093/mnras/staa2265},
archivePrefix = {arXiv},
       eprint = {2004.07979},
 primaryClass = {astro-ph.GA},
       adsurl = {https://ui.adsabs.harvard.edu/abs/2020MNRAS.497.4565E}
}

@ARTICLE{3sdir,
       author = {{Myles}, J. and {Alarcon}, A. and {Amon}, A. and {S{\'a}nchez}, C. and {Everett}, S. and {DeRose}, J. and {McCullough}, J. and {Gruen}, D. and {Bernstein}, G.~M. and {Troxel}, M.~A. and {Dodelson}, S. and {Campos}, A. and {MacCrann}, N. and {Yin}, B. and {Raveri}, M. and {Amara}, A. and {Becker}, M.~R. and {Choi}, A. and {Cordero}, J. and {Eckert}, K. and {Gatti}, M. and {Giannini}, G. and {Gschwend}, J. and {Gruendl}, R.~A. and {Harrison}, I. and {Hartley}, W.~G. and {Huff}, E.~M. and {Kuropatkin}, N. and {Lin}, H. and {Masters}, D. and {Miquel}, R. and {Prat}, J. and {Roodman}, A. and {Rykoff}, E.~S. and {Sevilla-Noarbe}, I. and {Sheldon}, E. and {Wechsler}, R.~H. and {Yanny}, B. and {Abbott}, T.~M.~C. and {Aguena}, M. and {Allam}, S. and {Annis}, J. and {Bacon}, D. and {Bertin}, E. and {Bhargava}, S. and {Bridle}, S.~L. and {Brooks}, D. and {Burke}, D.~L. and {Carnero Rosell}, A. and {Carrasco Kind}, M. and {Carretero}, J. and {Castander}, F.~J. and {Conselice}, C. and {Costanzi}, M. and {Crocce}, M. and {da Costa}, L.~N. and {Pereira}, M.~E.~S. and {Desai}, S. and {Diehl}, H.~T. and {Eifler}, T.~F. and {Elvin-Poole}, J. and {Evrard}, A.~E. and {Ferrero}, I. and {Fert{\'e}}, A. and {Flaugher}, B. and {Fosalba}, P. and {Frieman}, J. and {Garc{\'\i}a-Bellido}, J. and {Gaztanaga}, E. and {Giannantonio}, T. and {Hinton}, S.~R. and {Hollowood}, D.~L. and {Honscheid}, K. and {Hoyle}, B. and {Huterer}, D. and {James}, D.~J. and {Krause}, E. and {Kuehn}, K. and {Lahav}, O. and {Lima}, M. and {Maia}, M.~A.~G. and {Marshall}, J.~L. and {Martini}, P. and {Melchior}, P. and {Menanteau}, F. and {Mohr}, J.~J. and {Morgan}, R. and {Muir}, J. and {Ogando}, R.~L.~C. and {Palmese}, A. and {Paz-Chinch{\'o}n}, F. and {Plazas}, A.~A. and {Rodriguez-Monroy}, M. and {Samuroff}, S. and {Sanchez}, E. and {Scarpine}, V. and {Secco}, L.~F. and {Serrano}, S. and {Smith}, M. and {Soares-Santos}, M. and {Suchyta}, E. and {Swanson}, M.~E.~C. and {Tarle}, G. and {Thomas}, D. and {To}, C. and {Varga}, T.~N. and {Weller}, J. and {Wester}, W.},
        title = "{Dark Energy Survey Year 3 results: redshift calibration of the weak lensing source galaxies}",
      journal = {\mnras},
         year = 2021,
        month = aug,
       volume = {505},
       number = {3},
        pages = {4249-4277},
          doi = {10.1093/mnras/stab1515},
archivePrefix = {arXiv},
       eprint = {2012.08566},
 primaryClass = {astro-ph.CO},
       adsurl = {https://ui.adsabs.harvard.edu/abs/2021MNRAS.505.4249M}
}

@ARTICLE{Scottez,
       author = {{Scottez}, V. and {Benoit-L{\'e}vy}, A. and {Coupon}, J. and {Ilbert}, O. and {Mellier}, Y.},
        title = "{Testing the accuracy of clustering redshifts with simulations}",
      journal = {\mnras},
         year = 2018,
        month = mar,
       volume = {474},
       number = {3},
        pages = {3921-3930},
          doi = {10.1093/mnras/stx3056},
archivePrefix = {arXiv},
       eprint = {1705.02629},
 primaryClass = {astro-ph.CO},
       adsurl = {https://ui.adsabs.harvard.edu/abs/2018MNRAS.474.3921S}
}

@ARTICLE{lsst_book,
       author = {{LSST Science Collaboration} and {Abell}, Paul A. and {Allison}, Julius and {Anderson}, Scott F. and {Andrew}, John R. and {Angel}, J. Roger P. and {Armus}, Lee and {Arnett}, David and {Asztalos}, S.~J. and {Axelrod}, Tim S. and {Bailey}, Stephen and {Ballantyne}, D.~R. and {Bankert}, Justin R. and {Barkhouse}, Wayne A. and {Barr}, Jeffrey D. and {Barrientos}, L. Felipe and {Barth}, Aaron J. and {Bartlett}, James G. and {Becker}, Andrew C. and {Becla}, Jacek and {Beers}, Timothy C. and {Bernstein}, Joseph P. and {Biswas}, Rahul and {Blanton}, Michael R. and {Bloom}, Joshua S. and {Bochanski}, John J. and {Boeshaar}, Pat and {Borne}, Kirk D. and {Bradac}, Marusa and {Brandt}, W.~N. and {Bridge}, Carrie R. and {Brown}, Michael E. and {Brunner}, Robert J. and {Bullock}, James S. and {Burgasser}, Adam J. and {Burge}, James H. and {Burke}, David L. and {Cargile}, Phillip A. and {Chandrasekharan}, Srinivasan and {Chartas}, George and {Chesley}, Steven R. and {Chu}, You-Hua and {Cinabro}, David and {Claire}, Mark W. and {Claver}, Charles F. and {Clowe}, Douglas and {Connolly}, A.~J. and {Cook}, Kem H. and {Cooke}, Jeff and {Cooray}, Asantha and {Covey}, Kevin R. and {Culliton}, Christopher S. and {de Jong}, Roelof and {de Vries}, Willem H. and {Debattista}, Victor P. and {Delgado}, Francisco and {Dell'Antonio}, Ian P. and {Dhital}, Saurav and {Di Stefano}, Rosanne and {Dickinson}, Mark and {Dilday}, Benjamin and {Djorgovski}, S.~G. and {Dobler}, Gregory and {Donalek}, Ciro and {Dubois-Felsmann}, Gregory and {Durech}, Josef and {Eliasdottir}, Ardis and {Eracleous}, Michael and {Eyer}, Laurent and {Falco}, Emilio E. and {Fan}, Xiaohui and {Fassnacht}, Christopher D. and {Ferguson}, Harry C. and {Fernandez}, Yanga R. and {Fields}, Brian D. and {Finkbeiner}, Douglas and {Figueroa}, Eduardo E. and {Fox}, Derek B. and {Francke}, Harold and {Frank}, James S. and {Frieman}, Josh and {Fromenteau}, Sebastien and {Furqan}, Muhammad and {Galaz}, Gaspar and {Gal-Yam}, A. and {Garnavich}, Peter and {Gawiser}, Eric and {Geary}, John and {Gee}, Perry and {Gibson}, Robert R. and {Gilmore}, Kirk and {Grace}, Emily A. and {Green}, Richard F. and {Gressler}, William J. and {Grillmair}, Carl J. and {Habib}, Salman and {Haggerty}, J.~S. and {Hamuy}, Mario and {Harris}, Alan W. and {Hawley}, Suzanne L. and {Heavens}, Alan F. and {Hebb}, Leslie and {Henry}, Todd J. and {Hileman}, Edward and {Hilton}, Eric J. and {Hoadley}, Keri and {Holberg}, J.~B. and {Holman}, Matt J. and {Howell}, Steve B. and {Infante}, Leopoldo and {Ivezic}, Zeljko and {Jacoby}, Suzanne H. and {Jain}, Bhuvnesh and {R} and {Jedicke} and {Jee}, M. James and {Garrett Jernigan}, J. and {Jha}, Saurabh W. and {Johnston}, Kathryn V. and {Jones}, R. Lynne and {Juric}, Mario and {Kaasalainen}, Mikko and {Styliani} and {Kafka} and {Kahn}, Steven M. and {Kaib}, Nathan A. and {Kalirai}, Jason and {Kantor}, Jeff and {Kasliwal}, Mansi M. and {Keeton}, Charles R. and {Kessler}, Richard and {Knezevic}, Zoran and {Kowalski}, Adam and {Krabbendam}, Victor L. and {Krughoff}, K. Simon and {Kulkarni}, Shrinivas and {Kuhlman}, Stephen and {Lacy}, Mark and {Lepine}, Sebastien and {Liang}, Ming and {Lien}, Amy and {Lira}, Paulina and {Long}, Knox S. and {Lorenz}, Suzanne and {Lotz}, Jennifer M. and {Lupton}, R.~H. and {Lutz}, Julie and {Macri}, Lucas M. and {Mahabal}, Ashish A. and {Mandelbaum}, Rachel and {Marshall}, Phil and {May}, Morgan and {McGehee}, Peregrine M. and {Meadows}, Brian T. and {Meert}, Alan and {Milani}, Andrea and {Miller}, Christopher J. and {Miller}, Michelle and {Mills}, David and {Minniti}, Dante and {Monet}, David and {Mukadam}, Anjum S. and {Nakar}, Ehud and {Neill}, Douglas R. and {Newman}, Jeffrey A. and {Nikolaev}, Sergei and {Nordby}, Martin and {O'Connor}, Paul and {Oguri}, Masamune and {Oliver}, John and {Olivier}, Scot S. and {Olsen}, Julia K. and {Olsen}, Knut and {Olszewski}, Edward W. and {Oluseyi}, Hakeem and {Padilla}, Nelson D. and {Parker}, Alex and {Pepper}, Joshua and {Peterson}, John R. and {Petry}, Catherine and {Pinto}, Philip A. and {Pizagno}, James L. and {Popescu}, Bogdan and {Prsa}, Andrej and {Radcka}, Veljko and {Raddick}, M. Jordan and {Rasmussen}, Andrew and {Rau}, Arne and {Rho}, Jeonghee and {Rhoads}, James E. and {Richards}, Gordon T. and {Ridgway}, Stephen T. and {Robertson}, Brant E. and {Roskar}, Rok and {Saha}, Abhijit and {Sarajedini}, Ata and {Scannapieco}, Evan and {Schalk}, Terry and {Schindler}, Rafe and {Schmidt}, Samuel},
        title = "{LSST Science Book, Version 2.0}",
         year = 2009,
        month = dec,
          eid = {arXiv:0912.0201},
        pages = {arXiv:0912.0201},
          doi = {10.48550/arXiv.0912.0201},
       eprint = {0912.0201},
 primaryClass = {astro-ph.IM},
       adsurl = {https://ui.adsabs.harvard.edu/abs/2009arXiv0912.0201L}
}

@ARTICLE{how_accurate_is_limber,
       author = {{Simon}, P.},
        title = "{How accurate is Limber's equation?}",
      journal = {\aap},
         year = 2007,
        month = oct,
       volume = {473},
       number = {3},
        pages = {711-714},
          doi = {10.1051/0004-6361:20066352},
archivePrefix = {arXiv},
       eprint = {astro-ph/0609165},
 primaryClass = {astro-ph},
       adsurl = {https://ui.adsabs.harvard.edu/abs/2007A&A...473..711S}
}

@ARTICLE{Davis_Peebles,
       author = {{Davis}, M. and {Peebles}, P.~J.~E.},
        title = "{A survey of galaxy redshifts. V. The two-point position and velocity correlations.}",
      journal = {\apj},
         year = 1983,
        month = apr,
       volume = {267},
        pages = {465-482},
          doi = {10.1086/160884},
       adsurl = {https://ui.adsabs.harvard.edu/abs/1983ApJ...267..465D}
}

@ARTICLE{interloper,
       author = {{Addison}, G.~E. and {Bennett}, C.~L. and {Jeong}, D. and {Komatsu}, E. and {Weiland}, J.~L.},
        title = "{The Impact of Line Misidentification on Cosmological Constraints from Euclid and Other Spectroscopic Galaxy Surveys}",
      journal = {\apj},
         year = 2019,
        month = jul,
       volume = {879},
       number = {1},
          eid = {15},
        pages = {15},
          doi = {10.3847/1538-4357/ab22a0},
archivePrefix = {arXiv},
       eprint = {1811.10668},
 primaryClass = {astro-ph.CO},
       adsurl = {https://ui.adsabs.harvard.edu/abs/2019ApJ...879...15A}
}

@ARTICLE{cosmohub,
       author = {{Tallada}, P. and {Carretero}, J. and {Casals}, J. and {Acosta-Silva}, C. and {Serrano}, S. and {Caubet}, M. and {Castander}, F.~J. and {C{\'e}sar}, E. and {Crocce}, M. and {Delfino}, M. and {Eriksen}, M. and {Fosalba}, P. and {Gazta{\~n}aga}, E. and {Merino}, G. and {Neissner}, C. and {Tonello}, N.},
        title = "{CosmoHub: Interactive exploration and distribution of astronomical data on Hadoop}",
      journal = {Astronomy and Computing},
         year = 2020,
        month = jul,
       volume = {32},
          eid = {100391},
        pages = {100391},
          doi = {10.1016/j.ascom.2020.100391},
archivePrefix = {arXiv},
       eprint = {2003.03217},
 primaryClass = {astro-ph.IM},
       adsurl = {https://ui.adsabs.harvard.edu/abs/2020A&C....3200391T}
}

@ARTICLE{FS_2024,
       author = {{Euclid Collaboration: Castander}, F. and {Fosalba}, P. and {Stadel}, J. and others},
        title = "{Euclid: V. The Flagship galaxy mock catalogue: A comprehensive simulation for the Euclid mission}",
      journal = {\aap},
         year = 2025,
        month = may,
       volume = {697},
          eid = {A5},
        pages = {A5},
          doi = {10.1051/0004-6361/202450853},
archivePrefix = {arXiv},
       eprint = {2405.13495},
 primaryClass = {astro-ph.CO},
       adsurl = {https://ui.adsabs.harvard.edu/abs/2025A&A...697A...5E}
}

@ARTICLE{halofit,
       author = {{Takahashi}, Ryuichi and {Sato}, Masanori and {Nishimichi}, Takahiro and {Taruya}, Atsushi and {Oguri}, Masamune},
        title = "{Revising the Halofit Model for the Nonlinear Matter Power Spectrum}",
      journal = {\apj},
         year = 2012,
        month = dec,
       volume = {761},
       number = {2},
          eid = {152},
        pages = {152},
          doi = {10.1088/0004-637X/761/2/152},
archivePrefix = {arXiv},
       eprint = {1208.2701},
 primaryClass = {astro-ph.CO},
       adsurl = {https://ui.adsabs.harvard.edu/abs/2012ApJ...761..152T}
}

@article{Diego_Blas_2011_class,
   title={The Cosmic Linear Anisotropy Solving System (CLASS).
 Part II: Approximation schemes},
   ISSN={1475-7516},
   url={http://dx.doi.org/10.1088/1475-7516/2011/07/034},
   DOI={10.1088/1475-7516/2011/07/034},
   number={07},
   journal={JCAP},
   publisher={IOP Publishing},
   author={Diego Blas and Julien Lesgourgues and Thomas Tram},
   year={2011},
   month=jul, pages={07, 034} }

@ARTICLE{CCL,
       author = {{Chisari}, Nora Elisa and {Alonso}, David and {Krause}, Elisabeth and {Leonard}, C. Danielle and {Bull}, Philip and {Neveu}, J{\'e}r{\'e}my and {Villarreal}, Antonia Sierra and {Singh}, Sukhdeep and {McClintock}, Thomas and {Ellison}, John and {Du}, Zilong and {Zuntz}, Joe and {Mead}, Alexander and {Joudaki}, Shahab and {Lorenz}, Christiane S. and {Tr{\"o}ster}, Tilman and {Sanchez}, Javier and {Lanusse}, Francois and {Ishak}, Mustapha and {Hlozek}, Ren{\'e}e and {Blazek}, Jonathan and {Campagne}, Jean-Eric and {Almoubayyed}, Husni and {Eifler}, Tim and {Kirby}, Matthew and {Kirkby}, David and {Plaszczynski}, St{\'e}phane and {Slosar}, An{\v{z}}e and {Vrastil}, Michal and {Wagoner}, Erika L. and {LSST Dark Energy Science Collaboration}},
        title = "{Core Cosmology Library: Precision Cosmological Predictions for LSST}",
      journal = {\apjs},
         year = 2019,
        month = may,
       volume = {242},
       number = {1},
          eid = {2},
        pages = {2},
          doi = {10.3847/1538-4365/ab1658},
archivePrefix = {arXiv},
       eprint = {1812.05995},
 primaryClass = {astro-ph.CO},
       adsurl = {https://ui.adsabs.harvard.edu/abs/2019ApJS..242....2C}
}

@ARTICLE{Rockstar_2013,
       author = {{Behroozi}, Peter S. and {Wechsler}, Risa H. and {Wu}, Hao-Yi},
        title = "{The ROCKSTAR Phase-space Temporal Halo Finder and the Velocity Offsets of Cluster Cores}",
      journal = {\apj},
         year = 2013,
        month = jan,
       volume = {762},
       number = {2},
          eid = {109},
        pages = {109},
          doi = {10.1088/0004-637X/762/2/109},
archivePrefix = {arXiv},
       eprint = {1110.4372},
 primaryClass = {astro-ph.CO},
       adsurl = {https://ui.adsabs.harvard.edu/abs/2013ApJ...762..109B}
}

@ARTICLE{DESI_2016,
       author = {{DESI Collaboration} and {Aghamousa}, Amir and {Aguilar}, Jessica and {Ahlen}, Steve and {Alam}, Shadab and {Allen}, Lori E. and {Allende Prieto}, Carlos and {Annis}, James and {Bailey}, Stephen and {Balland}, Christophe and {Ballester}, Otger and {Baltay}, Charles and {Beaufore}, Lucas and {Bebek}, Chris and {Beers}, Timothy C. and {Bell}, Eric F. and {Bernal}, Jos{\'e} Luis and {Besuner}, Robert and {Beutler}, Florian and {Blake}, Chris and {Bleuler}, Hannes and {Blomqvist}, Michael and {Blum}, Robert and {Bolton}, Adam S. and {Briceno}, Cesar and {Brooks}, David and {Brownstein}, Joel R. and {Buckley-Geer}, Elizabeth and {Burden}, Angela and {Burtin}, Etienne and {Busca}, Nicolas G. and {Cahn}, Robert N. and {Cai}, Yan-Chuan and {Cardiel-Sas}, Laia and {Carlberg}, Raymond G. and {Carton}, Pierre-Henri and {Casas}, Ricard and {Castander}, Francisco J. and {Cervantes-Cota}, Jorge L. and {Claybaugh}, Todd M. and {Close}, Madeline and {Coker}, Carl T. and {Cole}, Shaun and {Comparat}, Johan and {Cooper}, Andrew P. and {Cousinou}, M. -C. and {Crocce}, Martin and {Cuby}, Jean-Gabriel and {Cunningham}, Daniel P. and {Davis}, Tamara M. and {Dawson}, Kyle S. and {de la Macorra}, Axel and {De Vicente}, Juan and {Delubac}, Timoth{\'e}e and {Derwent}, Mark and {Dey}, Arjun and {Dhungana}, Govinda and {Ding}, Zhejie and {Doel}, Peter and {Duan}, Yutong T. and {Ealet}, Anne and {Edelstein}, Jerry and {Eftekharzadeh}, Sarah and {Eisenstein}, Daniel J. and {Elliott}, Ann and {Escoffier}, St{\'e}phanie and {Evatt}, Matthew and {Fagrelius}, Parker and {Fan}, Xiaohui and {Fanning}, Kevin and {Farahi}, Arya and {Farihi}, Jay and {Favole}, Ginevra and {Feng}, Yu and {Fernandez}, Enrique and {Findlay}, Joseph R. and {Finkbeiner}, Douglas P. and {Fitzpatrick}, Michael J. and {Flaugher}, Brenna and {Flender}, Samuel and {Font-Ribera}, Andreu and {Forero-Romero}, Jaime E. and {Fosalba}, Pablo and {Frenk}, Carlos S. and {Fumagalli}, Michele and {Gaensicke}, Boris T. and {Gallo}, Giuseppe and {Garcia-Bellido}, Juan and {Gaztanaga}, Enrique and {Pietro Gentile Fusillo}, Nicola and {Gerard}, Terry and {Gershkovich}, Irena and {Giannantonio}, Tommaso and {Gillet}, Denis and {Gonzalez-de-Rivera}, Guillermo and {Gonzalez-Perez}, Violeta and {Gott}, Shelby and {Graur}, Or and {Gutierrez}, Gaston and {Guy}, Julien and {Habib}, Salman and {Heetderks}, Henry and {Heetderks}, Ian and {Heitmann}, Katrin and {Hellwing}, Wojciech A. and {Herrera}, David A. and {Ho}, Shirley and {Holland}, Stephen and {Honscheid}, Klaus and {Huff}, Eric and {Hutchinson}, Timothy A. and {Huterer}, Dragan and {Hwang}, Ho Seong and {Illa Laguna}, Joseph Maria and {Ishikawa}, Yuzo and {Jacobs}, Dianna and {Jeffrey}, Niall and {Jelinsky}, Patrick and {Jennings}, Elise and {Jiang}, Linhua and {Jimenez}, Jorge and {Johnson}, Jennifer and {Joyce}, Richard and {Jullo}, Eric and {Juneau}, St{\'e}phanie and {Kama}, Sami and {Karcher}, Armin and {Karkar}, Sonia and {Kehoe}, Robert and {Kennamer}, Noble and {Kent}, Stephen and {Kilbinger}, Martin and {Kim}, Alex G. and {Kirkby}, David and {Kisner}, Theodore and {Kitanidis}, Ellie and {Kneib}, Jean-Paul and {Koposov}, Sergey and {Kovacs}, Eve and {Koyama}, Kazuya and {Kremin}, Anthony and {Kron}, Richard and {Kronig}, Luzius and {Kueter-Young}, Andrea and {Lacey}, Cedric G. and {Lafever}, Robin and {Lahav}, Ofer and {Lambert}, Andrew and {Lampton}, Michael and {Landriau}, Martin and {Lang}, Dustin and {Lauer}, Tod R. and {Le Goff}, Jean-Marc and {Le Guillou}, Laurent and {Le Van Suu}, Auguste and {Lee}, Jae Hyeon and {Lee}, Su-Jeong and {Leitner}, Daniela and {Lesser}, Michael and {Levi}, Michael E. and {L'Huillier}, Benjamin and {Li}, Baojiu and {Liang}, Ming and {Lin}, Huan and {Linder}, Eric and {Loebman}, Sarah R. and {Luki{\'c}}, Zarija and {Ma}, Jun and {MacCrann}, Niall and {Magneville}, Christophe and {Makarem}, Laleh and {Manera}, Marc and {Manser}, Christopher J. and {Marshall}, Robert and {Martini}, Paul and {Massey}, Richard and {Matheson}, Thomas and {McCauley}, Jeremy and {McDonald}, Patrick and {McGreer}, Ian D. and {Meisner}, Aaron and {Metcalfe}, Nigel and {Miller}, Timothy N. and {Miquel}, Ramon and {Moustakas}, John and {Myers}, Adam and {Naik}, Milind and {Newman}, Jeffrey A. and {Nichol}, Robert C. and {Nicola}, Andrina and {Nicolati da Costa}, Luiz and {Nie}, Jundan and {Niz}, Gustavo and {Norberg}, Peder and {Nord}, Brian and {Norman}, Dara and {Nugent}, Peter and {O'Brien}, Thomas and {Oh}, Minji and {Olsen}, Knut A.~G. and {Padilla}, Cristobal and {Padmanabhan}, Hamsa and {Padmanabhan}, Nikhil and {Palanque-Delabrouille}, Nathalie and {Palmese}, Antonella and {Pappalardo}, Daniel and {P{\^a}ris}, Isabelle and {Park}, Changbom and {Patej}, Anna and {Peacock}, John A. and {Peiris}, Hiranya V. and {Peng}, Xiyan and {Percival}, Will J. and {Perruchot}, Sandrine and {Pieri}, Matthew M. and {Pogge}, Richard and {Pollack}, Jennifer E. and {Poppett}, Claire and {Prada}, Francisco and {Prakash}, Abhishek and {Probst}, Ronald G. and {Rabinowitz}, David and {Raichoor}, Anand and {Ree}, Chang Hee and {Refregier}, Alexandre and {Regal}, Xavier and {Reid}, Beth and {Reil}, Kevin and {Rezaie}, Mehdi and {Rockosi}, Constance M. and {Roe}, Natalie and {Ronayette}, Samuel and {Roodman}, Aaron and {Ross}, Ashley J. and {Ross}, Nicholas P. and {Rossi}, Graziano and {Rozo}, Eduardo and {Ruhlmann-Kleider}, Vanina and {Rykoff}, Eli S. and {Sabiu}, Cristiano and {Samushia}, Lado and {Sanchez}, Eusebio and {Sanchez}, Javier and {Schlegel}, David J. and {Schneider}, Michael and {Schubnell}, Michael and {Secroun}, Aur{\'e}lia and {Seljak}, Uros and {Seo}, Hee-Jong and {Serrano}, Santiago and {Shafieloo}, Arman and {Shan}, Huanyuan and {Sharples}, Ray and {Sholl}, Michael J. and {Shourt}, William V. and {Silber}, Joseph H. and {Silva}, David R. and {Sirk}, Martin M. and {Slosar}, Anze and {Smith}, Alex and {Smoot}, George F. and {Som}, Debopam and {Song}, Yong-Seon and {Sprayberry}, David and {Staten}, Ryan and {Stefanik}, Andy and {Tarle}, Gregory and {Sien Tie}, Suk and {Tinker}, Jeremy L. and {Tojeiro}, Rita and {Valdes}, Francisco and {Valenzuela}, Octavio and {Valluri}, Monica and {Vargas-Magana}, Mariana and {Verde}, Licia and {Walker}, Alistair R. and {Wang}, Jiali and {Wang}, Yuting and {Weaver}, Benjamin A. and {Weaverdyck}, Curtis and {Wechsler}, Risa H. and {Weinberg}, David H. and {White}, Martin and {Yang}, Qian and {Yeche}, Christophe and {Zhang}, Tianmeng and {Zhao}, Gong-Bo and {Zheng}, Yi and {Zhou}, Xu and {Zhou}, Zhimin and {Zhu}, Yaling and {Zou}, Hu and {Zu}, Ying},
        title = "{The DESI Experiment Part II: Instrument Design}",
         year = 2016,
        month = oct,
          eid = {arXiv:1611.00037},
        pages = {arXiv:1611.00037},
          doi = {10.48550/arXiv.1611.00037},
archivePrefix = {arXiv},
       eprint = {1611.00037},
 primaryClass = {astro-ph.IM},
       adsurl = {https://ui.adsabs.harvard.edu/abs/2016arXiv161100037D}
}

@ARTICLE{BOSS_color,
       author = {{Dawson}, Kyle S. and {Schlegel}, David J. and {Ahn}, Christopher P. and {Anderson}, Scott F. and {Aubourg}, {\'E}ric and {Bailey}, Stephen and {Barkhouser}, Robert H. and {Bautista}, Julian E. and {Beifiori}, Alessandra and {Berlind}, Andreas A. and {Bhardwaj}, Vaishali and {Bizyaev}, Dmitry and {Blake}, Cullen H. and {Blanton}, Michael R. and {Blomqvist}, Michael and {Bolton}, Adam S. and {Borde}, Arnaud and {Bovy}, Jo and {Brandt}, W.~N. and {Brewington}, Howard and {Brinkmann}, Jon and {Brown}, Peter J. and {Brownstein}, Joel R. and {Bundy}, Kevin and {Busca}, N.~G. and {Carithers}, William and {Carnero}, Aurelio R. and {Carr}, Michael A. and {Chen}, Yanmei and {Comparat}, Johan and {Connolly}, Natalia and {Cope}, Frances and {Croft}, Rupert A.~C. and {Cuesta}, Antonio J. and {da Costa}, Luiz N. and {Davenport}, James R.~A. and {Delubac}, Timoth{\'e}e and {de Putter}, Roland and {Dhital}, Saurav and {Ealet}, Anne and {Ebelke}, Garrett L. and {Eisenstein}, Daniel J. and {Escoffier}, S. and {Fan}, Xiaohui and {Filiz Ak}, N. and {Finley}, Hayley and {Font-Ribera}, Andreu and {G{\'e}nova-Santos}, R. and {Gunn}, James E. and {Guo}, Hong and {Haggard}, Daryl and {Hall}, Patrick B. and {Hamilton}, Jean-Christophe and {Harris}, Ben and {Harris}, David W. and {Ho}, Shirley and {Hogg}, David W. and {Holder}, Diana and {Honscheid}, Klaus and {Huehnerhoff}, Joe and {Jordan}, Beatrice and {Jordan}, Wendell P. and {Kauffmann}, Guinevere and {Kazin}, Eyal A. and {Kirkby}, David and {Klaene}, Mark A. and {Kneib}, Jean-Paul and {Le Goff}, Jean-Marc and {Lee}, Khee-Gan and {Long}, Daniel C. and {Loomis}, Craig P. and {Lundgren}, Britt and {Lupton}, Robert H. and {Maia}, Marcio A.~G. and {Makler}, Martin and {Malanushenko}, Elena and {Malanushenko}, Viktor and {Mandelbaum}, Rachel and {Manera}, Marc and {Maraston}, Claudia and {Margala}, Daniel and {Masters}, Karen L. and {McBride}, Cameron K. and {McDonald}, Patrick and {McGreer}, Ian D. and {McMahon}, Richard G. and {Mena}, Olga and {Miralda-Escud{\'e}}, Jordi and {Montero-Dorta}, Antonio D. and {Montesano}, Francesco and {Muna}, Demitri and {Myers}, Adam D. and {Naugle}, Tracy and {Nichol}, Robert C. and {Noterdaeme}, Pasquier and {Nuza}, Sebasti{\'a}n E. and {Olmstead}, Matthew D. and {Oravetz}, Audrey and {Oravetz}, Daniel J. and {Owen}, Russell and {Padmanabhan}, Nikhil and {Palanque-Delabrouille}, Nathalie and {Pan}, Kaike and {Parejko}, John K. and {P{\^a}ris}, Isabelle and {Percival}, Will J. and {P{\'e}rez-Fournon}, Ismael and {P{\'e}rez-R{\`a}fols}, Ignasi and {Petitjean}, Patrick and {Pfaffenberger}, Robert and {Pforr}, Janine and {Pieri}, Matthew M. and {Prada}, Francisco and {Price-Whelan}, Adrian M. and {Raddick}, M. Jordan and {Rebolo}, Rafael and {Rich}, James and {Richards}, Gordon T. and {Rockosi}, Constance M. and {Roe}, Natalie A. and {Ross}, Ashley J. and {Ross}, Nicholas P. and {Rossi}, Graziano and {Rubi{\~n}o-Martin}, J.~A. and {Samushia}, Lado and {S{\'a}nchez}, Ariel G. and {Sayres}, Conor and {Schmidt}, Sarah J. and {Schneider}, Donald P. and {Sc{\'o}ccola}, C.~G. and {Seo}, Hee-Jong and {Shelden}, Alaina and {Sheldon}, Erin and {Shen}, Yue and {Shu}, Yiping and {Slosar}, An{\v{z}}e and {Smee}, Stephen A. and {Snedden}, Stephanie A. and {Stauffer}, Fritz and {Steele}, Oliver and {Strauss}, Michael A. and {Streblyanska}, Alina and {Suzuki}, Nao and {Swanson}, Molly E.~C. and {Tal}, Tomer and {Tanaka}, Masayuki and {Thomas}, Daniel and {Tinker}, Jeremy L. and {Tojeiro}, Rita and {Tremonti}, Christy A. and {Vargas Maga{\~n}a}, M. and {Verde}, Licia and {Viel}, Matteo and {Wake}, David A. and {Watson}, Mike and {Weaver}, Benjamin A. and {Weinberg}, David H. and {Weiner}, Benjamin J. and {West}, Andrew A. and {White}, Martin and {Wood-Vasey}, W.~M. and {Yeche}, Christophe and {Zehavi}, Idit and {Zhao}, Gong-Bo and {Zheng}, Zheng},
        title = "{The Baryon Oscillation Spectroscopic Survey of SDSS-III}",
      journal = {\aj},
         year = 2013,
        month = jan,
       volume = {145},
       number = {1},
          eid = {10},
        pages = {10},
          doi = {10.1088/0004-6256/145/1/10},
archivePrefix = {arXiv},
       eprint = {1208.0022},
 primaryClass = {astro-ph.CO},
       adsurl = {https://ui.adsabs.harvard.edu/abs/2013AJ....145...10D}
}

@ARTICLE{DESI_validation,
       author = {{Adame}, A.~G. and {Aguilar}, J. and {Ahlen}, S. and {Alam}, S. and {Aldering}, G. and {Alexander}, D.~M. and {Alfarsy}, R. and {Allende Prieto}, C. and {Alvarez}, M. and {Alves}, O. and {Anand}, A. and {Andrade-Oliveira}, F. and {Armengaud}, E. and {Asorey}, J. and {Avila}, S. and {Aviles}, A. and {Bailey}, S. and {Balaguera-Antol{\'\i}nez}, A. and {Ballester}, O. and {Baltay}, C. and {Bault}, A. and {Bautista}, J. and {Behera}, J. and {Beltran}, S.~F. and {BenZvi}, S. and {Beraldo e Silva}, L. and {Bermejo-Climent}, J.~R. and {Berti}, A. and {Besuner}, R. and {Beutler}, F. and {Bianchi}, D. and {Blake}, C. and {Blum}, R. and {Bolton}, A.~S. and {Brieden}, S. and {Brodzeller}, A. and {Brooks}, D. and {Brown}, Z. and {Buckley-Geer}, E. and {Burtin}, E. and {Cabayol-Garcia}, L. and {Cai}, Z. and {Canning}, R. and {Cardiel-Sas}, L. and {Carnero Rosell}, A. and {Castander}, F.~J. and {Cervantes-Cota}, J.~L. and {Chabanier}, S. and {Chaussidon}, E. and {Chaves-Montero}, J. and {Chen}, S. and {Chen}, X. and {Chuang}, C. and {Claybaugh}, T. and {Cole}, S. and {Cooper}, A.~P. and {Cuceu}, A. and {Davis}, T.~M. and {Dawson}, K. and {de Belsunce}, R. and {de la Cruz}, R. and {de la Macorra}, A. and {de Mattia}, A. and {Demina}, R. and {Demirbozan}, U. and {DeRose}, J. and {Dey}, A. and {Dey}, B. and {Dhungana}, G. and {Ding}, J. and {Ding}, Z. and {Doel}, P. and {Doshi}, R. and {Douglass}, K. and {Edge}, A. and {Eftekharzadeh}, S. and {Eisenstein}, D.~J. and {Elliott}, A. and {Escoffier}, S. and {Fagrelius}, P. and {Fan}, X. and {Fanning}, K. and {Fawcett}, V.~A. and {Ferraro}, S. and {Ereza}, J. and {Flaugher}, B. and {Font-Ribera}, A. and {Forero-S{\'a}nchez}, D. and {Forero-Romero}, J.~E. and {Frenk}, C.~S. and {G{\"a}nsicke}, B.~T. and {Garc{\'\i}a}, L. {\'A}. and {Garc{\'\i}a-Bellido}, J. and {Garcia-Quintero}, C. and {Garrison}, L.~H. and {Gil-Mar{\'\i}n}, H. and {Golden-Marx}, J. and {Gontcho A Gontcho}, S. and {Gonzalez-Morales}, A.~X. and {Gonzalez-Perez}, V. and {Gordon}, C. and {Graur}, O. and {Green}, D. and {Gruen}, D. and {Guy}, J. and {Hadzhiyska}, B. and {Hahn}, C. and {Han}, J.~J. and {Hanif}, M.~M.~S. and {Herrera-Alcantar}, H.~K. and {Honscheid}, K. and {Hou}, J. and {Howlett}, C. and {Huterer}, D. and {Ir{\v{s}}i{\v{c}}}, V. and {Ishak}, M. and {Jana}, A. and {Jiang}, L. and {Jimenez}, J. and {Jing}, Y.~P. and {Joudaki}, S. and {Jullo}, E. and {Joyce}, R. and {Juneau}, S. and {Kizhuprakkat}, N. and {Kara{\c{c}}ayl{\i}}, N.~G. and {Karim}, T. and {Kehoe}, R. and {Kent}, S. and {Khederlarian}, A. and {Kim}, S. and {Kirkby}, D. and {Kisner}, T. and {Kitaura}, F. and {Kneib}, J. and {Koposov}, S.~E. and {Kov{\'a}cs}, A. and {Kremin}, A. and {Krolewski}, A. and {L'Huillier}, B. and {Lahav}, O. and {Lambert}, A. and {Lamman}, C. and {Lan}, T. -W. and {Landriau}, M. and {Lang}, D. and {Lange}, J.~U. and {Lasker}, J. and {Le Guillou}, L. and {Leauthaud}, A. and {Levi}, M.~E. and {Li}, T.~S. and {Linder}, E. and {Lyons}, A. and {Magneville}, C. and {Manera}, M. and {Manser}, C.~J. and {Margala}, D. and {Martini}, P. and {McDonald}, P. and {Medina}, G.~E. and {Medina-Varela}, L. and {Meisner}, A. and {Mena-Fern{\'a}ndez}, J. and {Meneses-Rizo}, J. and {Mezcua}, M. and {Miquel}, R. and {Montero-Camacho}, P. and {Moon}, J. and {Moore}, S. and {Moustakas}, J. and {Mueller}, E. and {Mundet}, J. and {Mu{\~n}oz-Guti{\'e}rrez}, A. and {Myers}, A.~D. and {Nadathur}, S. and {Napolitano}, L. and {Neveux}, R. and {Newman}, J.~A. and {Nie}, J. and {Niz}, G. and {Norberg}, P. and {Noriega}, H.~E. and {Paillas}, E. and {Palanque-Delabrouille}, N. and {Palmese}, A. and {Zhiwei}, P. and {Parkinson}, D. and {Penmetsa}, S. and {Percival}, W.~J. and {P{\'e}rez-Fern{\'a}ndez}, A. and {P{\'e}rez-R{\`a}fols}, I. and {Pieri}, M. and {Poppett}, C. and {Porredon}, A. and {Prada}, F. and {Pucha}, R. and {Raichoor}, A. and {Ram{\'\i}rez-P{\'e}rez}, C. and {Ramirez-Solano}, S. and {Rashkovetskyi}, M. and {Ravoux}, C. and {Rocher}, A. and {Rockosi}, C. and {Ross}, A.~J. and {Rossi}, G. and {Ruggeri}, R. and {Ruhlmann-Kleider}, V. and {Sabiu}, C.~G. and {Said}, K. and {Saintonge}, A. and {Samushia}, L. and {Sanchez}, E. and {Saulder}, C. and {Schaan}, E. and {Schlafly}, E.~F. and {Schlegel}, D. and {Scholte}, D. and {Schubnell}, M. and {Seo}, H. and {Shafieloo}, A. and {Sharples}, R. and {Sheu}, W. and {Silber}, J. and {Sinigaglia}, F. and {Siudek}, M. and {Slepian}, Z. and {Smith}, A. and {Sprayberry}, D. and {Stephey}, L. and {Su{\'a}rez-P{\'e}rez}, J. and {Sun}, Z. and {Tan}, T. and {Tarl{\'e}}, G. and {Tojeiro}, R. and {Ure{\~n}a-L{\'o}pez}, L.~A. and {Vaisakh}, R. and {Valcin}, D. and {Valdes}, F. and {Valluri}, M. and {Vargas-Maga{\~n}a}, M. and {Variu}, A. and {Verde}, L. and {Walther}, M. and {Wang}, B. and {Wang}, M.~S. and {Weaver}, B.~A. and {Weaverdyck}, N. and {Wechsler}, R.~H. and {White}, M. and {Xie}, Y. and {Yang}, J. and {Y{\`e}che}, C. and {Yu}, J. and {Yuan}, S. and {Zhang}, H. and {Zhang}, Z. and {Zhao}, C. and {Zheng}, Z. and {Zhou}, R. and {Zhou}, Z. and {Zou}, H. and {Zou}, S. and {Zu}, Y. and {DESI Collaboration}},
        title = "{Validation of the Scientific Program for the Dark Energy Spectroscopic Instrument}",
      journal = {\aj},
         year = 2024,
        month = feb,
       volume = {167},
       number = {2},
          eid = {62},
        pages = {62},
          doi = {10.3847/1538-3881/ad0b08},
archivePrefix = {arXiv},
       eprint = {2306.06307},
 primaryClass = {astro-ph.CO},
       adsurl = {https://ui.adsabs.harvard.edu/abs/2024AJ....167...62A}
}

@ARTICLE{DESY3_MAGLIM_z,
       author = {{Giannini}, G. and {Alarcon}, A. and {Gatti}, M. and {Porredon}, A. and {Crocce}, M. and {Bernstein}, G.~M. and {Cawthon}, R. and {S{\'a}nchez}, C. and {Doux}, C. and {Elvin-Poole}, J. and {Raveri}, M. and {Myles}, J. and {Lin}, H. and {Amon}, A. and {Allam}, S. and {Alves}, O. and {Andrade-Oliveira}, F. and {Baxter}, E. and {Bechtol}, K. and {Becker}, M.~R. and {Blazek}, J. and {Camacho}, H. and {Campos}, A. and {Carnero Rosell}, A. and {Carrasco Kind}, M. and {Choi}, A. and {Cordero}, J. and {De Vicente}, J. and {DeRose}, J. and {Diehl}, H.~T. and {Dodelson}, S. and {Drlica-Wagner}, A. and {Eckert}, K. and {Fang}, X. and {Farahi}, A. and {Fosalba}, P. and {Friedrich}, O. and {Gruen}, D. and {Gruendl}, R.~A. and {Gschwend}, J. and {Harrison}, I. and {Hartley}, W.~G. and {Huff}, E.~M. and {Jarvis}, M. and {Krause}, E. and {Kuropatkin}, N. and {Lemos}, P. and {MacCrann}, N. and {McCullough}, J. and {Muir}, J. and {Pandey}, S. and {Prat}, J. and {Rodriguez-Monroy}, M. and {Ross}, A.~J. and {Rykoff}, E.~S. and {Samuroff}, S. and {Secco}, L.~F. and {Sevilla-Noarbe}, I. and {Sheldon}, E. and {Troxel}, M.~A. and {Tucker}, D.~L. and {Weaverdyck}, N. and {Yanny}, B. and {Yin}, B. and {Zhang}, Y. and {Abbott}, T.~M.~C. and {Aguena}, M. and {Bacon}, D. and {Bertin}, E. and {Bocquet}, S. and {Brooks}, D. and {Burke}, D.~L. and {Carretero}, J. and {Castander}, F.~J. and {Costanzi}, M. and {da Costa}, L.~N. and {Pereira}, M.~E.~S. and {Desai}, S. and {Doel}, P. and {Ferrero}, I. and {Flaugher}, B. and {Friedel}, D. and {Frieman}, J. and {Garc{\'\i}a-Bellido}, J. and {Gerdes}, D.~W. and {Gutierrez}, G. and {Hinton}, S.~R. and {Hollowood}, D.~L. and {Honscheid}, K. and {James}, D.~J. and {Kent}, S. and {Kuehn}, K. and {Lahav}, O. and {Lidman}, C. and {Lima}, M. and {Melchior}, P. and {Mena-Fern{\'a}ndez}, J. and {Menanteau}, F. and {Miquel}, R. and {Ogando}, R.~L.~C. and {Paterno}, M. and {Paz-Chinch{\'o}n}, F. and {Pieres}, A. and {Plazas Malag{\'o}n}, A.~A. and {Roodman}, A. and {Sanchez}, E. and {Scarpine}, V. and {Smith}, M. and {Suchyta}, E. and {Swanson}, M.~E.~C. and {Tarle}, G. and {Thomas}, D. and {To}, C. and {Vincenzi}, M. and {DES Collaboration}},
        title = "{Dark Energy Survey Year 3 results: redshift calibration of the MAGLIM lens sample from the combination of SOMPZ and clustering and its impact on cosmology}",
      journal = {\mnras},
         year = 2024,
        month = jan,
       volume = {527},
       number = {2},
        pages = {2010-2036},
          doi = {10.1093/mnras/stad2945},
archivePrefix = {arXiv},
       eprint = {2209.05853},
 primaryClass = {astro-ph.CO},
       adsurl = {https://ui.adsabs.harvard.edu/abs/2024MNRAS.527.2010G}
}

@ARTICLE{bias_pert_lsst,
       author = {{Nicola}, Andrina and {Hadzhiyska}, Boryana and {Findlay}, Nathan and {Garc{\'\i}a-Garc{\'\i}a}, Carlos and {Alonso}, David and {Slosar}, An{\v{z}}e and {Guo}, Zhiyuan and {Kokron}, Nickolas and {Angulo}, Ra{\'u}l and {Aviles}, Alejandro and {Blazek}, Jonathan and {Dunkley}, Jo and {Jain}, Bhuvnesh and {Pellejero}, Marcos and {Sullivan}, James and {Walter}, Christopher W. and {Zennaro}, Matteo and {LSST Dark Energy Science Collaboration}},
        title = "{Galaxy bias in the era of LSST: perturbative bias expansions}",
      journal = {JCAP},
         year = 2024,
        month = feb,

       number = {2},
          eid = {015},
        pages = {02, 015},
          doi = {10.1088/1475-7516/2024/02/015},
archivePrefix = {arXiv},
       eprint = {2307.03226},
 primaryClass = {astro-ph.CO},
       adsurl = {https://ui.adsabs.harvard.edu/abs/2024JCAP...02..015N}
}

@ARTICLE{lensing_is_low,
       author = {{Chaves-Montero}, Jon{\'a}s and {Angulo}, Raul E. and {Contreras}, Sergio},
        title = "{The galaxy formation origin of the lensing is low problem}",
      journal = {\mnras},
         year = 2023,
        month = may,
       volume = {521},
       number = {1},
        pages = {937-951},
          doi = {10.1093/mnras/stad243},
archivePrefix = {arXiv},
       eprint = {2211.01744},
 primaryClass = {astro-ph.CO},
       adsurl = {https://ui.adsabs.harvard.edu/abs/2023MNRAS.521..937C}
}

@ARTICLE{theWizz,
       author = {{Morrison}, C.~B. and {Hildebrandt}, H. and {Schmidt}, S.~J. and {Baldry}, I.~K. and {Bilicki}, M. and {Choi}, A. and {Erben}, T. and {Schneider}, P.},
        title = "{the-wizz: clustering redshift estimation for everyone}",
      journal = {\mnras},
         year = 2017,
        month = may,
       volume = {467},
       number = {3},
        pages = {3576-3589},
          doi = {10.1093/mnras/stx342},
archivePrefix = {arXiv},
       eprint = {1609.09085},
 primaryClass = {astro-ph.CO},
       adsurl = {https://ui.adsabs.harvard.edu/abs/2017MNRAS.467.3576M}
}

@ARTICLE{KIDS_redshift_dis,
       author = {{Hildebrandt}, H. and {van den Busch}, J.~L. and {Wright}, A.~H. and {Blake}, C. and {Joachimi}, B. and {Kuijken}, K. and {Tr{\"o}ster}, T. and {Asgari}, M. and {Bilicki}, M. and {de Jong}, J.~T.~A. and {Dvornik}, A. and {Erben}, T. and {Getman}, F. and {Giblin}, B. and {Heymans}, C. and {Kannawadi}, A. and {Lin}, C. -A. and {Shan}, H. -Y.},
        title = "{KiDS-1000 catalogue: Redshift distributions and their calibration}",
      journal = {\aap},
         year = 2021,
        month = mar,
       volume = {647},
          eid = {A124},
        pages = {A124},
          doi = {10.1051/0004-6361/202039018},
archivePrefix = {arXiv},
       eprint = {2007.15635},
 primaryClass = {astro-ph.CO},
       adsurl = {https://ui.adsabs.harvard.edu/abs/2021A&A...647A.124H}
}

@ARTICLE{ROCHER_conform,
       author = {{Rocher}, Antoine and {Ruhlmann-Kleider}, Vanina and {Burtin}, Etienne and {Yuan}, Sihan and {de Mattia}, Arnaud and {Ross}, Ashley J. and {Aguilar}, Jessica and {Ahlen}, Steven and {Alam}, Shadab and {Bianchi}, Davide and {Brooks}, David and {Cole}, Shaun and {Dawson}, Kyle and {de la Macorra}, Axel and {Doel}, Peter and {Eisenstein}, Daniel J. and {Fanning}, Kevin and {Forero-Romero}, Jaime E. and {Garrison}, Lehman H. and {Gontcho A Gontcho}, Satya and {Gonzalez-Perez}, Violeta and {Guy}, Julien and {Hadzhiyska}, Boryana and {Hahn}, ChangHoon and {Honscheid}, Klaus and {Kisner}, Theodore and {Landriau}, Martin and {Lasker}, James and {E. Levi}, Michael and {Manera}, Marc and {Meisner}, Aaron and {Miquel}, Ramon and {Moustakas}, John and {Mueller}, Eva-Maria and {Newman}, Jeffrey A. and {Nie}, Jundan and {Percival}, Will J. and {Poppett}, Claire and {Qin}, Fei and {Rossi}, Graziano and {Samushia}, Lado and {Sanchez}, Eusebio and {Schlegel}, David and {Schubnell}, Michael and {Seo}, Hee-Jong and {Tarl{\'e}}, Gregory and {Vargas-Maga{\~n}a}, Mariana and {Weaver}, Benjamin A. and {Yu}, Jiaxi and {Zhang}, Hanyu and {Zheng}, Zheng and {Zhou}, Zhimin and {Zou}, Hu},
        title = "{The DESI One-Percent survey: exploring the Halo Occupation Distribution of Emission Line Galaxies with ABACUSSUMMIT simulations}",
      journal = {JCAP},
         year = 2023,
        month = oct,
       
       number = {10},
          eid = {016},
        pages = {10, 016},
          doi = {10.1088/1475-7516/2023/10/016},
archivePrefix = {arXiv},
       eprint = {2306.06319},
 primaryClass = {astro-ph.CO},
       adsurl = {https://ui.adsabs.harvard.edu/abs/2023JCAP...10..016R}
}

@ARTICLE{conformity_DESI,
       author = {{Gao}, Hongyu and {Jing}, Y.~P. and {Xu}, Kun and {Zhao}, Donghai and {Gui}, Shanquan and {Zheng}, Yun and {Luo}, Xiaolin and {Aguilar}, Jessica Nicole and {Ahlen}, Steven and {Brooks}, David and {Claybaugh}, Todd and {Cole}, Shaun and {de la Macorra}, Axel and {Forero-Romero}, Jaime E. and {Gontcho A Gontcho}, Satya and {Ishak}, Mustapha and {Lambert}, Andrew and {Landriau}, Martin and {Manera}, Marc and {Meisner}, Aaron and {Miquel}, Ramon and {Nie}, Jundan and {Rezaie}, Mehdi and {Rossi}, Graziano and {Sanchez}, Eusebio and {Schubnell}, Michael and {Seo}, Hee-Jong and {Tarl{\'e}}, Gregory and {Weaver}, Benjamin Alan and {Zhou}, Zhimin},
        title = "{The DESI One-Percent Survey: A Concise Model for the Galactic Conformity of Emission-line Galaxies}",
      journal = {\apj},
         year = 2024,
        month = jan,
       volume = {961},
       number = {1},
          eid = {74},
        pages = {74},
          doi = {10.3847/1538-4357/ad09d6},
archivePrefix = {arXiv},
       eprint = {2309.03802},
 primaryClass = {astro-ph.GA},
       adsurl = {https://ui.adsabs.harvard.edu/abs/2024ApJ...961...74G}
}

@ARTICLE{emcee,
       author = {{Foreman-Mackey}, Daniel and {Hogg}, David W. and {Lang}, Dustin and {Goodman}, Jonathan},
        title = "{emcee: The MCMC Hammer}",
      journal = {\pasp},
         year = 2013,
        month = mar,
       volume = {125},
       number = {925},
        pages = {306},
          doi = {10.1086/670067},
archivePrefix = {arXiv},
       eprint = {1202.3665},
 primaryClass = {astro-ph.IM},
       adsurl = {https://ui.adsabs.harvard.edu/abs/2013PASP..125..306F}
}

@ARTICLE{George,
        author = {{Ambikasaran}, Sivaram and {Foreman-Mackey}, Daniel and {Greengard}, Leslie and {Hogg}, David W. and {O'Neil}, Michael},
         title = "{Fast Direct Methods for Gaussian Processes}",
       journal = {IEEE Transactions on Pattern Analysis and Machine Intelligence},
          year = 2015,
         month = jun,
        volume = {38},
         pages = {252},
           doi = {10.1109/TPAMI.2015.2448083},
 archivePrefix = {arXiv},
        eprint = {1403.6015},
  primaryClass = {math.NA},
        adsurl = {https://ui.adsabs.harvard.edu/abs/2015ITPAM..38..252A}
}

@ARTICLE{Hartlap,
       author = {{Hartlap}, J. and {Simon}, P. and {Schneider}, P.},
        title = "{Why your model parameter confidences might be too optimistic. Unbiased estimation of the inverse covariance matrix}",
      journal = {\aap},
         year = 2007,
        month = mar,
       volume = {464},
       number = {1},
        pages = {399-404},
          doi = {10.1051/0004-6361:20066170},
archivePrefix = {arXiv},
       eprint = {astro-ph/0608064},
 primaryClass = {astro-ph},
       adsurl = {https://ui.adsabs.harvard.edu/abs/2007A&A...464..399H}
}

@ARTICLE{photo-z-perf_cosmo,
       author = {{Bordoloi}, R. and {Lilly}, S.~J. and {Amara}, A.},
        title = "{Photo-z performance for precision cosmology}",
      journal = {\mnras},
         year = 2010,
        month = aug,
       volume = {406},
       number = {2},
        pages = {881-895},
          doi = {10.1111/j.1365-2966.2010.16765.x},
archivePrefix = {arXiv},
       eprint = {0910.5735},
 primaryClass = {astro-ph.CO},
       adsurl = {https://ui.adsabs.harvard.edu/abs/2010MNRAS.406..881B}
}

@ARTICLE{Wright_2020,
       author = {{Wright}, Angus H. and {Hildebrandt}, Hendrik and {van den Busch}, Jan Luca and {Heymans}, Catherine},
        title = "{Photometric redshift calibration with self-organising maps}",
      journal = {\aap},
         year = 2020,
        month = may,
       volume = {637},
          eid = {A100},
        pages = {A100},
          doi = {10.1051/0004-6361/201936782},
archivePrefix = {arXiv},
       eprint = {1909.09632},
 primaryClass = {astro-ph.CO},
       adsurl = {https://ui.adsabs.harvard.edu/abs/2020A&A...637A.100W}
}

@ARTICLE{photo-z_HSC,
       author = {{Tanaka}, Masayuki and {Coupon}, Jean and {Hsieh}, Bau-Ching and {Mineo}, Sogo and {Nishizawa}, Atsushi J. and {Speagle}, Joshua and {Furusawa}, Hisanori and {Miyazaki}, Satoshi and {Murayama}, Hitoshi},
        title = "{Photometric redshifts for Hyper Suprime-Cam Subaru Strategic Program Data Release 1}",
      journal = {\pasj},
         year = 2018,
        month = jan,
       volume = {70},
          eid = {S9},
        pages = {S9},
          doi = {10.1093/pasj/psx077},
archivePrefix = {arXiv},
       eprint = {1704.05988},
 primaryClass = {astro-ph.GA},
       adsurl = {https://ui.adsabs.harvard.edu/abs/2018PASJ...70S...9T}
}

@INPROCEEDINGS{Carretero_cosmo_hub,
       author = {{Carretero}, J. and {Tallada}, P. and {Casals}, J. and {Caubet}, M. and {Castander}, F. and {Blot}, L. and {Alarc{\'o}n}, A. and {Serrano}, S. and {Fosalba}, P. and {Acosta-Silva}, C. and {Tonello}, N. and {Torradeflot}, F. n. and {Eriksen}, M. and {Neissner}, C. and {Delfino}, M.},
        title = "{CosmoHub and SciPIC: Massive cosmological data analysis, distribution and generation using a Big Data platform}",
    booktitle = {Proceedings of the European Physical Society Conference on High Energy Physics. 5-12 July},
         year = 2017,
        month = jul,
          eid = {488},
        pages = {488},
          doi = {10.22323/1.314.0488},
       adsurl = {https://ui.adsabs.harvard.edu/abs/2017ehep.confE.488C}
}

@ARTICLE{HSC_clustering_z,
       author = {{Rau}, Markus Michael and {Dalal}, Roohi and {Zhang}, Tianqing and {Li}, Xiangchong and {Nishizawa}, Atsushi J. and {More}, Surhud and {Mandelbaum}, Rachel and {Miyatake}, Hironao and {Strauss}, Michael A. and {Takada}, Masahiro},
        title = "{Weak lensing tomographic redshift distribution inference for the Hyper Suprime-Cam Subaru Strategic Program three-year shape catalogue}",
      journal = {\mnras},
         year = 2023,
        month = oct,
       volume = {524},
       number = {4},
        pages = {5109-5131},
          doi = {10.1093/mnras/stad1962},
archivePrefix = {arXiv},
       eprint = {2211.16516},
 primaryClass = {astro-ph.CO},
       adsurl = {https://ui.adsabs.harvard.edu/abs/2023MNRAS.524.5109R}
}

@ARTICLE{Wright_2020b,
       author = {{Wright}, Angus H. and {Hildebrandt}, Hendrik and {van den Busch}, Jan Luca and {Heymans}, Catherine and {Joachimi}, Benjamin and {Kannawadi}, Arun and {Kuijken}, Konrad},
        title = "{KiDS+VIKING-450: Improved cosmological parameter constraints from redshift calibration with self-organising maps}",
      journal = {\aap},
         year = 2020,
        month = aug,
       volume = {640},
          eid = {L14},
        pages = {L14},
          doi = {10.1051/0004-6361/202038389},
archivePrefix = {arXiv},
       eprint = {2005.04207},
 primaryClass = {astro-ph.CO},
       adsurl = {https://ui.adsabs.harvard.edu/abs/2020A&A...640L..14W}
}

@ARTICLE{Hurterer2006,
       author = {{Huterer}, Dragan and {Takada}, Masahiro and {Bernstein}, Gary and {Jain}, Bhuvnesh},
        title = "{Systematic errors in future weak-lensing surveys: requirements and prospects for self-calibration}",
      journal = {\mnras},
         year = 2006,
        month = feb,
       volume = {366},
       number = {1},
        pages = {101-114},
          doi = {10.1111/j.1365-2966.2005.09782.x},
archivePrefix = {arXiv},
       eprint = {astro-ph/0506030},
 primaryClass = {astro-ph},
       adsurl = {https://ui.adsabs.harvard.edu/abs/2006MNRAS.366..101H}
}

@ARTICLE{Hartley2020,
       author = {{Hartley}, W.~G. and {Chang}, C. and {Samani}, S. and {Carnero Rosell}, A. and {Davis}, T.~M. and {Hoyle}, B. and {Gruen}, D. and {Asorey}, J. and {Gschwend}, J. and {Lidman}, C. and {Kuehn}, K. and {King}, A. and {Rau}, M.~M. and {Wechsler}, R.~H. and {DeRose}, J. and {Hinton}, S.~R. and {Whiteway}, L. and {Abbott}, T.~M.~C. and {Aguena}, M. and {Allam}, S. and {Annis}, J. and {Avila}, S. and {Bernstein}, G.~M. and {Bertin}, E. and {Bridle}, S.~L. and {Brooks}, D. and {Burke}, D.~L. and {Carrasco Kind}, M. and {Carretero}, J. and {Castander}, F.~J. and {Cawthon}, R. and {Costanzi}, M. and {da Costa}, L.~N. and {Desai}, S. and {Diehl}, H.~T. and {Dietrich}, J.~P. and {Flaugher}, B. and {Fosalba}, P. and {Frieman}, J. and {Garc{\'\i}a-Bellido}, J. and {Gaztanaga}, E. and {Gerdes}, D.~W. and {Gruendl}, R.~A. and {Gutierrez}, G. and {Hollowood}, D.~L. and {Honscheid}, K. and {James}, D.~J. and {Kent}, S. and {Krause}, E. and {Kuropatkin}, N. and {Lahav}, O. and {Lima}, M. and {Maia}, M.~A.~G. and {Marshall}, J.~L. and {Melchior}, P. and {Menanteau}, F. and {Miquel}, R. and {Ogando}, R.~L.~C. and {Palmese}, A. and {Paz-Chinch{\'o}n}, F. and {Plazas}, A.~A. and {Roodman}, A. and {Rykoff}, E.~S. and {Sanchez}, E. and {Scarpine}, V. and {Schubnell}, M. and {Serrano}, S. and {Sevilla-Noarbe}, I. and {Smith}, M. and {Soares-Santos}, M. and {Suchyta}, E. and {Tarle}, G. and {Troxel}, M.~A. and {Tucker}, D.~L. and {Varga}, T.~N. and {Weller}, J. and {Wilkinson}, R.~D. and {DES Collaboration}},
        title = "{The impact of spectroscopic incompleteness in direct calibration of redshift distributions for weak lensing surveys}",
      journal = {\mnras},
         year = 2020,
        month = aug,
       volume = {496},
       number = {4},
        pages = {4769-4786},
          doi = {10.1093/mnras/staa1812},
archivePrefix = {arXiv},
       eprint = {2003.10454},
 primaryClass = {astro-ph.GA},
       adsurl = {https://ui.adsabs.harvard.edu/abs/2020MNRAS.496.4769H}
}

@ARTICLE{newman2008,
       author = {{Newman}, Jeffrey A.},
        title = "{Calibrating Redshift Distributions beyond Spectroscopic Limits with Cross-Correlations}",
      journal = {\apj},
         year = 2008,
        month = sep,
       volume = {684},
       number = {1},
        pages = {88-101},
          doi = {10.1086/589982},
archivePrefix = {arXiv},
       eprint = {0805.1409},
 primaryClass = {astro-ph},
       adsurl = {https://ui.adsabs.harvard.edu/abs/2008ApJ...684...88N}
}

@ARTICLE{Campos2024,
       author = {{Campos}, A. and {Yin}, B. and {Dodelson}, S. and {Amon}, A. and {Alarcon}, A. and {S{\'a}nchez}, C. and {Bernstein}, G.~M. and {Giannini}, G. and {Myles}, J. and {Samuroff}, S. and {Alves}, O. and {Andrade-Oliveira}, F. and {Bechtol}, K. and {Becker}, M.~R. and {Blazek}, J. and {Camacho}, H. and {Carnero Rosell}, A. and {Carrasco Kind}, M. and {Cawthon}, R. and {Chang}, C. and {Chen}, R. and {Choi}, A. and {Cordero}, J. and {Davis}, C. and {DeRose}, J. and {Diehl}, H.~T. and {Doux}, C. and {Drlica-Wagner}, A. and {Eckert}, K. and {Eifler}, T.~F. and {Elvin-Poole}, J. and {Everett}, S. and {Fang}, X. and {Fert{\'e}}, A. and {Friedrich}, O. and {Gatti}, M. and {Gruen}, D. and {Gruendl}, R.~A. and {Harrison}, I. and {Hartley}, W.~G. and {Herner}, K. and {Huang}, H. and {Huff}, E.~M. and {Jarvis}, M. and {Krause}, E. and {Kuropatkin}, N. and {Leget}, P. -F. and {MacCrann}, N. and {McCullough}, J. and {Navarro-Alsina}, A. and {Pandey}, S. and {Prat}, J. and {Raveri}, M. and {Rollins}, R.~P. and {Roodman}, A. and {Rosenfeld}, R. and {Ross}, A.~J. and {Rykoff}, E.~S. and {Sanchez}, J. and {Secco}, L.~F. and {Sevilla-Noarbe}, I. and {Sheldon}, E. and {Shin}, T. and {Troxel}, M.~A. and {Tutusaus}, I. and {Varga}, T.~N. and {Wechsler}, R.~H. and {Yanny}, B. and {Zhang}, Y. and {Zuntz}, J. and {Aguena}, M. and {Annis}, J. and {Bacon}, D. and {Bocquet}, S. and {Brooks}, D. and {Burke}, D.~L. and {Carretero}, J. and {Castander}, F.~J. and {Costanzi}, M. and {da Costa}, L.~N. and {De Vicente}, J. and {Doel}, P. and {Ferrero}, I. and {Flaugher}, B. and {Frieman}, J. and {Garc{\'\i}a-Bellido}, J. and {Gaztanaga}, E. and {Gutierrez}, G. and {Hinton}, S.~R. and {Hollowood}, D.~L. and {Honscheid}, K. and {James}, D.~J. and {Kuehn}, K. and {Lima}, M. and {Lin}, H. and {Marshall}, J.~L. and {Mena-Fern{\'a}ndez}, J. and {Menanteau}, F. and {Miquel}, R. and {Ogando}, R.~L.~C. and {Paterno}, M. and {Pereira}, M.~E.~S. and {Pieres}, A. and {Plazas Malag{\'o}n}, A.~A. and {Porredon}, A. and {Sanchez}, E. and {Sanchez Cid}, D. and {Smith}, M. and {Suchyta}, E. and {Swanson}, M.~E.~C. and {Tarle}, G. and {To}, C. and {Vikram}, V. and {Weaverdyck}, N.},
        title = "{Enhancing weak lensing redshift distribution characterization by optimizing the Dark Energy Survey Self-Organizing Map Photo-z method}",
         year = 2024,
        month = aug,
          eid = {arXiv:2408.00922},
        pages = {arXiv:2408.00922},
          doi = {10.48550/arXiv.2408.00922},
archivePrefix = {arXiv},
       eprint = {2408.00922},
 primaryClass = {astro-ph.CO},
       adsurl = {https://ui.adsabs.harvard.edu/abs/2024arXiv240800922C}
}

@ARTICLE{Euclid_overview,
       author = {{Euclid Collaboration: Mellier}, Y. and {Abdurro'uf} and {Acevedo~Barroso}, J.A. and others},
        title = "{Euclid: I. Overview of the Euclid mission}",
      journal = {\aap},
         year = 2025,
        month = may,
       volume = {697},
          eid = {A1},
        pages = {A1},
          doi = {10.1051/0004-6361/202450810},
archivePrefix = {arXiv},
       eprint = {2405.13491},
 primaryClass = {astro-ph.CO},
       adsurl = {https://ui.adsabs.harvard.edu/abs/2025A&A...697A...1E}
}

@ARTICLE{Percival_factor,
       author = {{Percival}, Will J. and {Friedrich}, Oliver and {Sellentin}, Elena and {Heavens}, Alan},
        title = "{Matching Bayesian and frequentist coverage probabilities when using an approximate data covariance matrix}",
      journal = {\mnras},
         year = 2022,
        month = mar,
       volume = {510},
       number = {3},
        pages = {3207-3221},
          doi = {10.1093/mnras/stab3540},
archivePrefix = {arXiv},
       eprint = {2108.10402},
 primaryClass = {astro-ph.IM},
       adsurl = {https://ui.adsabs.harvard.edu/abs/2022MNRAS.510.3207P}
}

@software{Treecorr,
       author = {{Jarvis}, Mike},
        title = "{TreeCorr: Two-point correlation functions}",
 howpublished = {Astrophysics Source Code Library, record ascl:1508.007},
         year = 2015,
        month = aug,
          eid = {ascl:1508.007},
       adsurl = {https://ui.adsabs.harvard.edu/abs/2015ascl.soft08007J}
}

@ARTICLE{Stolzner_GaussianFitting,
       author = {{St{\"o}lzner}, B. and {Joachimi}, B. and {Korn}, A. and {Hildebrandt}, H. and {Wright}, A.~H.},
        title = "{Self-calibration and robust propagation of photometric redshift distribution uncertainties in weak gravitational lensing}",
      journal = {\aap},
         year = 2021,
        month = jun,
       volume = {650},
          eid = {A148},
        pages = {A148},
          doi = {10.1051/0004-6361/202040130},
archivePrefix = {arXiv},
       eprint = {2012.07707},
 primaryClass = {astro-ph.CO},
       adsurl = {https://ui.adsabs.harvard.edu/abs/2021A&A...650A.148S}
}

@ARTICLE{DESI_ELG_LRG,
       author = {{Yuan}, Sihan and {Wechsler}, Risa H. and {Wang}, Yunchong and {de los Reyes}, Mithi A.~C. and {Myles}, Justin and {Rocher}, Antoine and {Hadzhiyska}, Boryana and {Aguilar}, Jessica Nicole and {Ahlen}, Steven and {Brooks}, David and {Claybaugh}, Todd and {Cole}, Shaun and {de la Macorra}, Axel and {Forero-Romero}, Jaime E. and {Gontcho}, Satya Gontcho A. and {Guy}, Julien and {Honscheid}, Klaus and {Kisner}, Theodore and {Levi}, Michael and {Manera}, Marc and {Meisner}, Aaron and {Miquel}, Ramon and {Moustakas}, John and {Nie}, Jundan and {Palanque-Delabrouille}, Nathalie and {Poppett}, Claire and {Rezaie}, Mehdi and {Ross}, Ashley J. and {Rossi}, Graziano and {Sanchez}, Eusebio and {Schubnell}, Michael and {Seo}, Hee-Jong and {Tarl{\'e}}, Gregory and {Weaver}, Benjamin Alan and {Zhou}, Zhimin},
        title = "{Unraveling emission line galaxy conformity at z \raisebox{-0.5ex}\textasciitilde 1 with DESI early data}",
      journal = {\mnras},
         year = 2025,
        month = apr,
       volume = {538},
       number = {2},
        pages = {1216-1240},
          doi = {10.1093/mnras/staf368},
archivePrefix = {arXiv},
       eprint = {2310.09329},
 primaryClass = {astro-ph.CO},
       adsurl = {https://ui.adsabs.harvard.edu/abs/2025MNRAS.538.1216Y}
}

@ARTICLE{clust_z_mcQuinn_white,
       author = {{McQuinn}, Matthew and {White}, Martin},
        title = "{On using angular cross-correlations to determine source redshift distributions}",
      journal = {\mnras},
         year = 2013,
        month = aug,
       volume = {433},
       number = {4},
        pages = {2857-2883},
          doi = {10.1093/mnras/stt914},
archivePrefix = {arXiv},
       eprint = {1302.0857},
 primaryClass = {astro-ph.CO},
       adsurl = {https://ui.adsabs.harvard.edu/abs/2013MNRAS.433.2857M}
}

@ARTICLE{anisotropic_redshift_distribution,
       author = {{Baleato Lizancos}, Ant{\'o}n and {White}, Martin},
        year = 2023,
        title = "{The impact of anisotropic redshift distributions on angular clustering}",
      journal = {JCAP},
        month = jul,
       number = {7},
          eid = {044},
        pages = {07, 044},
          doi = {10.1088/1475-7516/2023/07/044},
archivePrefix = {arXiv},
       eprint = {2305.15406},
 primaryClass = {astro-ph.CO},
       adsurl = {https://ui.adsabs.harvard.edu/abs/2023JCAP...07..044B}
}

@ARTICLE{propagating_photo_z_uncertainty,
       author = {{Reischke}, Robert},
        title = "{Propagating photo-z uncertainties: a functional derivative approach}",
      journal = {\mnras},
         year = 2024,
        month = jun,
       volume = {530},
       number = {4},
        pages = {4412-4421},
          doi = {10.1093/mnras/stad3791},
archivePrefix = {arXiv},
       eprint = {2301.04085},
 primaryClass = {astro-ph.CO},
       adsurl = {https://ui.adsabs.harvard.edu/abs/2024MNRAS.530.4412R}
}

@ARTICLE{lsst_bias_magn_Sanchez,
       author = {{Pandey}, Shivam and {S{\'a}nchez}, Carles and {Jain}, Bhuvnesh and {LSST Dark Energy Science Collaboration}},
        title = "{Cosmology with imaging galaxy surveys: The impact of evolving galaxy bias and magnification}",
      journal = {\prd},
         year = 2025,
        month = jan,
       volume = {111},
       number = {2},
          eid = {023527},
        pages = {023527},
          doi = {10.1103/PhysRevD.111.023527},
archivePrefix = {arXiv},
       eprint = {2310.01315},
 primaryClass = {astro-ph.CO},
       adsurl = {https://ui.adsabs.harvard.edu/abs/2025PhRvD.111b3527P}
}

@ARTICLE{Davis2018,
       author = {{Davis}, C. and {Rozo}, E. and {Roodman}, A. and {Alarcon}, A. and {Cawthon}, R. and {Gatti}, M. and {Lin}, H. and {Miquel}, R. and {Rykoff}, E.~S. and {Troxel}, M.~A. and {Vielzeuf}, P. and {Abbott}, T.~M.~C. and {Abdalla}, F.~B. and {Allam}, S. and {Annis}, J. and {Bechtol}, K. and {Benoit-L{\'e}vy}, A. and {Bertin}, E. and {Brooks}, D. and {Buckley-Geer}, E. and {Burke}, D.~L. and {Carnero Rosell}, A. and {Carrasco Kind}, M. and {Carretero}, J. and {Castander}, F.~J. and {Crocce}, M. and {Cunha}, C.~E. and {D'Andrea}, C.~B. and {da Costa}, L.~N. and {Desai}, S. and {Diehl}, H.~T. and {Doel}, P. and {Drlica-Wagner}, A. and {Fausti Neto}, A. and {Flaugher}, B. and {Fosalba}, P. and {Frieman}, J. and {Garc{\'\i}a-Bellido}, J. and {Gaztanaga}, E. and {Gerdes}, D.~W. and {Giannantonio}, T. and {Gruen}, D. and {Gruendl}, R.~A. and {Gutierrez}, G. and {Honscheid}, K. and {Jain}, B. and {James}, D.~J. and {Jeltema}, T. and {Krause}, E. and {Kuehn}, K. and {Kuhlmann}, S. and {Kuropatkin}, N. and {Lahav}, O. and {Li}, T.~S. and {Lima}, M. and {March}, M. and {Marshall}, J.~L. and {Martini}, P. and {Melchior}, P. and {Ogando}, R.~L.~C. and {Plazas}, A.~A. and {Romer}, A.~K. and {Sanchez}, E. and {Scarpine}, V. and {Schindler}, R. and {Schubnell}, M. and {Sevilla-Noarbe}, I. and {Smith}, M. and {Soares-Santos}, M. and {Sobreira}, F. and {Suchyta}, E. and {Swanson}, M.~E.~C. and {Tarle}, G. and {Thomas}, D. and {Vikram}, V. and {Walker}, A.~R. and {Wechsler}, R.~H. and {DES Collaboration}},
        title = "{Cross-correlation redshift calibration without spectroscopic calibration samples in DES Science Verification Data}",
      journal = {\mnras},
         year = 2018,
        month = jun,
       volume = {477},
       number = {2},
        pages = {2196-2208},
          doi = {10.1093/mnras/sty787},
archivePrefix = {arXiv},
       eprint = {1707.08256},
 primaryClass = {astro-ph.CO},
       adsurl = {https://ui.adsabs.harvard.edu/abs/2018MNRAS.477.2196D}
}

@ARTICLE{Gatti_DESY1,
       author = {{Gatti}, M. and {Vielzeuf}, P. and {Davis}, C. and {Cawthon}, R. and {Rau}, M.~M. and {DeRose}, J. and {De Vicente}, J. and {Alarcon}, A. and {Rozo}, E. and {Gaztanaga}, E. and {Hoyle}, B. and {Miquel}, R. and {Bernstein}, G.~M. and {Bonnett}, C. and {Carnero Rosell}, A. and {Castander}, F.~J. and {Chang}, C. and {da Costa}, L.~N. and {Gruen}, D. and {Gschwend}, J. and {Hartley}, W.~G. and {Lin}, H. and {MacCrann}, N. and {Maia}, M.~A.~G. and {Ogando}, R.~L.~C. and {Roodman}, A. and {Sevilla-Noarbe}, I. and {Troxel}, M.~A. and {Wechsler}, R.~H. and {Asorey}, J. and {Davis}, T.~M. and {Glazebrook}, K. and {Hinton}, S.~R. and {Lewis}, G. and {Lidman}, C. and {Macaulay}, E. and {M{\"o}ller}, A. and {O'Neill}, C.~R. and {Sommer}, N.~E. and {Uddin}, S.~A. and {Yuan}, F. and {Zhang}, B. and {Abbott}, T.~M.~C. and {Allam}, S. and {Annis}, J. and {Bechtol}, K. and {Brooks}, D. and {Burke}, D.~L. and {Carollo}, D. and {Carrasco Kind}, M. and {Carretero}, J. and {Cunha}, C.~E. and {D'Andrea}, C.~B. and {DePoy}, D.~L. and {Desai}, S. and {Eifler}, T.~F. and {Evrard}, A.~E. and {Flaugher}, B. and {Fosalba}, P. and {Frieman}, J. and {Garc{\'\i}a-Bellido}, J. and {Gerdes}, D.~W. and {Goldstein}, D.~A. and {Gruendl}, R.~A. and {Gutierrez}, G. and {Honscheid}, K. and {Hoormann}, J.~K. and {Jain}, B. and {James}, D.~J. and {Jarvis}, M. and {Jeltema}, T. and {Johnson}, M.~W.~G. and {Johnson}, M.~D. and {Krause}, E. and {Kuehn}, K. and {Kuhlmann}, S. and {Kuropatkin}, N. and {Li}, T.~S. and {Lima}, M. and {Marshall}, J.~L. and {Melchior}, P. and {Menanteau}, F. and {Nichol}, R.~C. and {Nord}, B. and {Plazas}, A.~A. and {Reil}, K. and {Rykoff}, E.~S. and {Sako}, M. and {Sanchez}, E. and {Scarpine}, V. and {Schubnell}, M. and {Sheldon}, E. and {Smith}, M. and {Smith}, R.~C. and {Soares-Santos}, M. and {Sobreira}, F. and {Suchyta}, E. and {Swanson}, M.~E.~C. and {Tarle}, G. and {Thomas}, D. and {Tucker}, B.~E. and {Tucker}, D.~L. and {Vikram}, V. and {Walker}, A.~R. and {Weller}, J. and {Wester}, W. and {Wolf}, R.~C.},
        title = "{Dark Energy Survey Year 1 results: cross-correlation redshifts - methods and systematics characterization}",
      journal = {\mnras},
         year = 2018,
        month = jun,
       volume = {477},
       number = {2},
        pages = {1664-1682},
          doi = {10.1093/mnras/sty466},
archivePrefix = {arXiv},
       eprint = {1709.00992},
 primaryClass = {astro-ph.CO},
       adsurl = {https://ui.adsabs.harvard.edu/abs/2018MNRAS.477.1664G}
}

@ARTICLE{Menard2013,
       author = {{M{\'e}nard}, Brice and {Scranton}, Ryan and {Schmidt}, Samuel and {Morrison}, Chris and {Jeong}, Donghui and {Budavari}, Tamas and {Rahman}, Mubdi},
        title = "{Clustering-based redshift estimation: method and application to data}",
         year = 2013,
        month = mar,
          eid = {arXiv:1303.4722},
        pages = {arXiv:1303.4722},
          doi = {10.48550/arXiv.1303.4722},
archivePrefix = {arXiv},
       eprint = {1303.4722},
 primaryClass = {astro-ph.CO},
       adsurl = {https://ui.adsabs.harvard.edu/abs/2013arXiv1303.4722M}
}

@ARTICLE{Gatti_Giulia_DESY3,
       author = {{Gatti}, M. and {Giannini}, G. and {Bernstein}, G.~M. and {Alarcon}, A. and {Myles}, J. and {Amon}, A. and {Cawthon}, R. and {Troxel}, M. and {DeRose}, J. and {Everett}, S. and {Ross}, A.~J. and {Rykoff}, E.~S. and {Elvin-Poole}, J. and {Cordero}, J. and {Harrison}, I. and {Sanchez}, C. and {Prat}, J. and {Gruen}, D. and {Lin}, H. and {Crocce}, M. and {Rozo}, E. and {Abbott}, T.~M.~C. and {Aguena}, M. and {Allam}, S. and {Annis}, J. and {Avila}, S. and {Bacon}, D. and {Bertin}, E. and {Brooks}, D. and {Burke}, D.~L. and {Rosell}, A. Carnero and {Kind}, M. Carrasco and {Carretero}, J. and {Castander}, F.~J. and {Choi}, A. and {Conselice}, C. and {Costanzi}, M. and {Crocce}, M. and {da Costa}, L.~N. and {Pereira}, M.~E.~S. and {Dawson}, K. and {Desai}, S. and {Diehl}, H.~T. and {Eckert}, K. and {Eifler}, T.~F. and {Evrard}, A.~E. and {Ferrero}, I. and {Flaugher}, B. and {Fosalba}, P. and {Frieman}, J. and {Garc{\'\i}a-Bellido}, J. and {Gaztanaga}, E. and {Giannantonio}, T. and {Gruendl}, R.~A. and {Gschwend}, J. and {Hinton}, S.~R. and {Hollowood}, D.~L. and {Honscheid}, K. and {Hoyle}, B. and {Huterer}, D. and {James}, D.~J. and {Kuehn}, K. and {Kuropatkin}, N. and {Lahav}, O. and {Lima}, M. and {MacCrann}, N. and {Maia}, M.~A.~G. and {March}, M. and {Marshall}, J.~L. and {Melchior}, P. and {Menanteau}, F. and {Miquel}, R. and {Mohr}, J.~J. and {Morgan}, R. and {Ogando}, R.~L.~C. and {Palmese}, A. and {Paz-Chinch{\'o}n}, F. and {Percival}, W.~J. and {Plazas}, A.~A. and {Rodriguez-Monroy}, M. and {Roodman}, A. and {Rossi}, G. and {Samuroff}, S. and {Sanchez}, E. and {Scarpine}, V. and {Secco}, L.~F. and {Serrano}, S. and {Sevilla-Noarbe}, I. and {Smith}, M. and {Soares-Santos}, M. and {Suchyta}, E. and {Swanson}, M.~E.~C. and {Tarle}, G. and {Thomas}, D. and {To}, C. and {Varga}, T.~N. and {Weller}, J. and {Wilkinson}, R.~D. and {Wilkinson}, R.~D. and {DES Collaboration}},
        title = "{Dark Energy Survey Year 3 Results: clustering redshifts - calibration of the weak lensing source redshift distributions with redMaGiC and BOSS/eBOSS}",
      journal = {\mnras},
         year = 2022,
        month = feb,
       volume = {510},
       number = {1},
        pages = {1223-1247},
          doi = {10.1093/mnras/stab3311},
archivePrefix = {arXiv},
       eprint = {2012.08569},
 primaryClass = {astro-ph.CO},
       adsurl = {https://ui.adsabs.harvard.edu/abs/2022MNRAS.510.1223G}
}

@ARTICLE{Cawton2022,
       author = {{Cawthon}, R. and {Elvin-Poole}, J. and {Porredon}, A. and {Crocce}, M. and {Giannini}, G. and {Gatti}, M. and {Ross}, A.~J. and {Rykoff}, E.~S. and {Carnero Rosell}, A. and {DeRose}, J. and {Lee}, S. and {Rodriguez-Monroy}, M. and {Amon}, A. and {Bechtol}, K. and {De Vicente}, J. and {Gruen}, D. and {Morgan}, R. and {Sanchez}, E. and {Sanchez}, J. and {Sevilla-Noarbe}, I. and {Abbott}, T.~M.~C. and {Aguena}, M. and {Allam}, S. and {Annis}, J. and {Avila}, S. and {Bacon}, D. and {Bertin}, E. and {Brooks}, D. and {Burke}, D.~L. and {Carrasco Kind}, M. and {Carretero}, J. and {Castander}, F.~J. and {Choi}, A. and {Costanzi}, M. and {da Costa}, L.~N. and {Pereira}, M.~E.~S. and {Dawson}, K. and {Desai}, S. and {Diehl}, H.~T. and {Eckert}, K. and {Everett}, S. and {Ferrero}, I. and {Fosalba}, P. and {Frieman}, J. and {Garc{\'\i}a-Bellido}, J. and {Gaztanaga}, E. and {Gruendl}, R.~A. and {Gschwend}, J. and {Gutierrez}, G. and {Hinton}, S.~R. and {Hollowood}, D.~L. and {Honscheid}, K. and {Huterer}, D. and {James}, D.~J. and {Kim}, A.~G. and {Kneib}, J. -P. and {Kuehn}, K. and {Kuropatkin}, N. and {Lahav}, O. and {Lima}, M. and {Lin}, H. and {Maia}, M.~A.~G. and {Melchior}, P. and {Menanteau}, F. and {Miquel}, R. and {Mohr}, J.~J. and {Muir}, J. and {Myles}, J. and {Palmese}, A. and {Pandey}, S. and {Paz-Chinch{\'o}n}, F. and {Percival}, W.~J. and {Plazas}, A.~A. and {Roodman}, A. and {Rossi}, G. and {Scarpine}, V. and {Serrano}, S. and {Smith}, M. and {Soares-Santos}, M. and {Suchyta}, E. and {Swanson}, M.~E.~C. and {Tarle}, G. and {To}, C. and {Troxel}, M.~A. and {Wilkinson}, R.~D. and {DES Collaboration}},
        title = "{Dark Energy Survey Year 3 results: calibration of lens sample redshift distributions using clustering redshifts with BOSS/eBOSS}",
      journal = {\mnras},
         year = 2022,
        month = jul,
       volume = {513},
       number = {4},
        pages = {5517-5539},
          doi = {10.1093/mnras/stac1160},
archivePrefix = {arXiv},
       eprint = {2012.12826},
 primaryClass = {astro-ph.CO},
       adsurl = {https://ui.adsabs.harvard.edu/abs/2022MNRAS.513.5517C}
}

@ARTICLE{Gwyn_UNIONS,
       author = {{Gwyn}, Stephen and {McConnachie}, Alan W. and {Cuillandre}, Jean-Charles and {Chambers}, Ken C. and {Magnier}, Eugene A. and {Hudson}, Michael J. and {Oguri}, Masamune and {Furusawa}, Hisanori and {Hildebrandt}, Hendrik and {Carlberg}, Raymond and {Ellison}, Sara L. and {Furusawa}, Junko and {Gavazzi}, Rapha{\"e}l and {Ibata}, Rodrigo and {Mellier}, Yannick and {Osato}, Ken and {Aussel}, H. and {Baumont}, Lucie and {Bayer}, Manuel and {Boulade}, Olivier and {C{\^o}t{\'e}}, Patrick and {Chemaly}, David and {Daley}, Cail and {Duc}, Pierre-Alain and {Ellien}, A. and {Fabbro}, S{\'e}bastien and {Ferreira}, Leonardo and {Fitriana}, Itsna K. and {Le Floc'h}, Emeric and {Hammer} and {Francois} and {Fudamoto}, Yoshinobu and {Gao}, Hua and {Goh}, L.~W.~K. and {Goto}, Tomotsugu and {Guerrini}, Sacha and {Guinot}, Axel and {H{\'e}nault-Brunet}, Vincent and {Harikane}, Yuichi and {Hayashi}, Kohei and {Heesters}, Nick and {Ichikawa}, Kohei and {Kilbinger}, Martin and {Kuzma}, P.~B. and {Li}, Qinxun and {Liaudat}, Tob{\'\i}as I. and {Lin}, Chien-Cheng and {M{\"u}ller}, Oliver and {Martin}, Nicolas F. and {Matsuoka}, Yoshiki and {Medina}, Gustavo E. and {Miyatake}, Hironao and {Miyazaki}, Satoshi and {Mpetha}, Charlie T. and {Nagao}, Tohru and {Navarro}, Julio F. and {Niwano}, Masafumi and {Ogami}, Itsuki and {Okabe}, Nobuhiro and {Onoue}, Masafusa and {Paek}, Gregory S.~H. and {Parker}, Laura C. and {Patton}, David R. and {Hervas Peters}, Fabian and {Prunet}, Simon and {S{\'a}nchez-Janssen}, Rub{\'e}n and {Schultheis}, M. and {Sestito}, Federico and {Smith}, Simon E.~T. and {Starck}, J. -L. and {Starkenburg}, Else and {Stone}, Connor and {Storfer}, Christopher and {Suzuki}, Yoshihisa and {Erben} and {T.} and {Taibi}, Salvatore and {Thomas}, G.~F. and {TianFang}, Zhang and {Toba}, Yoshiki and {Uchiyama}, Hisakazu and {Valls-Gabaud}, David and {Venn}, Kim A. and {Van Waerbeke}, Ludovic and {Wainscoat}, Richard J. and {Wilkinson}, Scott and {Wittje}, Anna and {Yoshida}, Taketo and {Zhong}, Yuxing},
        title = "{UNIONS: The Ultraviolet Near-Infrared Optical Northern Survey}",
      journal = {arXiv e-prints},
         year = 2025,
        month = mar,
          eid = {arXiv:2503.13783},
        pages = {arXiv:2503.13783},
          doi = {10.48550/arXiv.2503.13783},
archivePrefix = {arXiv},
       eprint = {2503.13783},
 primaryClass = {astro-ph.GA},
       adsurl = {https://ui.adsabs.harvard.edu/abs/2025arXiv250313783G}
}

@ARTICLE{Schmidt2013,
       author = {{Schmidt}, Samuel J. and {M{\'e}nard}, Brice and {Scranton}, Ryan and {Morrison}, Christopher and {McBride}, Cameron K.},
        title = "{Recovering redshift distributions with cross-correlations: pushing the boundaries}",
      journal = {\mnras},
         year = 2013,
        month = jun,
       volume = {431},
       number = {4},
        pages = {3307-3318},
          doi = {10.1093/mnras/stt410},
archivePrefix = {arXiv},
       eprint = {1303.0292},
 primaryClass = {astro-ph.CO},
       adsurl = {https://ui.adsabs.harvard.edu/abs/2013MNRAS.431.3307S}
}

@ARTICLE{van_den_Busch_2020,
       author = {{van den Busch}, J.~L. and {Hildebrandt}, H. and {Wright}, A.~H. and {Morrison}, C.~B. and {Blake}, C. and {Joachimi}, B. and {Erben}, T. and {Heymans}, C. and {Kuijken}, K. and {Taylor}, E.~N.},
        title = "{Testing KiDS cross-correlation redshifts with simulations}",
      journal = {\aap},
         year = 2020,
        month = oct,
       volume = {642},
          eid = {A200},
        pages = {A200},
          doi = {10.1051/0004-6361/202038835},
archivePrefix = {arXiv},
       eprint = {2007.01846},
 primaryClass = {astro-ph.CO},
       adsurl = {https://ui.adsabs.harvard.edu/abs/2020A&A...642A.200V}
}

@ARTICLE{Landy_Szalay,
       author = {{Landy}, Stephen D. and {Szalay}, Alexander S.},
        title = "{Bias and Variance of Angular Correlation Functions}",
      journal = {\apj},
         year = 1993,
        month = jul,
       volume = {412},
        pages = {64},
          doi = {10.1086/172900},
       adsurl = {https://ui.adsabs.harvard.edu/abs/1993ApJ...412...64L}
}

@ARTICLE{magnification_DESY3,
       author = {{Elvin-Poole}, J. and {MacCrann}, N. and {Everett}, S. and {Prat}, J. and {Rykoff}, E.~S. and {De Vicente}, J. and {Yanny}, B. and {Herner}, K. and {Fert{\'e}}, A. and {Di Valentino}, E. and {Choi}, A. and {Burke}, D.~L. and {Sevilla-Noarbe}, I. and {Alarcon}, A. and {Alves}, O. and {Amon}, A. and {Andrade-Oliveira}, F. and {Baxter}, E. and {Bechtol}, K. and {Becker}, M.~R. and {Bernstein}, G.~M. and {Blazek}, J. and {Camacho}, H. and {Campos}, A. and {Carnero Rosell}, A. and {Carrasco Kind}, M. and {Cawthon}, R. and {Chang}, C. and {Chen}, R. and {Cordero}, J. and {Crocce}, M. and {Davis}, C. and {DeRose}, J. and {Diehl}, H.~T. and {Dodelson}, S. and {Doux}, C. and {Drlica-Wagner}, A. and {Eckert}, K. and {Eifler}, T.~F. and {Elsner}, F. and {Fang}, X. and {Fosalba}, P. and {Friedrich}, O. and {Gatti}, M. and {Giannini}, G. and {Gruen}, D. and {Gruendl}, R.~A. and {Harrison}, I. and {Hartley}, W.~G. and {Huang}, H. and {Huff}, E.~M. and {Huterer}, D. and {Krause}, E. and {Kuropatkin}, N. and {Leget}, P. -F. and {Lemos}, P. and {Liddle}, A.~R. and {McCullough}, J. and {Muir}, J. and {Myles}, J. and {Navarro-Alsina}, A. and {Pandey}, S. and {Park}, Y. and {Porredon}, A. and {Raveri}, M. and {Rodriguez-Monroy}, M. and {Rollins}, R.~P. and {Roodman}, A. and {Rosenfeld}, R. and {Ross}, A.~J. and {S{\'a}nchez}, C. and {Sanchez}, J. and {Secco}, L.~F. and {Sheldon}, E. and {Shin}, T. and {Troxel}, M.~A. and {Tutusaus}, I. and {Varga}, T.~N. and {Weaverdyck}, N. and {Wechsler}, R.~H. and {Yin}, B. and {Zhang}, Y. and {Zuntz}, J. and {Aguena}, M. and {Avila}, S. and {Bacon}, D. and {Bertin}, E. and {Bocquet}, S. and {Brooks}, D. and {Garc{\'\i}a-Bellido}, J. and {Honscheid}, K. and {Jarvis}, M. and {Li}, T.~S. and {Mena-Fern{\'a}ndez}, J. and {To}, C. and {Wilkinson}, R.~D. and {DES Collaboration}},
        title = "{Dark Energy Survey Year 3 results: magnification modelling and impact on cosmological constraints from galaxy clustering and galaxy-galaxy lensing}",
      journal = {\mnras},
         year = 2023,
        month = aug,
       volume = {523},
       number = {3},
        pages = {3649-3670},
          doi = {10.1093/mnras/stad1594},
archivePrefix = {arXiv},
       eprint = {2209.09782},
 primaryClass = {astro-ph.CO},
       adsurl = {https://ui.adsabs.harvard.edu/abs/2023MNRAS.523.3649E}
}

@ARTICLE{Y3_DES_Krause,
       author = {{Krause}, E. and {Fang}, X. and {Pandey}, S. and {Secco}, L.~F. and {Alves}, O. and {Huang}, H. and {Blazek}, J. and {Prat}, J. and {Zuntz}, J. and {Eifler}, T.~F. and {MacCrann}, N. and {DeRose}, J. and {Crocce}, M. and {Porredon}, A. and {Jain}, B. and {Troxel}, M.~A. and {Dodelson}, S. and {Huterer}, D. and {Liddle}, A.~R. and {Leonard}, C.~D. and {Amon}, A. and {Chen}, A. and {Elvin-Poole}, J. and {Fert{\'e}}, A. and {Muir}, J. and {Park}, Y. and {Samuroff}, S. and {Brandao-Souza}, A. and {Weaverdyck}, N. and {Zacharegkas}, G. and {Rosenfeld}, R. and {Campos}, A. and {Chintalapati}, P. and {Choi}, A. and {Di Valentino}, E. and {Doux}, C. and {Herner}, K. and {Lemos}, P. and {Mena-Fern{\'a}ndez}, J. and {Omori}, Y. and {Paterno}, M. and {Rodriguez-Monroy}, M. and {Rogozenski}, P. and {Rollins}, R.~P. and {Troja}, A. and {Tutusaus}, I. and {Wechsler}, R.~H. and {Abbott}, T.~M.~C. and {Aguena}, M. and {Allam}, S. and {Andrade-Oliveira}, F. and {Annis}, J. and {Bacon}, D. and {Baxter}, E. and {Bechtol}, K. and {Bernstein}, G.~M. and {Brooks}, D. and {Buckley-Geer}, E. and {Burke}, D.~L. and {Carnero Rosell}, A. and {Carrasco Kind}, M. and {Carretero}, J. and {Castander}, F.~J. and {Cawthon}, R. and {Chang}, C. and {Costanzi}, M. and {da Costa}, L.~N. and {Pereira}, M.~E.~S. and {De Vicente}, J. and {Desai}, S. and {Diehl}, H.~T. and {Doel}, P. and {Everett}, S. and {Evrard}, A.~E. and {Ferrero}, I. and {Flaugher}, B. and {Fosalba}, P. and {Frieman}, J. and {Garc{\'\i}a-Bellido}, J. and {Gaztanaga}, E. and {Gerdes}, D.~W. and {Giannantonio}, T. and {Gruen}, D. and {Gruendl}, R.~A. and {Gschwend}, J. and {Gutierrez}, G. and {Hartley}, W.~G. and {Hinton}, S.~R. and {Hollowood}, D.~L. and {Honscheid}, K. and {Hoyle}, B. and {Huff}, E.~M. and {James}, D.~J. and {Kuehn}, K. and {Kuropatkin}, N. and {Lahav}, O. and {Lima}, M. and {Maia}, M.~A.~G. and {Marshall}, J.~L. and {Martini}, P. and {Melchior}, P. and {Menanteau}, F. and {Miquel}, R. and {Mohr}, J.~J. and {Morgan}, R. and {Myles}, J. and {Palmese}, A. and {Paz-Chinch{\'o}n}, F. and {Petravick}, D. and {Pieres}, A. and {Plazas Malag{\'o}n}, A.~A. and {Sanchez}, E. and {Scarpine}, V. and {Schubnell}, M. and {Serrano}, S. and {Sevilla-Noarbe}, I. and {Smith}, M. and {Soares-Santos}, M. and {Suchyta}, E. and {Tarle}, G. and {Thomas}, D. and {To}, C. and {Varga}, T.~N. and {Weller}, J.},
        title = "{Dark Energy Survey Year 3 Results: Multi-Probe Modeling Strategy and Validation}",
         year = 2021,
        month = may,
          eid = {arXiv:2105.13548},
        pages = {arXiv:2105.13548},
          doi = {10.48550/arXiv.2105.13548},
archivePrefix = {arXiv},
       eprint = {2105.13548},
 primaryClass = {astro-ph.CO},
       adsurl = {https://ui.adsabs.harvard.edu/abs/2021arXiv210513548K}
}

@ARTICLE{Q1_inter,
       author = {{Euclid Collaboration: Le Brun}, V. and {Bethermin}, M. and {Moresco}, M. and others},
       title = "{Euclid Quick Data Release (Q1) - Characteristics and limitations of the spectroscopic measurements}",
      journal = {A\&A, submitted},
         year = 2025,
        month = mar,
          eid = {arXiv:2503.15308},
        pages = {arXiv:2503.15308},
archivePrefix = {arXiv},
       eprint = {2503.15308},
 primaryClass = {astro-ph.CO},
       adsurl = {https://ui.adsabs.harvard.edu/abs/2025arXiv250315308E}
}

@ARTICLE{EuclidSkyVIS,
author = {{Euclid Collaboration: Cropper}, M. and {Al-Bahlawan}, A. and {Amiaux}, J. and others},
	title = {Euclid - II. The VIS instrument},
	DOI= "10.1051/0004-6361/202450996",
	url= "https://doi.org/10.1051/0004-6361/202450996",
	journal = {A\&A},
	year = 2025,
	volume = 697,
	pages = "A2",
}

@ARTICLE{EuclidSkyNISP,
author = {{Euclid Collaboration: Jahnke}, K. and {Gillard}, W. and {Schirmer}, M. and others},
	title = {Euclid - III. The NISP Instrument},
	DOI= "10.1051/0004-6361/202450786",
	url= "https://doi.org/10.1051/0004-6361/202450786",
	journal = {A\&A},
	year = 2025,
	volume = 697,
	pages = "A3",
}

@ARTICLE{Pocino-EP12,
       author = {{Euclid Collaboration: Pocino}, A. and {Tutusaus}, I. and {Castander}, F.~J. and others},
        title = "{Euclid preparation. XII. Optimizing the photometric sample of the Euclid survey for galaxy clustering and galaxy-galaxy lensing analyses}",
      journal = {\aap},
         year = 2021,
        month = nov,
       volume = {655},
          eid = {A44},
        pages = {A44},
          doi = {10.1051/0004-6361/202141061},
archivePrefix = {arXiv},
       eprint = {2104.05698},
 primaryClass = {astro-ph.CO},
       adsurl = {https://ui.adsabs.harvard.edu/abs/2021A&A...655A..44E}
}

@ARTICLE{Ilbert-EP11,
       author = {{Euclid Collaboration: Ilbert}, O. and {de la Torre}, S. and {Martinet}, N. and others},
        title = "{Euclid preparation. XI. Mean redshift determination from galaxy redshift probabilities for cosmic shear tomography}",
      journal = {\aap},
         year = 2021,
        month = mar,
       volume = {647},
          eid = {A117},
        pages = {A117},
          doi = {10.1051/0004-6361/202040237},
archivePrefix = {arXiv},
       eprint = {2101.02228},
 primaryClass = {astro-ph.CO},
       adsurl = {https://ui.adsabs.harvard.edu/abs/2021A&A...647A.117E}
}

@ARTICLE{Desprez-EP10,
       author = {{Euclid Collaboration: Desprez}, G. and {Paltani}, S. and {Coupon}, J. and others},
        title = "{Euclid preparation. X. The Euclid photometric-redshift challenge}",
      journal = {\aap},
         year = 2020,
        month = dec,
       volume = {644},
          eid = {A31},
        pages = {A31},
          doi = {10.1051/0004-6361/202039403},
archivePrefix = {arXiv},
       eprint = {2009.12112},
 primaryClass = {astro-ph.GA},
       adsurl = {https://ui.adsabs.harvard.edu/abs/2020A&A...644A..31E}
}

@ARTICLE{Blanchard-EP7,
       author = {{Euclid Collaboration: Blanchard}, A. and {Camera}, S. and {Carbone}, C. and others},
        title = "{Euclid preparation. VII. Forecast validation for Euclid cosmological probes}",
      journal = {\aap},
         year = 2020,
        month = oct,
       volume = {642},
          eid = {A191},
        pages = {A191},
          doi = {10.1051/0004-6361/202038071},
archivePrefix = {arXiv},
       eprint = {1910.09273},
 primaryClass = {astro-ph.CO},
       adsurl = {https://ui.adsabs.harvard.edu/abs/2020A&A...642A.191E}
}

@ARTICLE{Lepori24,
       author = {{Lepori}, F. and {Schulz}, S. and {Tutusaus}, I. and others},
        title = "{Euclid: Relativistic effects in the dipole of the two-point correlation function}",
      journal = {\aap},
         year = 2025,
        month = feb,
       volume = {694},
          eid = {A321},
        pages = {A321},
          doi = {10.1051/0004-6361/202452531},
archivePrefix = {arXiv},
       eprint = {2410.06268},
 primaryClass = {astro-ph.CO},
       adsurl = {https://ui.adsabs.harvard.edu/abs/2025A&A...694A.321L}
}

@ARTICLE{Naidoo23,
       author = {{Naidoo}, K. and {Johnston}, H. and {Joachimi}, B. and others},
        title = "{Euclid: Calibrating photometric redshifts with spectroscopic cross-correlations}",
      journal = {\aap},
         year = 2023,
        month = feb,
       volume = {670},
          eid = {A149},
        pages = {A149},
          doi = {10.1051/0004-6361/202244795},
archivePrefix = {arXiv},
       eprint = {2208.10503},
 primaryClass = {astro-ph.CO},
       adsurl = {https://ui.adsabs.harvard.edu/abs/2023A&A...670A.149N}
}

@ARTICLE{Contarini22,
       author = {{Contarini}, S. and {Verza}, G. and {Pisani}, A. and others},
        title = "{Euclid: Cosmological forecasts from the void size function}",
      journal = {\aap},
         year = 2022,
        month = nov,
       volume = {667},
          eid = {A162},
        pages = {A162},
          doi = {10.1051/0004-6361/202244095},
archivePrefix = {arXiv},
       eprint = {2205.11525},
 primaryClass = {astro-ph.CO},
       adsurl = {https://ui.adsabs.harvard.edu/abs/2022A&A...667A.162C}
}

@ARTICLE{Laureijs11,
       author = {{Laureijs}, R. and {Amiaux}, J. and {Arduini}, S. and {Augu{\`e}res}, J. -L. and {Brinchmann}, J. and {Cole}, R. and {Cropper}, M. and {Dabin}, C. and {Duvet}, L. and {Ealet}, A. and {Garilli}, B. and {Gondoin}, P. and {Guzzo}, L. and {Hoar}, J. and {Hoekstra}, H. and {Holmes}, R. and {Kitching}, T. and {Maciaszek}, T. and {Mellier}, Y. and {Pasian}, F. and {Percival}, W. and {Rhodes}, J. and {Saavedra Criado}, G. and {Sauvage}, M. and {Scaramella}, R. and {Valenziano}, L. and {Warren}, S. and {Bender}, R. and {Castander}, F. and {Cimatti}, A. and {Le F{\`e}vre}, O. and {Kurki-Suonio}, H. and {Levi}, M. and {Lilje}, P. and {Meylan}, G. and {Nichol}, R. and {Pedersen}, K. and {Popa}, V. and {Rebolo Lopez}, R. and {Rix}, H. -W. and {Rottgering}, H. and {Zeilinger}, W. and {Grupp}, F. and {Hudelot}, P. and {Massey}, R. and {Meneghetti}, M. and {Miller}, L. and {Paltani}, S. and {Paulin-Henriksson}, S. and {Pires}, S. and {Saxton}, C. and {Schrabback}, T. and {Seidel}, G. and {Walsh}, J. and {Aghanim}, N. and {Amendola}, L. and {Bartlett}, J. and {Baccigalupi}, C. and {Beaulieu}, J. -P. and {Benabed}, K. and {Cuby}, J. -G. and {Elbaz}, D. and {Fosalba}, P. and {Gavazzi}, G. and {Helmi}, A. and {Hook}, I. and {Irwin}, M. and {Kneib}, J. -P. and {Kunz}, M. and {Mannucci}, F. and {Moscardini}, L. and {Tao}, C. and {Teyssier}, R. and {Weller}, J. and {Zamorani}, G. and {Zapatero Osorio}, M.~R. and {Boulade}, O. and {Foumond}, J.~J. and {Di Giorgio}, A. and {Guttridge}, P. and {James}, A. and {Kemp}, M. and {Martignac}, J. and {Spencer}, A. and {Walton}, D. and {Bl{\"u}mchen}, T. and {Bonoli}, C. and {Bortoletto}, F. and {Cerna}, C. and {Corcione}, L. and {Fabron}, C. and {Jahnke}, K. and {Ligori}, S. and {Madrid}, F. and {Martin}, L. and {Morgante}, G. and {Pamplona}, T. and {Prieto}, E. and {Riva}, M. and {Toledo}, R. and {Trifoglio}, M. and {Zerbi}, F. and {Abdalla}, F. and {Douspis}, M. and {Grenet}, C. and {Borgani}, S. and {Bouwens}, R. and {Courbin}, F. and {Delouis}, J. -M. and {Dubath}, P. and {Fontana}, A. and {Frailis}, M. and {Grazian}, A. and {Koppenh{\"o}fer}, J. and {Mansutti}, O. and {Melchior}, M. and {Mignoli}, M. and {Mohr}, J. and {Neissner}, C. and {Noddle}, K. and {Poncet}, M. and {Scodeggio}, M. and {Serrano}, S. and {Shane}, N. and {Starck}, J. -L. and {Surace}, C. and {Taylor}, A. and {Verdoes-Kleijn}, G. and {Vuerli}, C. and {Williams}, O.~R. and {Zacchei}, A. and {Altieri}, B. and {Escudero Sanz}, I. and {Kohley}, R. and {Oosterbroek}, T. and {Astier}, P. and {Bacon}, D. and {Bardelli}, S. and {Baugh}, C. and {Bellagamba}, F. and {Benoist}, C. and {Bianchi}, D. and {Biviano}, A. and {Branchini}, E. and {Carbone}, C. and {Cardone}, V. and {Clements}, D. and {Colombi}, S. and {Conselice}, C. and {Cresci}, G. and {Deacon}, N. and {Dunlop}, J. and {Fedeli}, C. and {Fontanot}, F. and {Franzetti}, P. and {Giocoli}, C. and {Garcia-Bellido}, J. and {Gow}, J. and {Heavens}, A. and {Hewett}, P. and {Heymans}, C. and {Holland}, A. and {Huang}, Z. and {Ilbert}, O. and {Joachimi}, B. and {Jennins}, E. and {Kerins}, E. and {Kiessling}, A. and {Kirk}, D. and {Kotak}, R. and {Krause}, O. and {Lahav}, O. and {van Leeuwen}, F. and {Lesgourgues}, J. and {Lombardi}, M. and {Magliocchetti}, M. and {Maguire}, K. and {Majerotto}, E. and {Maoli}, R. and {Marulli}, F. and {Maurogordato}, S. and {McCracken}, H. and {McLure}, R. and {Melchiorri}, A. and {Merson}, A. and {Moresco}, M. and {Nonino}, M. and {Norberg}, P. and {Peacock}, J. and {Pello}, R. and {Penny}, M. and {Pettorino}, V. and {Di Porto}, C. and {Pozzetti}, L. and {Quercellini}, C. and {Radovich}, M. and {Rassat}, A. and {Roche}, N. and {Ronayette}, S. and {Rossetti}, E. and {Sartoris}, B. and {Schneider}, P. and {Semboloni}, E. and {Serjeant}, S. and {Simpson}, F. and {Skordis}, C. and {Smadja}, G. and {Smartt}, S. and {Spano}, P. and {Spiro}, S. and {Sullivan}, M. and {Tilquin}, A. and {Trotta}, R. and {Verde}, L. and {Wang}, Y. and {Williger}, G. and {Zhao}, G. and {Zoubian}, J. and {Zucca}, E.},
        title = "{Euclid Definition Study Report}",
      journal = {ESA/SRE(2011)12},
         year = 2011,
        month = oct,
          eid = {arXiv:1110.3193},
        pages = {arXiv:1110.3193},
          doi = {10.48550/arXiv.1110.3193},
archivePrefix = {arXiv},
       eprint = {1110.3193},
 primaryClass = {astro-ph.CO},
       adsurl = {https://ui.adsabs.harvard.edu/abs/2011arXiv1110.3193L}
}

%
%

\begin{appendix}
  \onecolumn 
\section{Higher-order development of the galaxy bias}

We introduced a first-order bias expansion in Sect. \ref{sec:galaxy_bias}
\begin{equation}
    \delta_x=b_{x}^{(1)}\,\delta_{\rm m} +{\smallO{\delta_{\rm m}}}\,.
\end{equation}
The two-point correlation between two samples $a$ and $b$ localised at the same redshift is
\begin{equation}
    \xi_{ab}(\theta):=\l \delta_a\,\delta_b\r_{\theta}= b_a\,b_b\,\l \delta_{\rm m}\delta_{\rm m}\r_{\theta}+\smallO{\l \delta_{\rm m}\delta_{\rm m}\r}\,.
\end{equation}
We are using the ratio of correlations to infer $n(z)$, as it is equal to unity at main order (cf. Sect. \ref{sc:method_clust_z}), 
\begin{equation}
    \frac{\xi_{ab}(\theta)}{\sqrt{\xi_{aa}(\theta)\,\xi_{bb}(\theta)}}=1+\smallO{1}\,.\label{eq:order_0}
\end{equation}
In this appendix we investigate the impact of second-order terms in the bias expansion on the ratio Eq. \eqref{eq:order_0}.
We can express  the galaxy overdensity $\delta_{\rm g}$ perturbatively up to second-order with 
\begin{align}
    \delta_{\rm g}&=b_{\rm g}^{(1)}\,\delta_{\rm m} +\sum_{j} b_{j,\,{\rm g}}^{(2)}\,\mathcal{O}_j + \smallO{\textstyle\sum_j \mathcal{O}_j}\,,
\end{align}
where $\mathcal{O}_j$ are a set of second-order fields, with $\l \mathcal{O}_j \r =0$. We note that we do not consider the stochastic contribution, as we are working with real-space correlations. For example, in \cite{scales_bias} and \cite{ bias_pert_lsst} they used
\begin{equation}
    \sum_{j} b_{j,\,{\rm g}}^{(2)}\,\mathcal{O}_j =\frac{b_{\rm g}^{(2)}}{2}(\delta_{\rm m}^2-\l\delta_{\rm m}^2\r)+\frac{s_{\rm g}}{2}(s^2-\l s^2\r)+d_{\rm g}\,\nabla^2\delta_{\rm m}\,,
\end{equation}
where $b_{\rm g}^{(2)}$ is the quadratic bias, $s_{\rm g}$ is the tidal bias, $s^2$ is the trace squared of the tidal tensor, and $d_{\rm gal}$ is the non-local bias.
The correlation between $a$ and $b$ (with possibly $a=b$) at second order is: 
\begin{align}
    \xi_{ab}&= b_{a}^{(1)}\,b_{b}^{(1)}\l\delta_{\rm m} \delta_{\rm m} \r+\sum_j \left(b_{j,\,a}^{(2)}\,b_{b}^{(1)}+b_{a}^{(1)}\,b_{j,\,b}^{(2)}\right)\l\delta_{\rm m}\, \mathcal{O}_j\r+\smallO{\textstyle\sum\l\delta_{\rm m}\,\mathcal{O}_j\r}\\
    &= b_{a}^{(1)}\,b_{b}^{(1)}\l \delta_{\rm m} \delta_{\rm m} \r \left(1+\sum_j\left(\frac{b_{j,\,a}^{(2)}}{b_{a}^{(1)}}+\frac{b_{j,\,b}^{(2)}}{b_{b}^{(1)}}\right)\frac{\l\delta_{\rm m}\,\mathcal{O}_j\r}{\l \delta_{\rm m}\delta_{\rm m}\r}+\smallO{\textstyle\sum\frac{\l\delta_{\rm m}\,\mathcal{O}_j\r}{\l \delta_{\rm m}\delta_{\rm m}\r}}\right)\,.
\end{align}
We note we did drop the $\theta$ index for clarity. Using this relation for the product of the auto-correlation, we get at second order,
\begin{align}
    \left(\xi_{aa}\,\xi_{bb}\right)^{-1/2}&=\left( b_{a}^{(1)}b_{b}^{(1)}\l \delta_{\rm m} \delta_{\rm m}\r\right)^{-1} \left(1+\sum_j\left(2\frac{b_{j,\,a}^{(2)}}{b_{a}^{(1)}}+2\frac{b_{j,\,b}^{(2)}}{b_{b}^{(1)}}\right)\frac{\l\delta_{\rm m}\,\mathcal{O}_j\r}{\l \delta_{\rm m}\delta_{\rm m}\r}+\smallO{\textstyle\sum\frac{\l\delta_{\rm m}\,\mathcal{O}_j\r}{\l \delta_{\rm m}\delta_{\rm m}\r}}\right)^{-1/2}\\
    &=\left(b_{a}^{(1)}b_{b}^{(1)}\l \delta_{\rm m} \delta_{\rm m}\r\right)^{-1} \left(1-\sum_j\left(\frac{b_{j,\,a}^{(2)}}{b_{a}^{(1)}}+\frac{b_{j,\,b}^{(2)}}{b_{b}^{(1)}}\right)\frac{\l\delta_{\rm m}\,\mathcal{O}_j\r}{\l \delta_{\rm m}\delta_{\rm m}\r}+\smallO{\textstyle\sum\frac{\l\delta_{\rm m}\,\mathcal{O}_j\r}{\l \delta_{\rm m}\delta_{\rm m}\r}}\right)\,.\\
\end{align}
Thus, there is no next-to-leading order correction due to non-linearities:
\begin{equation}
    \frac{\xi_{ab}}{\sqrt{\xi_{aa}\,\xi_{bb}}}=1+\smallO{\textstyle\sum\frac{\l\delta_{\rm m}\,\mathcal{O}_j\r}{\l \delta_{\rm m}\delta_{\rm m}\r}}\,.\label{eq:secondorder_1}
\end{equation}

\label{app:bias_next_order}

\section{Magnification modelling}\label{app:magnification}
\begin{figure*}
    \centering
    \includegraphics[width=1\linewidth]{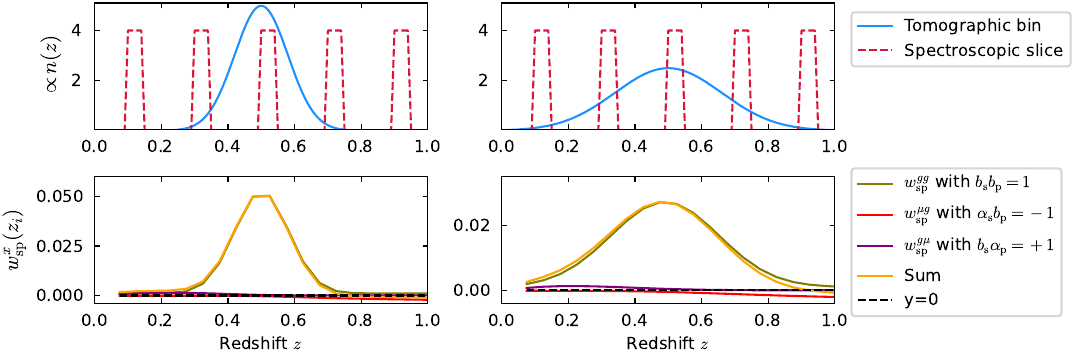}
    \caption{\textit{Top panel}: The redshift distributions of the photometric sample (blue) and five out of the 20 spectroscopic samples (red) are shown. The photometric distribution on the left follows a Gaussian with parameters $(\mu,\,\sigma) = (0.5,\,0.08)$, while on the right, it follows $(\mu,\,\sigma) = (0.5,\,0.16)$.
\textit{Bottom panel}: The angular correlation is displayed for clustering only (green), magnification only (red and purple), and their sum (orange). The values of the coefficients $b$ and $\alpha$ are provided in the legend. Magnification impacts the shape of the angular correlation (sum vs. clustering only) and may influence the mean redshift, particularly for larger tomographic bins (right vs. left).}
    \label{fig:wgg_wgmu}
\end{figure*}

Here we describe the modelling of the magnification terms cf. Eq. \eqref{eq:2magn_terms}.
The impact of the magnification change in flux depends on a parameter $s_{\mu}$ related to the selection function applied to the sample \citep{magnification_DESY3}. In the case of a magnitude threshold, it is the well-known derivative of the cumulative number count $N_{\mu}^{{\rm c}}$,
evaluated at the maximum magnitude,
\begin{equation}
    s_{\mu}= \frac{\diff}{\diff m}[\ln{N^{\rm c}_\mu}]_{m_{\rm max}}\,.\label{eq:s=dmu}
\end{equation}
Usually, changes in flux and solid angle are concatenated through the coefficient\footnote{There are different and confusing conventions in the literature. For instance, sometimes $\alpha$ is given as $s$, $2.5 s$, or $5s-2$.} $ \alpha=2.5s_{\mu}-1\,.$

Evaluating the $\alpha$s is in practice very challenging, since we are rarely considering only one magnitude cut as in Eq. \eqref{eq:s=dmu},  and it requires dedicated works \citep[e.g. cf. the case of two magnitude cuts in][]{Lepori24}, but is necessary for weak lensing \citep{hildebrandt_magn,magnification_DESY3}. 

The spectroscopic sample is magnified by the matter traced by low-$z$ photometric galaxies, and the matter traced by spectroscopic galaxies is magnifying high-$z$ photometric galaxies. 
Neglecting RSD, using the conventions from \cite{Y3_DES_Krause} and the Limber approximation, we have 
\begin{align}
    w_{\rm sp}^{\rm full}(\theta)
    &=\sum_{ij\in\{{\rm DD},\,{\rm D}\mu,\, \mu {\rm D}\}}c^i_{\rm s}\,c^j_{\rm p}\int \mathrm{d}\chi\; \frac{\mathcal{W}_i^{\rm s}\mathcal{W}_j^{\rm p}}{\chi^2}\,\sum_\ell \frac{2\ell+1}{4\pi}\mathcal{L}_\ell(\cos \theta)\,P_{\rm m}\left(k(\chi,\ell ),\,z(\chi)\right)\,,\label{eq:magn_WW}
\end{align}
where  $c_{a}^{\rm D}=b_{a}$ is the galaxy bias, $c_{a}^\mu=\alpha_a$ is the magnification coefficient, $\mathcal{L}_\ell$ is the Legendre polynomials of order $\ell$, $P_{\rm m}$ is the matter power spectrum, $k(\chi,\ell)=\frac{\ell+0.5}{\chi}$,  and 
\begin{align}
    &\mathcal{W}_{\rm D}^a=n_a(z)\frac{\diff z}{\diff\chi}\hspace{0.5cm} \text{and} \hspace{0.5cm}  \mathcal{W}_\mu^a=\frac{3\, \Omega_{\rm m}\,H_0^2}{2\,c^2}\int_\chi^{\infty}\mathrm{d}\chi^\prime \;n_a(\chi^{\prime})\frac{\chi}{a(\chi)}\frac{\chi^{\prime}-\chi}{\chi^{\prime}}\,.
\end{align}

These expressions are simplified if the spectroscopic sample is localised in a small redshift slice ($\Delta z$ small enough to neglect redshift variations) centred in $z_{\rm s}$. In that case, we can model the distribution of the photometric sample by a step function 
\begin{equation}
n_{\rm p}(z)=\sum_i n_{\rm p}(z_i)\; {\rm H}( \Delta z/2-\vert z-z_i\vert)\,,
\end{equation}
where ${\rm H}$ is the Heaviside function, equal to unity for $\Delta z/2-\vert z-z_i\vert>0$ and 0 otherwise. We then have
\begin{align}
    &w_{\rm sp}^{{\rm D}{\rm D}}(\theta) \approx  b_{\rm s}(z_{\rm s})\,b_{\rm p}(z_{\rm s})\, n_{\rm p}(z_{\rm s})\,\xi_{\rm m}(\chi_{\rm s})\,,\\
    &w_{\rm sp}^{{\rm D}\mu}(\theta)=\frac{3\,\Omega_{\rm m}\,H_0 }{c} b_{\rm s}(z_i)\,n_{\rm s}(z_i)\,\frac{H_0\,(1+z_i)}{H(z_i)}\, \xi_{\rm m}(z_i,\,\theta )\sum_{j=i+1,...} \Delta z \,\alpha_{\rm p}(z_j)\, n_{\rm p}(z_j)\,\frac{\chi(z_j)-\chi(z_i)}{\chi(z_j)}\,\chi(z_i)\,,\\
    &w_{\rm sp}^{\mu {\rm D}}(\theta)=\frac{3\,\Omega_{\rm m}\,H_0 }{c}\sum_{j=0,..i-1} \Delta z\, b_{\rm p}(z_j)\,n_{\rm p}(z_j)\,\frac{H_0\,(1+z_j)}{H(z_j)}\, \xi_{\rm m}(z_j,\,\theta )\,\alpha_{\rm s}(z_j)\, \frac{\chi(z_i)-\chi(z_j)}{\chi(z_i)}\,\chi(z_j)\,.
\end{align}
We can rewrite our magnification term as 
\begin{equation}
\begin{aligned}
    M(\theta)=\alpha_{\rm s}(z_{\rm s})\sum_{j=,...,i-1} \,\,&\bp(z_j)\,n_{{\rm p}}(z_{j})\,D(z_j,\,z_{ i},\,\theta)+\bs (z_i)\sum_{j=i+1,...}\,\alpha_{{\rm p}}(z_j)\,n_{{\rm p}}(z_j)\,D(z_i,\,z_j,\,\theta)\,,
\end{aligned}
\end{equation}
with 
\begin{align}
    D(z_k,\,z_l,\,\theta)=\frac{3\,H_0\,\Omega_{\rm m}}{c}\,\Delta z\,\frac{H_0}{H(z_k)}\,(1+z_k)\;\xi_{\rm m}( z_k,\,\theta)\,\frac{\chi(z_l)-\chi(z_k)}{\chi(z_l)}\,\chi(z_k)
    \,.
\end{align}
Our final expression contains several differences with \cite{Gatti_Giulia_DESY3}. We are taking into account the redshift variation of spectroscopic and photometric biases, and $\alpha$. Our conventions and the sum indices are slightly different (we try to be more explicit).

We illustrate the impact of magnification on angular correlation in Fig. \ref{fig:wgg_wgmu} using toy models. Two tomographic bins are introduced, with Gaussian redshift distributions centred at $z=0.5$ with widths $\sigma = 0.08$ and $\sigma = 0.16$, respectively. The clustering and the two magnification terms are evaluated using spectroscopic samples divided into 20 redshift slices with $\Delta z = 0.05$. We adopt $b_{\rm s} = b_{\rm p} = 1$, $\alpha_{\rm s} = 1$, and $\alpha_{\rm p} = -1$.

A noticeable difference is observed between clustering alone and the sum of the three effects, indicating that magnification is not entirely negligible. Since different $\alpha$ values are used for the spectroscopic and photometric samples, the combined effect appears to be shifted towards lower redshifts, potentially introducing a bias in the mean redshift. Finally, the impact is very limited for the narrower tomographic bin (left) but becomes more pronounced for the wider bin (right).

This highlights the importance of accounting for magnification effects when using broad tomographic bins.  With UNION and LSST photometry, our bins are projected to be narrower and more localised than DES ones, and thus magnification may be negligible for \Euclid clustering redshifts, even if it was not for DES.

\section{Pair counts}
\label{App:pair_count}

The pair-count between two samples is theoretically given by 
\begin{align}
{\rm D}_a{\rm D}_b(\rp)&=\iint \dz_1\dz_2\, \l N_a N_b \r_{\rp }=\iint \dz_1\dz_2\,N_a(z_1)\,N_b(z_2)\,\l (1+\delta_a)(1+\delta_b)\r_{\rp }^{\rm data}\,,\label{eq:pair_count}
\end{align}
where  $N(z)=N_{\rm tot} \, n(z)$ is the number galaxy distribution. We use galaxy weights to correct selection effects, anisotropies due to observation conditions or background etc. 

\subsection{From continuous to discrete}
In practice, the pair count is evaluated for a certain distance range $\rp \pm \Delta \rp$.  It means that for a galaxy of sample $a$, we are counting all the sample $b$ galaxies within an annulus with radius $\rp ^i$ and width $\Delta \rp ^i $. For instance, we use logarithmic binning, for which $\Delta \rp ^i\propto \rp ^i$. Thus, the correlation (including galaxy weights) of Eq. \eqref{eq:pair_count} is related to the theoretical one through,
\begin{align}
    \l (1+\delta_a)(1+\delta_b)\r_{r_{\rm p}^i}^{\rm data}= \int_0^{2\pi}\mathrm{d} \theta\, \int_{r_{\rm p}^i\pm \Delta r_{\rm p}^i/2}\mathrm{d} r_{\rm p}\l (1+\delta_a)(1+\delta_b)\r_{r_{\rm p},\, \theta} \approx 2\pi\, r_{\rm p}^i\,\Delta r_{\rm p}^i\,\l (1+\delta_a)(1+\delta_b)\r_{r_{\rm p}^i}\label{eq:2pirp}\,.
\end{align}

\subsection{Mask and estimator}
There is an additional geometrical effect to Eq. \eqref{eq:2pirp} which is the mask: we do not observe the full sky. The annuli are not perfect and there is a geometrical correction 
\begin{equation}
\l (1+\delta_a)(1+\delta_b)\r_{r_{\rm p}^i}^{\rm data}\approx 2\pi\, r_{\rm p}^i\,\Delta r_{\rm p}^i\,\eta_{\rm mask}(\rp)\,\l (1+\delta_a)(1+\delta_b)\r_{r_{\rm p}^i}\label{eq:2pirp_masked}\,.
\end{equation}
To correct it, we compare the pair count of the correlated samples, with the ones of uncorrelated samples, submitted to the same mask, the so-called ‘‘randoms’’,
\begin{equation}
    {\rm R}_a{\rm R}_b(\rp)=N_{a}^{\rm rand}N_{b}^{\rm rand}\,2\pi\, r_{\rm p}^i\,\Delta r_{\rm p}^i\,\eta_{\rm mask}(\rp)\,.
\end{equation}
Comparing different pair counts, we always correct the amplitude, which scales as the number of objects. We usually use 10 times more randoms than objects, so  we multiply $\rm DD/RR$  by 100. This step is done by default in \texttt{Treecorr}.

Computing random-galaxy pair counts, $\rm DR$ [which scales as $\,2\pi\, r_{\rm p}^i\,\Delta r_{\rm p}^i\,\eta_{\rm mask}(\rp)$ as well] is optimal for variance reduction \citep{Landy_Szalay}.
That is why our fiducial analysis involves the full Landy--Szalay estimator,
\begin{align}
    w_{ab}^{\rm data}(\rp )&=\frac{{\rm D}_a{\rm D}_b-{\rm R}_a{\rm D}_b-{\rm D}_a{\rm R}_b+{\rm R}_a{\rm R}_b}{{\rm R}_a{\rm R}_b}(\rp)\label{eq:fullw}=\l(1+\delta_a)(1+\delta_b)-(1+\delta_a)-(1+\delta_b)+1\r_{r_{\rm p}^i}=\l\delta_a\,\delta_b\r_{r_{\rm p}^i}\,,
\end{align}
where we omitted the galaxy number normalisation factors introduced below. 
A common choice for clustering-redshifts is to use the Davis and Peebles estimator
\begin{equation}
    w_{ab}(\rp )\approx\frac{{\rm D}_a{\rm D}_b-{\rm R}_a{\rm D}_b}{{\rm R}_a{\rm D}_b}(\rp)\label{eq:DP}\,,
\end{equation}
mainly because there is no need for a random catalogue for the second sample \citep{Davis_Peebles,Gatti_DESY1,van_den_Busch_2020}.  For \Euclid, a random catalogue would be created for the different analyses, so there is no motivation for using Eq. \eqref{eq:DP} rather than Eq. \eqref{eq:fullw}.

\section{Systematic errors and bias correction} \label{app:bin_size_sys}
\begin{table*}
\caption{Mean redshift and standard deviation bias for the different systematic effects (including none, and all) are reported;  Req. refers to the requirement that needs to be fulfilled. }
\begin{tabular}{cccc|cc}
\hline
& & & & &  \\[-9pt]
Bin &Configuration& \multicolumn{2}{c}{Shifted-Stretched Model} &\multicolumn{2}{|c}{Suppressed-GP}\\

&  & $\langle z \rangle -\langle z \rangle_{\rm true}$ ($\times 1000$) & $\sigma_z/\sigma_z^{\rm true}-1$ ($\times 100$)& $\langle z \rangle -\langle z \rangle_{\rm true}$ ($\times 1000$) & $\sigma_z/\sigma_z^{\rm true}-1$ ($\times 100$)  \\
 & & & & &  \\[-9pt]
 \hline
& & & & &  \\[-9pt]
Low-$z$ & \begin{tabular}[c]{@{}c@{}}Independent\\ Full bin \\ Full bin +RSD \\ Full bin +Magn \\ Total  \\ \textbf{Req.} \end{tabular}
& \begin{tabular}[c]{@{}c@{}}$-1.3\pm0.9$ \\ $-0.6\pm1.2$ \\ $-0.6\pm1.2$ \\ $-1.3\pm1.1$ \\ $-0.4\pm1.2$ \\ 2.8 \end{tabular}& \begin{tabular}[c]{@{}c@{}}$0.6\pm1.4$ \\ $1.6\pm1.7$ \\ $2.9\pm1.8$ \\ $0.4\pm1.6$ \\ $1.0\pm1.8$ \\ 10 \end{tabular}
& \begin{tabular}[c]{@{}c@{}}$0.7\pm1.9$ \\ $0.4\pm2.2$ \\ $1.0\pm2.2$ \\ $1.9\pm2.5$ \\ $2.4\pm2.6$ \\ 2.8 \end{tabular}& \begin{tabular}[c]{@{}c@{}}$-0.2\pm4.6$ \\ $1.2\pm4.9$ \\ $-2.1\pm4.8$ \\ $5.8\pm7.1$ \\ $5.6\pm7.3$ \\ 10 \end{tabular}\\
 & & & & &  \\[-9pt]
\hline
& & & & &  \\[-9pt]
Mid-$z$ & \begin{tabular}[c]{@{}c@{}}Independent\\ Full bin \\ Full bin +RSD \\ Full bin +Magn \\ Total  \\ \textbf{Req.} \end{tabular}
 & \begin{tabular}[c]{@{}c@{}}$1.4\pm0.6$ \\ $1.1\pm0.8$ \\ $1.8\pm0.9$ \\ $1.3\pm0.8$ \\ $2.2\pm0.8$ \\ 3.8 \end{tabular}& \begin{tabular}[c]{@{}c@{}}$2.2\pm0.8$ \\ $1.4\pm1.0$ \\ $2.9\pm1.2$ \\ $0.4\pm1.1$ \\ $2.3\pm1.0$ \\ 10 \end{tabular}
& \begin{tabular}[c]{@{}c@{}}$1.7\pm1.6$ \\ $3.8\pm1.9$ \\ $3.8\pm1.9$ \\ $3.8\pm1.7$ \\ $3.8\pm1.7$ \\ 3.8 \end{tabular}& \begin{tabular}[c]{@{}c@{}}$-0.1\pm3.6$ \\ $-2.2\pm3.9$ \\ $-5.3\pm3.7$ \\ $-0.7\pm3.7$ \\ $-3.1\pm4.0$ \\ 10 \end{tabular}\\
 & & & & &  \\[-9pt]
\hline
& & & & &  \\[-9pt]
High-$z$ & \begin{tabular}[c]{@{}c@{}}Independent\\ Full bin \\ Full bin +RSD \\ Full bin +Magn \\ Total  \\ \textbf{Req.} \end{tabular}    
& \begin{tabular}[c]{@{}c@{}}$-1.9\pm1.0$ \\ $-1.0\pm1.3$ \\ $-1.4\pm1.3$ \\ $1.0\pm1.2$ \\ $1.1\pm1.3$ \\ 4.4 \end{tabular}& \begin{tabular}[c]{@{}c@{}}$0.4\pm0.6$ \\ $0.7\pm0.8$ \\ $1.5\pm0.8$ \\ $-1.8\pm0.7$ \\ $-1.3\pm0.8$ \\ 10 \end{tabular}
& \begin{tabular}[c]{@{}c@{}}$-0.7\pm2.2$ \\ $0.2\pm2.5$ \\ $-0.6\pm2.3$ \\ $2.3\pm2.7$ \\ $2.2\pm2.6$ \\ 4.4 \end{tabular}& \begin{tabular}[c]{@{}c@{}}$1.1\pm1.9$ \\ $0.5\pm2.1$ \\ $-0.6\pm1.9$ \\ $4.0\pm2.4$ \\ $3.6\pm2.2$ \\ 10 \end{tabular}\\
 & & & & &  \\[-5pt]
\hline

\end{tabular}
\label{tab:bin_dev_0.05}
\end{table*}

\begin{table*}
\caption{Mean redshift and standard deviation for the different bias corrections M0--M5, for SSM and SGP fitting, without systematics (independent case in Sect. \ref{sub:Res_indep}); Req. refers to the requirement to fulfil. }
\begin{tabular}{cccc|cc}
\hline
 & & & & &  \\[-9pt]
Bin &Configuration& \multicolumn{2}{c}{Shifted-Stretched Model} &\multicolumn{2}{|c}{Suppressed-GP}\\
&  & $\langle z \rangle -\langle z \rangle_{\rm true}$ ($\times 1000$) & $\sigma_z/\sigma_z^{\rm true}-1$ ($\times 100$)& $\langle z \rangle -\langle z \rangle_{\rm true}$ ($\times 1000$) & $\sigma_z/\sigma_z^{\rm true}-1$ ($\times 100$)  \\ 
 & & & & &  \\[-9pt]
 \hline
 & & & & &  \\[-9pt]
Low-$z$ & \begin{tabular}[c]{@{}c@{}} M0 \\M1\\ M2\\M3\\M4\\M5  \\ \textbf{Req.} \end{tabular}
& \begin{tabular}[c]{@{}c@{}}$-0.3\pm1.4$ \\ $-1.1\pm1.0$ \\ $-1.1\pm1.0$ \\ $-1.2\pm1.6$ \\ $-1.5\pm1.6$ \\ $-1.3\pm0.9$ \\ 2.8 \end{tabular}& \begin{tabular}[c]{@{}c@{}}$0.8\pm2.0$ \\ $0.3\pm1.6$ \\ $0.1\pm1.5$ \\ $0.5\pm2.4$ \\ $0.3\pm2.4$ \\ $0.6\pm1.4$ \\ 10 \end{tabular}
& \begin{tabular}[c]{@{}c@{}}$1.0\pm2.1$ \\ $0.7\pm2.0$ \\ $0.7\pm1.9$ \\ $-0.1\pm2.4$ \\ $-0.5\pm2.4$ \\ $0.7\pm1.9$ \\ 2.8 \end{tabular}& \begin{tabular}[c]{@{}c@{}}$0.5\pm4.7$ \\ $0.0\pm4.6$ \\ $-0.0\pm4.2$ \\ $0.5\pm5.0$ \\ $0.8\pm5.1$ \\ $-0.2\pm4.6$ \\ 10 \end{tabular}\\
 & & & & &  \\[-9pt]
\hline
 & & & & &  \\[-9pt]
Mid-$z$ & \begin{tabular}[c]{@{}c@{}}M0 \\M1\\ M2\\M3\\M4\\M5\\\textbf{Req.} \end{tabular}
& \begin{tabular}[c]{@{}c@{}}$4.9\pm0.6$ \\ $2.7\pm0.6$ \\ $2.7\pm0.6$ \\ $1.4\pm0.4$ \\ $1.5\pm0.4$ \\ $1.4\pm0.6$ \\ 3.8 \end{tabular}& \begin{tabular}[c]{@{}c@{}}$2.7\pm0.8$ \\ $2.4\pm0.8$ \\ $2.4\pm0.8$ \\ $2.0\pm0.5$ \\ $2.0\pm0.5$ \\ $2.2\pm0.8$ \\ 10 \end{tabular}
& \begin{tabular}[c]{@{}c@{}}$5.2\pm1.6$ \\ $3.0\pm1.7$ \\ $2.9\pm1.7$ \\ $1.2\pm1.4$ \\ $1.3\pm1.4$ \\ $1.7\pm1.6$ \\ 3.8 \end{tabular}& \begin{tabular}[c]{@{}c@{}}$-0.7\pm3.4$ \\ $-0.2\pm3.7$ \\ $-0.1\pm3.7$ \\ $1.1\pm3.2$ \\ $1.1\pm3.2$ \\ $-0.1\pm3.6$ \\ 10 \end{tabular}\\
 & & & & &  \\[-9pt]
\hline
 & & & & &  \\[-9pt]
High-$z$ & \begin{tabular}[c]{@{}c@{}}M0 \\M1\\ M2\\M3\\M4\\M5\\ \textbf{Req.} \end{tabular}    
& \begin{tabular}[c]{@{}c@{}}$17.0\pm1.2$ \\ $5.1\pm1.0$ \\ $5.0\pm1.1$ \\ $-1.8\pm1.0$ \\ $-3.7\pm1.0$ \\ $-1.9\pm1.0$ \\ 4.4 \end{tabular}& \begin{tabular}[c]{@{}c@{}}$-5.6\pm0.7$ \\ $-1.8\pm0.7$ \\ $-1.8\pm0.7$ \\ $0.4\pm0.6$ \\ $1.1\pm0.6$ \\ $0.4\pm0.6$ \\ 10 \end{tabular}
& \begin{tabular}[c]{@{}c@{}}$17.8\pm2.6$ \\ $6.5\pm2.1$ \\ $6.5\pm2.3$ \\ $-0.4\pm2.1$ \\ $-2.3\pm2.0$ \\ $-0.7\pm2.2$ \\ 4.4 \end{tabular}& \begin{tabular}[c]{@{}c@{}}$3.9\pm2.1$ \\ $2.1\pm1.9$ \\ $2.1\pm1.9$ \\ $0.7\pm1.8$ \\ $0.3\pm1.8$ \\ $1.1\pm1.9$ \\ 10 \end{tabular}\\
 & & & & &  \\[-9pt]
\hline

\end{tabular}
\label{tab:dev_M05}
\end{table*}

In this appendix, we provide additional tables to Sects. \ref{sub:Res_indep} and \ref{sub:Res_bias_corr}. 
In Table \ref{tab:bin_dev_0.05}, we report the values of the deviations on the mean redshift and shape due to the systematic effects. The values are the means over the five sky patches, and correspond to $\Delta z=0.05$ and the M5 bias correction. We found similar results with other bias corrections. 
Finally in Table \ref{tab:dev_M05}, we report the values of the deviations on the mean redshift and standard deviation associated with the bias correction, in the absence of systematic effects.

\section{Mis-modelling of clustering-redshifts with the one-bin approximation}\label{app:m-bin}

\begin{figure*}
    \centering
    \includegraphics[width=1\linewidth]{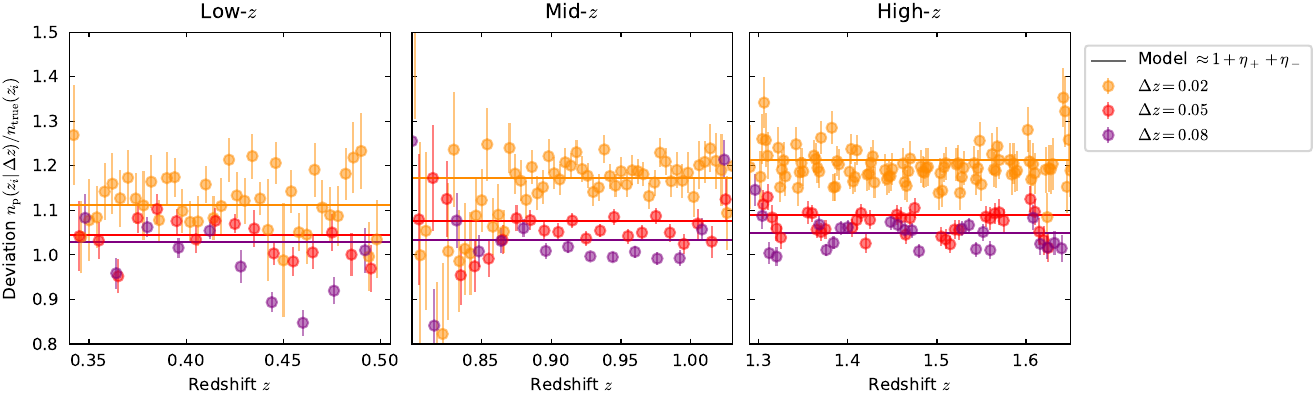}
    \caption{Deviation of the measured $n_{\rm p}(z_i)$ to the true distribution, for different slicing $\Delta z$ (points in colours), and the global offset predicted by our model, which includes correlation at the edge of the slices (solid lines in colours). The points correspond to the measurements for the five patches, hence the difference in redshift between them is not $\Delta z$. }
    \label{fig:Deviation_binning}
\end{figure*}

 In Sect. \ref{sec:comparison_corr}, we introduced an improvement to the modelling to expand beyond this one-bin approximation, which in the context of M5 bias correction is,
\begin{align}
    \frac{w_{\rm sp}}{\Delta z\sqrt{w_{\rm ss}\, w_{\rm pp}}}&(z_i)\approx n_b(z_i)\times \left( 1+\frac{n_b(z_i+\Delta z)}{n_b(z_i)}\,\eta_+(z_i)+\frac{n_b(z_i-\Delta z)}{n_b(z_i)}\,\eta_-(z_i) \right)\,.\nonumber
\end{align}
Since $\eta_\pm>0$, we expect to measure $n(z)$ with a norm higher than unity. The global offset should be approximately
\begin{equation}
    \sum_{z_i}\frac{w_{\rm sp}}{\Delta z\sqrt{w_{\rm ss}\, w_{\rm pp}}}(z_i)\approx 1+\eta_+(\l z \r)+\eta_-(\l z \r)\,.\label{eq:pred_offset}
\end{equation}
In Fig. \ref{fig:Deviation_binning} we show the ratio of the measured $n_{\rm p}(z_i)$ to the true distribution, for three slicing $\Delta z$, and the five sky patches. We also report the offset predicted from Eq. \eqref{eq:pred_offset} for each $\Delta z$. We observe consistency between the points and the rough prediction.   Concatenating the measurements from the different redshift slices, we can express the correlation as
\begin{equation}
    \Vec{w}_{\rm sp}(\theta)=B_{\rm s}\,B_{\rm p}\,\Lambda \, \Vec{n}\,,\label{eq:implement_mbins_n}
\end{equation}
where $\Vec{n}=\{n(z_i)\}_{i=1,..,n}$, $B_x$ are diagonal matrices representing the galaxy biases, and $\Lambda$ is a tridiagonal matrix, easily invertible, with unity on its diagonal and smaller terms (corrections) on the first off-diagonals.  We attempted to implement Eq. \eqref{eq:implement_mbins_n} for the inference of $n_{\rm p}(z)$ by replacing the one-bin modelling with three-bins. We did not observe a significant improvement in the bias on the mean redshift compared to the standard renormalisation procedure, however. In Fig. \ref{fig:Deviation_binning}, we note that points from different patches at similar redshifts (visible as groups of 5 points corresponding to the 5 patches in the high-$z$ bin) are spread over a wide range, despite the true distributions being very similar across the patches. This behaviour could be attributed to substantial sample variance, which might explain why three-bins perform similarly to one-bin. Additionally, we observe that three-bins still exhibit significant deviations in Fig. \ref{fig:Toy_model_wab}, highlighting the challenges of improving one-bin modelling despite our efforts. We note that in Fig. \ref{fig:nz_diff_bin_eta}, the measurements for $0.75<z<0.85$ are systematically under the true $n(z)$ for $\Delta z=0.02,\,0.05$, which might cause the positive shift in mean redshift.  This under-correlation is also visible in Fig. \ref{fig:Deviation_binning}. We did not identify its cause.

\end{appendix}

\end{document}